\documentclass{jfm}
\usepackage{graphicx}
\usepackage{newtxtext}
\usepackage{newtxmath}
\usepackage{natbib}
\usepackage{hyperref}
\hypersetup{
    colorlinks = true,
    urlcolor   = blue,
    citecolor  = black,
}

\newcommand{\RomanNumeralCaps}[1]
\linenumbers

\usepackage{booktabs}
\usepackage{color}
\usepackage{caption}
\usepackage{mathtools}
\usepackage{ amssymb }
\usepackage{subcaption}

\usepackage{float}
\usepackage{bm}
\usepackage{centernot}
\usepackage[OliveGreen,dvipsnames]{xcolor}

\newcommand{\bU}{{\bf U}}
\newcommand{\bC}{{\bf C}}

\newcommand{\bu}{{\bf u}}
\newcommand{\bfu}{{\bf \tilde{u}}}
\newcommand{\bhu}{{\bf \hat{u}}}
\newcommand{\tu}{\tilde{u}}
\newcommand{\tv}{\tilde{v}}
\newcommand{\tw}{\tilde{w}}
\newcommand{\tp}{\tilde{p}}

\newcommand{\bx}{{\bf x}}
\newcommand{\bhx}{{\bf \hat{x}}}
\newcommand{\bhy}{{\bf \hat{y}}}
\newcommand{\bhz}{{\bf \hat{z}}}
\newcommand{\bnab}{\bm{\nabla}}

\newcommand{\mq}{\overline{q}}

\newcommand{\fq}{q^\prime}

\newcommand{\dd}{\delta}
\newcommand{\D}{\Delta}
\newcommand{\bd}{\bm{\delta}}
\newcommand{\optL}{{\mathbb L}}
\newcommand{\J}{{\mathbb J}}

\newcommand{\half}{\tfrac{1}{2}}
\def\rk#1{{\textcolor{black}{#1}}}
\def\vm#1{\textcolor{black}{#1}}

\title{Improved assessment of the statistical stability of turbulent flows using extended Orr-Sommerfeld stability analysis}

\author{Vilda K. Markeviciute\aff{1}
   \and
  Rich  R. Kerswell\aff{1}\corresp{\email{r.r.kerswell@damtp.cam.ac.uk}}
 }

\affiliation{\aff{1}Department of Applied Mathematics and Theoretical Physics, University of Cambridge, Wilberforce Rd, Cambridge CB3 0WA, UK}

\begin{document}

\maketitle

\begin{abstract}

The concept of statistical stability is central to Malkus's 1956 attempt to predict the mean profile in shear flow turbulence. Here we discuss how his original attempt to assess this - an Orr-Sommerfeld analysis on the mean profile - can be improved by considering a cumulant expansion of the Navier-Stokes equations. Focusing on the simplest non-trivial closure (commonly referred to as CE2) which corresponds to the quasilinearized Navier-Stokes equations, we develop an extended Orr-Sommerfeld analysis (EOS) which also incorporates information about the fluctuation field. A more practical version of this  - minimally extended Orr-Sommerfeld analysis (mEOS) - is identified and tested on a number of statistically-steady and therefore statistically stable turbulent channel flows. Beyond the concept of statistical stability, this extended stability analysis should also improve the popular approach of mean-flow linear analysis in time-dependent shear flows by including more information about the underlying flow in its predictions as well as for other flows with additional physics such as convection.

\end{abstract}



\section{Introduction}

%
%
This paper revisits the seminal work of Malkus (1956) which attempted to build a theory of shear turbulence. This theory was based upon  maximising the momentum transport (or equivalently dissipation rate) achieved by the flow amongst all those with a marginally-stable mean profile. Malkus clearly had a statistical form of marginal stability in mind but, to make progress, had to resort to specifying marginality with respect to the then 50-year-old Orr-Sommerfeld equation \citep{Orr1907, Sommerfeld1908}. So posed, his marginal stability idea was quickly repudiated (Reynolds \& Tiederman 1967 and more recently Iyer {\it et al.} 2019) although further studies showed it could be made to work if {an anisotropic eddy viscosity model was used} \citep{ReynoldsHussain72, Sen00,Sen07,Malkus79}.  The concept of statistical stability was, however, central to Malkus's thinking (remaining so throughout his career, e.g. Malkus 1996 and 2003) and is clearly different from stability as viewed within the context of the governing Navier-Stokes equations (epitomised by the celebrated Orr-Sommerfeld equation). 
For example, it is fairly uncontentious to assert that the turbulent attractor in, say, pressure-driven channel flow {at high enough Reynolds number and large enough domain} has stationary  statistics (defined by averaging over one or more  homogeneous directions or in an ensemble sense) and so within the partial differential equations which govern how these statistics evolve, the realised turbulence is a stable fixed point - i.e. turbulence is {\em statistically} stable to infinitesimal perturbations of the statistics. This is in contrast to the time-dependent turbulent attractor as viewed in the Navier-Stokes equations where adding a small disturbance may well see that disturbance grow and never decay yet the original statistics still recover (an example is shown herein). The difference, of course, is that a flow disturbance can have a component along the turbulent trajectory and hence acts as a time-shift: this part of the disturbance never decays to zero but does not affect  the statistics \cite[e.g.][]{Nikitin18}. 
%
%
For his theory, Malkus wanted a statistical stability criterion based only on the lowest order statistic - the mean flow. Ideally more statistical information needs to be incorporated and it is our objective here to attempt this.  
At the very least, doing so should improve the now-standard approach of carrying out linear stability analysis of the mean profile of time-dependent flows in an attempt to understand observed coherent structures \cite[e.g.][]{CrightonGaster76, Gaster85, Roshko93, Lesshafft06,
Barkley06, SippLebedev07,Akervik08,Sipp10, Mantic-Lugo14, Beneddine16,Lefauve18}. 

%
%
The motivation for this work comes from two different directions. The first is the general approach of applying linear analysis around the  mean profile of a time-varying, possibly turbulent flow to deduce information about the likely fluctuations seen. Initially, this took the form  of linear stability analysis stimulated by  Malkus's work  which tends to work well in free-shear turbulent flows like jets where inviscid (inflectional) instabilities dominate \citep{CrightonGaster76,Gaster85,Roshko93} but less well in wall-bounded situations {such as channel flow} where viscosity can be important \citep[although there have been successes e.g.][]{Barkley06,SippLebedev07,Beneddine16}. Driven by the fact that shear flow mean profiles tend to be linearly stable, this approach subsequently diversified into non-modal analysis \citep{Butler93, DelAlamo06, Cossu09}  and input-output or resolvent analyses \citep{Jovanovic05, Cherny05, HwangCossu10a, HwangCossu10b, McKeonSharma10, Moarref12, Blesbois13, SharmaMcKeon13, Schmidt18}. The resolvent approach has been particularly illuminating in showing exactly how far linear analysis can go given the mean flow profile \citep{McKeon17, Jovanovic21}. Predictions can be made of the dominant fluctuation response at a given frequency and wavevector which typically resonates with observations at least up to amplitude and phase. In some sense this solves `half' the problem of turbulence - given a mean flow, linear analysis around this can extract the dominant  fluctuation structures - and refocuses attention on the `other' half - predicting  the mean profile. Identifying the required amplitudes and phases of the all fluctuation fields to support the observed mean profile, however, may approach the difficulty of 'just' solving the Navier-Stokes equations. One alternative is to appeal to  some simpler `organising' physical principle as epitomised in \cite{Malkus56} and then the idea of statistical stability seems a key concept. 

\begin{figure}
 \centerline{ \includegraphics{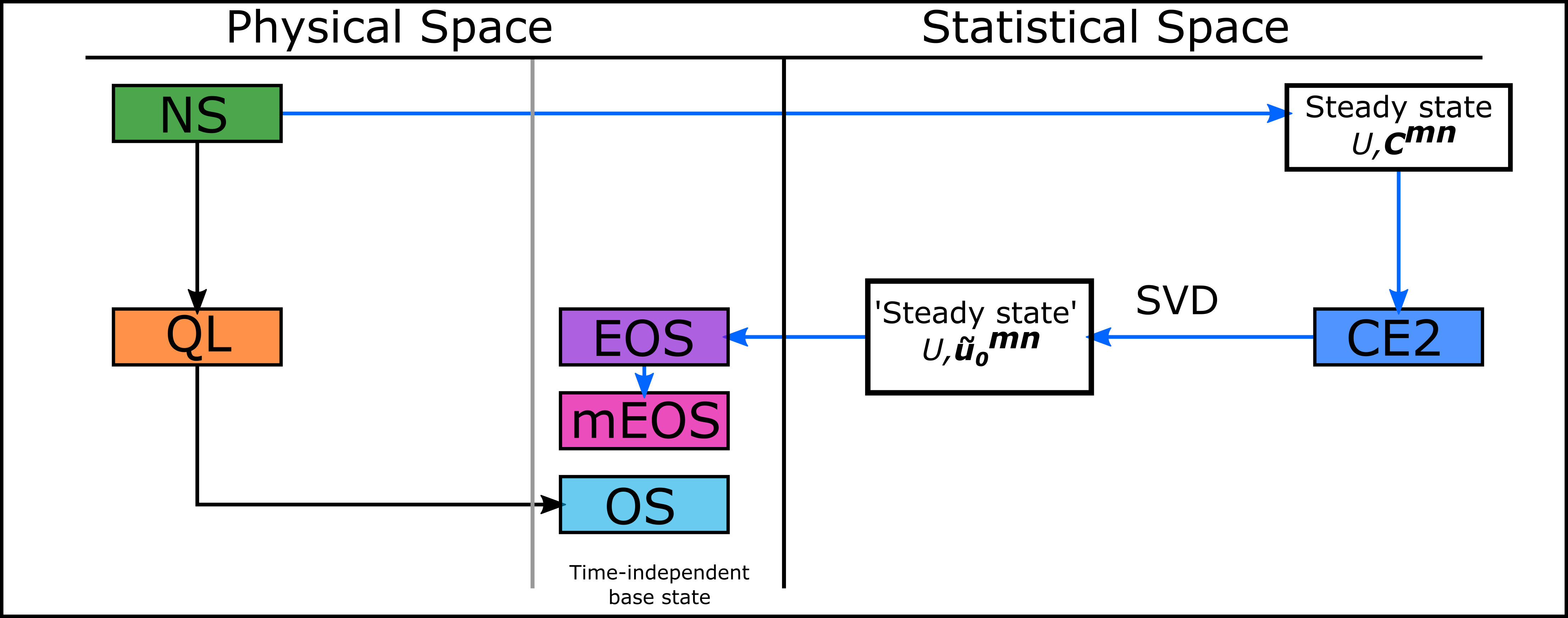}}%
 \caption{\label{fig:1} Comparison between standard and statistical considerations of linear stability of turbulent states. On the left are physical space equations (methods are ordered top to bottom by decreasing non-linearity): Full Navier-Stokes equations (NS), Quasilinear approximation (QL), Orr-Sommerfeld equation around a turbulent mean velocity profile (OS), Extended and minimal Extended Orr-Sommerfeld equations (EOS and mEOS). On the right are statistical space equations: CE2 equations which include up to second order statistics. Black arrows indicate the standard path of turbulent flow stability analysis leading to OS; blue arrows indicate the statistical approach to turbulent flow stability analysis leading to EOS and mEOS, emphasising how the steady statistical state $U,\bm{C^{mn}}$ can be used to obtain the steady state counterpart $U,\bm{\tilde{u}_0^{mn}}$ in the physical space.  
}
 \end{figure}

%
%
The second motivation is the resurgence of interest in dealing with statistical quantities directly through a cumulant expansion of time-varying flows \citep{Hopf52, Orszag73, Frisch95}. Generating the evolution equations for cumulants immediately highlights the closure problem of turbulence in which the time-derivative of a $n$th-order cumulant requires knowledge of the $(n$+$1)$th cumulant and so forth.  Present-day computational power has, however, renewed interest in the pursuit of simple cumulant-discard closures to see how well they do in modelling flows in a variety of contexts: e.g. in atmospheric dynamics \citep{FarrellIoannou03, Marston08, SrinivasanYoung12, Tobias13, ParkerKrommes14}; astrophysics \citep{Tobias11}; plasmas \citep{FarrellIoannou09,ParkerKrommes13} and wall-bounded shear flows \citep{FarrellIoannou12, Constantinou14, Farrell16}. The most popular closure ignores third and higher order cumulants - commonly referred to as CE2 - and has the nice property of being exact for a quasilinear (or historically a `mean field theory')  version of the Navier-Stokes equations \citep[e.g. see the recent review by][]{MarstonTobias22}.  Significantly for our purposes here, CE2 and higher closures (CE$n$ where $n \geq 3$ is the highest order cumulant retained)  present the most natural framework in which to extend Malkus's idea of statistical stability. In what follows, the focus will be on using CE2 and its relationship to the quasilinear version of the Navier-Stokes equations to develop an improved version of the usual Orr-Sommerfeld analysis (or `OS analysis' for short) of the mean profile: see Figure \ref{fig:1}. {The overall objective is to develop a way to judge whether a state is statistically stable or not using a subset of its statistics.}

%
%

It is worth emphasizing that the approach taken here is perfectly general and not confined to shear flows or even fluid mechanics. The key idea given a physical system described by a PDE is to generate evolution equations - `statistical' equations - for the first few statistics of the solution. If the solution of the original PDE is spatiotemporally complicated  but has steady statistics, the premise pursued here is that a better way to assess the stability of this solution is to look at the linear stability {\em within} the statistical equations rather than the original PDEs. The former is an approximation given a finite number of the system's statistics are considered but yields a spectral problem built upon a steady state in statistical space. The latter has to rely on a costly ensemble of simulations where a distribution of small perturbations are each added to the solution and their evolution monitored \citep{Iyer19,Duguet22}. A good complementary example of where this approach would be useful is convection \citep{Malkus54, Wen22}.

%
%

%
%
The plan of this paper is to illustrate the analysis within the context of channel flow described in section \ref{II}. The Reynolds number is assumed high enough for the  computational box used that the flow can be taken as approximately statistically stationary. 
\rk{An averaging procedure will be assumed below such that the mean flow can be assumed to only depend on the cross-stream variable $y$ i.e. $\bU=U(y)\bhx$. This could be ensemble averaging or averaging over the streamwise ($x$) and spanwise ($z$) directions.}
\rk{Using the latter spatially-averaging approach}, Malkus pointedly only chose a spanwise average so that  his mean profile $\bU=U(x,y) \bhx$ could depend on the cross-stream {\em and} streamwise variables. As a result, his Orr-Sommerfeld analysis targeted the stability of a streamwise-independent mean profile $U(y)\bhx$ to streamwise-dependent {\em mean flow} disturbances $\delta \bU(y) \exp(ik(x-ct))$ rather than fluctuation fields defined as having non-vanishing spanwise dependence. Contrarily, there are growing arguments to only streamwise average to retain the  spanwise structure of the mean flow, i.e.  $\bU=U(y,z)\bhx$ (e.g. see Table 1 of \cite{Lozano-Duran21} for a sample list of relevant works). Despite this, the focus here is on the simplest mean flow definition {for a statistically steady state, $\bU=U(y)\bhx$,} given the central role this plays in resolvent flow analysis but, there's no doubt, extending the mean flow definition is clearly an important direction to extend the approach discussed here. Standard Orr-Sommerfeld analysis is recalled in section \ref{II.A}.

Section \ref{II.B} then introduces the cumulant expansion approach \citep{Hopf52,Orszag73,Frisch95} and the  hierarchy of evolution equations for these statistical quantities. Solving these  equations to (hopefully) reach a steady state is an appealingly direct way to estimate the properties of statistically steady turbulent flows since it avoids having to average across large DNS-generated datasets. 
Here, however, the emphasis is on the concept of stability in this statistical framework and its relationship to (dynamic) stability within the Navier-Stokes equations (see section \ref{II.C}) {\em not} on the accuracy of suitable closures of the cumulant equations in capturing the reference flow state. {That is, our strategy is to apply  a statistical stability criterion from a statistical closure to a solution of the Navier-Stokes equations to approximate its statistical stability there. } 
Our particular focus will be on the simplest non-trivial cumulant-discard scheme CE2  given the rapidly increasing dimensionality of the approach: CE$n$ works with cumulants up to order $n$ which, before exploiting any symmetries, is typically a  $3n$-$2$ rank tensor (3 spatial coordinates per field reduced by 2 averaging directions). CE2 is exact for the quasilinearized Navier-Stokes equations - or QL equations - and translating what a statistically-steady state in CE2 means for the QL equations is a crucial step discussed in \S\ref{II.D}. Appendix \ref{appendixA} presents an  equally important discussion on how the stability predictions  within the CE2 and QL systems are related. Quasilinearization \citep{Vedenov61, Herring63, Herring64}  has enjoyed a resurgence of interest recently \cite[e.g.][]{Hernandez20, Skitka20, OConnor21, MarstonTobias22} given its accessibility, and together with its `generalised' elaboration \citep{Marston16} in which the definition of what constitutes a mean is  extended, has the ability to focus on different parts of the nonlinearity in the Navier-Stokes equations \cite[e.g.][]{Hwang21}. 

Section \ref{III.A} then introduces an extended version of Orr-Sommerfeld analysis - or `EOS' analysis -  based on translating the statistical stability problem in CE2 back to the dynamical equations. Intriguingly, this is {\em not} the same as just working within the QL approximation (as Appendix \ref{appendixA} makes clear). Applying EOS analysis carries a substantial overhead so we consider a reduced (practical) version referred to as `minimally extended Orr-Sommerfeld analysis' - or `mEOS' analysis - in section \ref{III.B} which is almost as cheap to apply as OS analysis. 

Interestingly, applying the same strategy of mapping a cumulant-based system back to the underlying dynamical equations can only go one level higher in sophistication and requires a jump directly to  CE$\infty$. This transforms EOS into the familiar  linearised Navier-Stokes equations albeit based around the steady base state derived by assuming stationary 2nd rank cumulants. This `infinitely extended Orr-Sommerfeld' analysis is described in \S\ref{III.B}. Section \ref{IV} then explores the performance of OS, EOS and mEOS analyses on 4 different turbulent states realised in 2D channel flow which {is used as an approximation of a statistically steady flow}. The limitations of the analysis are discussed in \S\ref{V} followed by a summary and  final thoughts  in section \S\ref{VI}.



%
%
\section{\label{II}Formulation: channel flow}

For context in this work we consider channel flow $\bu^*(\bx^*,t^*)$ of a fluid with density $\rho^*$ and kinematic viscosity $\nu^*$ between two parallel plates at $y^*=\pm h^*$ across which a \rk{time-dependent} pressure  gradient $9\rho^* U^{*2}G(t^*)/4h^*\,\bhx$ is imposed such that the bulk flow 
\begin{equation}
U^*:=  \frac{1}{4h^{*2}L_z} \int^{h^* L_z}_{-h^*L_z}\int^{h^*}_{-h^*} \,\bu^*\, dy^* dz^*
\end{equation}
is kept constant (unstarred/starred quantities are dimensionless/dimensional and periodicity is imposed across the spanwise domain $z^* \in h^*[-L_z,L_z]$). Non-dimensionalizing the Navier-Stokes equations using $h^*$, $3U^*/2$ (so Reynolds numbers based on the bulk speed and the laminar centreline speed $U^{*c}=3U^*/2$ correspond for the laminar parabolic flow) and $\rho^*$ leads to 
\begin{equation}
\begin{aligned}
\bu_t+ \bu \cdot \bnab \bu &= G(t)\bhx-\bnab p +\frac{1}{Re} \nabla^2 \bu,\\
\bnab \cdot \bu &= 0
\end{aligned}
\label{eq:NSfull}
\end{equation}
with $\frac{1}{4L_z}\int^{L_z}_{-L_z}\int^1_{-1} u\, dydz=2/3$ where $\bu=\bu^*/U^*=u \bhx+v \bhy+w \bhz$, $t=3U^*/(2h^*) t^*$ and $Re:=3U^* h^*/2\nu^*$.
We also impose streamwise periodicity of the flow over $x^* \in h^*[-L_x,L_x]$ so the  non-dimensionalised flow domain is $(x,y,z) \in [-L_x,L_x] \times [-1,1] \times [-L_z,L_z]$ with non-slip boundary conditions on the plates at $y=\pm 1$  (fundamental wavenumbers in $x$ and $z$ are labelled $\alpha:=\pi/L_x$ and $\beta:=\pi/L_z$ respectively). 
%
%
\rk{In terms of an averaging procedure, a number of choices present themselves: ensemble averaging, spatial averaging and time averaging (or even a combination thereof) which all should be equivalent for a statistically stationary system in a large enough domain. However, in what follows we want to treat a numerical experiment in a finite domain with finite $Re$ and then ensemble averaging is the most natural choice as will be come clear below. } This averaging process is denoted by an overbar, $\overline{(\,\cdot\,)}$,
%
%
and then the  flow can be decomposed into a mean $U(y,t) \bhx:=\overline{\bu(\bx,t)}$ and fluctuation part $\bfu:=\bu-\overline{\bu}=\tu \bhx+\tv \bhy + \tw \bhz$ (due to symmetry, a vanishing mean spanwise component is assumed so $( \,\overline{\tilde{v} \tilde{w}} \,)_y=0$). 
The Navier-Stokes equations can be similarly decomposed into a mean part, 
\begin{equation}
     U_t -\frac{1}{Re}U_{yy} = G - ( \,\overline{\tilde{u} \tilde{v}} \,)_y
     \label{U}
\end{equation}
and a fluctuation part,
\begin{equation}
 \bfu_t  = \frac{1}{Re} \nabla^2 \bfu -\bnab \tilde{p} -{U} \bfu_x - \tilde{v} U_y \bhx -(\bfu \cdot \bnab \bfu -\overline{\bfu \cdot \bnab \bfu}),
    \label{fu}
\end{equation}    
which is incompressible
\begin{equation}
\bm{\nabla}\cdot \bfu = 0  \label{divu}
\end{equation}
where subscripts denote derivatives (e.g. $U_y:=dU/dy$). 

%
%
\subsection{\label{II.A}Orr-Sommerfeld stability analysis}

Given a possibly turbulent flow $(U,\bfu)$, the `standard' linear stability analysis is to consider small (a.k.a. infinitesimal) perturbations $(0, \delta \bfu)$ to a base state $(U,{\bf 0})$ where the fluctuation field is ignored and the mean flow $U$ is assumed steady. As a result only the equations (\ref{fu}) and (\ref{divu}) need be perturbed (and hence linearised) and since $U=U(y)$, the ensuing eigenvalue calculation is parameterised by a streamwise and spanwise wavenumber. Squire's theorem \cite{Squire33} is usually invoked to focus the search for instability to spanwise-independent perturbations and leads to the celebrated Orr-Sommerfeld equation \citep{Orr1907,Sommerfeld1908}. In primitive variables, $\tw$ then decouples from $\tu$ and $\tv$ and can be ignored, leaving the reduced eigenproblem
%
\begin{align}
-im \alpha c\, \tu & = \frac{1}{Re}(\tu_{yy}-m^2\alpha^2 \tu)-im \alpha \tp -im \alpha U\tu-U_y \tv, \label{(7)}\\
-im \alpha c \,\tv & = \frac{1}{Re}(\tv_{yy}-m^2 \alpha^2\tv)-\tp_y -im \alpha U\tv,         \label{(8)}\\ 
      0 & = im \alpha \, \tu+\tv_y \label{(9)}
\end{align}
%
where $(\tu,\tv,\tp) \propto e^{im \alpha (x-ct)}$ and $c:=c_r+ic_i$ is the (complex) eigenvalue. The Orr-Sommerfeld equation is then reached by defining a  streamfunction $\psi$ (so $\tu=\psi_y$ and $\tv=-im \alpha \psi$) and eliminating the pressure so that 
\begin{equation}
(U-c)(\partial_y^2-m^2 \alpha^2)\psi-U_{yy} \psi=\frac{1}{i m \alpha Re} (\partial_y^2-m^2 \alpha^2)^2 \psi.
\end{equation}
In what follows, we actually work with the primitive variable formulation (\ref{(7)})-(\ref{(9)}) as it is clearer to interpret the origin of new terms added below, the eigenvalue problem is better conditioned (2 equations with  second order operators as opposed to one with a fourth order operator) {\em and} it is easily extended to 3D if needed (for 3D disturbances, the Orr-Sommerfeld equation must be augmented with Squire's equation for the cross-stream vorticity).
Nevertheless, we refer to this general approach of doing linear stability around a {turbulent} mean as `Orr-Sommerfeld analysis' - or `OS analysis' for short - in recognition of its conception in  Malkus's work.

The (matrix) size of the eigenproblem (\ref{(7)})-(\ref{(9)}) is only $ 3N_y \times 3N_y$ for each streamwise wavenumber $m$ and so needs to be repeated $N_x$ times ($N_x,N_y$ represent the streamwise and wall-normal truncations). When the mean profile is known from simulations or experiments, for typical resolutions, this is an easily accessible procedure which,  through the structure of the most unstable eigenvectors, can shed some light on the dominant structures of the turbulent flow.

A popular extension to the standard OS analysis involves using an eddy viscosity $E(y)$ instead of the molecular viscosity. An eddy viscosity can be determined self-consistently from the Reynolds stress needed to sustain the turbulent mean profile (e.g. \cite{ReynoldsTiederman67}) using the steady version of (\ref{U}):
\begin{equation}
\frac{1}{Re} [(\,1+E(y)\,)U_y]_y:=\frac{1}{Re}U_{yy}- ( \,\overline{\tilde{u} \tilde{v}} \,)_y=-G
\, \Rightarrow \,
E(y) := -Re \frac{  \overline{ \tilde{u} \tilde{v}} }{U_y}
=
-\frac{Re}{U_y} \int^y_0 G(\bar{y})d\bar{y}-1
\label{eq:eddy_viscosity}
\end{equation}
Then, the eigenproblem can be modified to:
\begin{align}
-im \alpha c\, \tu & = \frac{1}{Re} \left(    \left[(1+E)\tu_{y} \right]_y-m^2\alpha^2 (1+E) \tu \right)-im \alpha \tp -im \alpha U\tu-U_y \tv, \\
-im \alpha c \,\tv & = \frac{1}{Re}\left(    \left[(1+E)\tv_{y} \right]_y-m^2\alpha^2 (1+E) \tv \right)-\tp_y -im \alpha U\tv,         \\ 
      0 & = im \alpha \, \tu+\tv_y.
\end{align}
We will use this to assess the performance of our extended Orr-Sommerfeld analysis in \S\ref{IV.C}.

%
%
\subsection{\label{II.B}Statistics: cumulants}

In this section we consider a statistical framework for the flow  by working with the equal-time cumulants of the flow \citep{Hopf52,Orszag73, Frisch95}. Even within this framework, we specialise further to exclusively consider equal-$x$-and-$z$ cumulants which are the subset of cumulants which influence the mean flow. (In fact, only equal-$y$ cumulants are needed in  the mean flow equation - see (\ref{C-U}) below. However, the evolution equations for these are not available without also solving for `non-equal' $y$ cumulants.) The first cumulant is the mean $U(y,t)$. The second cumulant is the symmetric matrix
\begin{equation}
\bC(y_1,y_2,t):= \overline{\bfu(x,y_1,z,t) \otimes \bfu(x,y_2,z,t)}
=\left(\begin{array}{ccc}
     C_{11} & C_{12} & C_{13} \\
     C_{21} & C_{22} & C_{23} \\
     C_{31} & C_{32} & C_{33} 
\end{array} \right),
\end{equation}
where we introduce the notation $C_{ij}(1,2):= \overline{[\bfu(x,y_1,z,t)]_i [\bfu(x,y_2,z,t)]_j}\quad$ (here $[\bfu]_1=\tilde{u}$, $[\bfu]_2=\tilde{v}$ and $[\bfu]_3=\tilde{w}$)  to emphasize the $y$ arguments and de-emphasize the implicit time dependence, and the third cumulant is the third order tensor
\begin{equation}
C_{ijk}^{(3)}(1,2,3):=\overline{[\bfu(x,y_1,z,t)]_i [\bfu(x,y_2,z,t)]_j [\bfu(x,y_3,z,t)]_k}.
\end{equation}
These correspond to the second and third central moments of the flow respectively. Cumulants and central moments, however, diverge  at fourth order and beyond, e.g.
%
\begin{align}
C_{ijkl}^{(4)}(1,2,3,4)&:=
\overline{ [\bfu(x,y_1,z,t)]_i [\bfu(x,y_2,z,t)]_j [\bfu(x,y_3,z,t)]_k [\bfu(x,y_4,z,t)]_l
}
\\
& \qquad 
-C_{ij}(1,2)C_{kl}(3,4)
-C_{ik}(1,3)C_{jl}(2,4)-C_{il}(1,4)C_{jk}(2,3).
\end{align}
%
We will not go this high in the cumulant expansion used here  but just note that the $n^{th}$ order cumulant is $n$-dimensional in space so that storage when doing computations becomes prohibitive very quickly. Hence the onus is on applying some sort of closure as soon as possible. 

%
%
To derive evolution equations for the cumulants, we introduce a double Fourier series representation of the flow
\begin{equation}
\bfu(x,y,z,t):= \sum_m \sum_n \bfu^{mn}(y,t) e^{im \alpha x+in \beta z}
\end{equation}
where $m,n \in {\mathbb Z}$. Clearly $\bfu^{-m-n}=\bfu^{*mn}$ (the complex conjugate of $\bfu^{mn}$) for a real flow but it will be clearer not to build this into the notation in anticipation of deriving perturbation equations later. Hence, we write
\begin{align}
C_{ij}(1,2) &= \sum_m \sum_n \biggl\{ C_{ij}^{mn}(1,2):= [\bfu^{mn} (y_1,t)]_i [\bfu^{-m-n} (y_2,t)]_j \biggr\} \label{C2}\\
C^{(3)}_{ijk}(1,2,3) &= \sum_m \sum_n \biggl\{ C_{ijk}^{(3)mn}(1,2,3)
:=\sum_p \sum_q [\bfu^{mn}(y_1,t)]_i [\bfu^{pq}(y_2,t)]_j \nonumber\\ & \hspace{6cm} \times [\bfu^{-(m+p)-(n+q)}(y_3,t)]_k \biggr\} \label{C3}
\end{align}
(note e.g. $C_{ij}^{mn}(1,2)=C_{ji}^{-m-n}(2,1)$\,).
Equations to evolve the cumulants are obtained by temporally differentiating their definitions in (\ref{C2}) and (\ref{C3}) and using (\ref{fu}). For example, for the second order cumulant,
\begin{align}
\partial_t C_{ij}^{mn}(1,2) & = [\bfu^{mn}(y_1,t)]_i \partial_t  [\bfu^{-m-n}(y_2,t)]_j    
             +\partial_t [\bfu^{mn}(y_1,t)]_i  [\bfu^{-m-n}(y_2,t)]_j , \nonumber \\
                       & =  \frac{1}{Re} \left(\partial^2_1+\partial^2_2 -2m^2\alpha^2-2n^2\beta^2\right) C_{ij}^{mn}(1,2) \nonumber \\
                       & \hspace{2cm} -\left[\begin{array}{c} 
                       -im\alpha \\ \partial_2 \\ -in\beta \end{array}\right]_j C_{i4}^{mn}(1,2)
                       -\left[\begin{array}{c} 
                       im\alpha \\ \partial_1 \\ in\beta \end{array}\right]_i C^{mn}_{4j}(1,2)
                       \nonumber \\
                      & \hspace{2cm} +im\alpha [\,U(2)-U(1)\,]C_{ij}^{mn}(1,2)
                         -U_y(2)C_{i2}^{mn}(1,2) \delta_{1j}
                         \nonumber \\
                        & -U_y(1)C_{2j}^{mn}(1,2) \delta_{i1} 
                        -\left[\begin{array}{c} 
                       -im\alpha \\ \partial_2 \\ -in\beta \end{array}\right]_k C_{ijk}^{(3)mn}(1,2,2)
                           -\left[\begin{array}{c} 
                        im\alpha \\ \partial_1 \\ in\beta \end{array}\right]_k C_{ikj}^{(3)-m-n}(2,1,1)  
\label{C2a}                        
\end{align}
where $\partial_i:= \partial_{y_i}$, $U(i)=U(y_i)$ for $i=1,2$ and $C_{i4}^{mn}(1,2):= [\bfu^{mn} (y_1,t)]_i \tp^{-m-n}(y_2,t)=:C_{4i}^{-m-n}(2,1)$ are 3 extra `velocity-pressure' cumulants that get generated. Incompressibility conditions give the required 3 extra matrix constraints 
\begin{equation}
\left[\begin{array}{c} 
im \alpha\\ \partial_1 \\ in \beta \end{array}\right]_i C_{ij}^{mn}(1,2)=0 \qquad j \in \{1,2,3\}
\label{C-incompressible}
\end{equation}
along with the equation for the mean equation (\ref{U})
\begin{equation}
     U_t -\frac{1}{Re}U_{yy} = G - \sum_m \sum_n \partial_y C_{12}^{mn}(y,y,t)
\label{C-U}
\end{equation}
to close the system. The infamous closure problem of the Navier-Stokes equation is immediately evident here in that the evolution equation for the 2nd order cumulant depends on the 3rd order cumulant, a pattern which continues for higher order cumulants so the system never closes. A popular (lowest) closure - commonly called CE2 - is to simply ignore the 3rd order cumulant which is equivalent to ignoring the fluctuation-fluctuation term (last bracketed term on the rhs of (\ref{fu})\,). This is the quasilinear approximation or sometimes referred to as mean field theory \cite[e.g.][]{Vedenov61,Herring63, Herring64},
\begin{align}
     U_t &= \frac{1}{Re}U_{yy}+G - ( \,\overline{\tilde{u} \tilde{v}} \,)_y
     \label{QL-U}\\
 \bfu_t  &= \frac{1}{Re} \nabla^2 \bfu -\bnab \tilde{p} -{U} \bfu_x - \tilde{v} U_y \bhx, \label{QL_fu}\\
0 &=\bm{\nabla}\cdot \bfu .  \label{QL}
\end{align}
The defining feature of this approximation is that (\ref{QL_fu}) is linear in $\bfu$ so that fluctuations with different wavenumbers are only coupled in the mean flow equation (\ref{QL-U}). This linearity also means that any fluctuation field (parametrised by streamwise and spanwise wavenumber) can not be consistently in the stable manifold of $U$ as it varies with time. In particular, if $U$ is steady, only marginally-stable fluctuation fields \rk{(typically with a temporal frequency) can be non-vanishing}. Malkus argued for this model (and its marginal stability implications) on the basis that the fluctuation-fluctuation nonlinear term was only stabilising. This would be reasonable if bifurcations from unidirectional shear flows were always supercritical but, some decades later, subcriticality has been realised the more generic situation \citep{Kerswell05, Eckhardt07,Kawahara12, Graham21}.

Applying a closure at next order so $C^{(4)}$ is some assumed function of the lower cumulants or simply ignored (termed CE3) is less straightforward as the ensuing positive definiteness of the second cumulant is not automatic \citep{MarstonQiTobias14}. This difficulty explains the popularity of the lower-order CE2 approximation \rk{where, for example, the existence and stability of steady solutions  has recently been investigated for ODE systems \citep{Li21,Li22}.}

%
%
\subsection{\label{II.C}Approximations to statistical stability}

The approach here is to consider the stability within the cumulant framework as this presents a natural way to assess stability of the flow statistics. Ideally, this should be attempted for a cumulant expansion which is high enough order to show a robustness against including  even higher order cumulants. However, the rate at which the dimensionality of this procedure explodes means that only second and perhaps third order closures are currently practical. As a result, we focus on CE2 here and identify a clear way to progress to higher order (see \S\ref{III.B})

CE2 is the statistical equations (\ref{C2a})-(\ref{C-U}) with the third order cumulants in (\ref{C2a}) set to zero. The corresponding equations for perturbations $(\delta U, \delta C_{ij}^{mn} )$ upon a base statistical state $(U,C_{ij}^{mn})$ - hereafter referred to as the $\delta$CE2 problem - are
\begin{align}
\partial_t \,\dd C_{ij}^{mn}(1,2) 
                       & =  \frac{1}{Re} \left(\partial^2_1+\partial^2_2 -2m^2\alpha^2-2n^2\beta^2\right) \dd C_{ij}^{mn}(1,2) \nonumber \\
                        &-\left[\begin{array}{c} 
                       -im \alpha\\ \partial_2 \\ -in \beta \end{array}\right]_j \dd C_{i4}^{mn}(1,2)
                       -\left[\begin{array}{c} 
                       im\alpha \\ \partial_1 \\ in\beta \end{array}\right]_i \dd C^{mn}_{4j}(1,2)
                       \nonumber \\
                       &+im \alpha [\,U(2)-U(1)\,] \dd C_{ij}^{mn}(1,2)
                         -U_y(2) \dd C_{i2}^{mn}(1,2) \delta_{1j}
                         -U_y(1) \dd C_{2j}^{mn}(1,2) \delta_{i1} \nonumber\\
                        & +im \alpha [\,\dd U(2)- \dd U(1)\,] C_{ij}^{mn}(1,2)
                         -\dd U_y(2) C_{i2}^{mn}(1,2) \delta_{1j}
                         -\dd U_y(1) C_{2j}^{mn}(1,2) \delta_{i1} \label{dC_1}\\
0 &= \left[\begin{array}{c} 
im\alpha\\ \partial_1 \\ in \beta \end{array}\right]_i \dd C_{ij}^{mn}(1,2),\label{dC_incompress}\\
     \partial_t \dd U &= \frac{1}{Re}\dd U_{yy} +\dd G - \sum_m \sum_n \partial_y \dd C_{12}^{mn}(y,y,t).
\label{C-U1}
\end{align}
An equivalent equation arises in non-modal  stability theory \citep[e.g.][]{FarrellIoannou93, Jovanovic05}.
If the pressure gradient is kept fixed $\dd G=0$, otherwise the constant volume flux condition $\int^1_{-1} \dd U dy=0$ is the extra constraint required. 
Crucially the ansatz $(\dd U, \dd C_{ij}^{mn}) \propto e^{ \lambda t}$ is possible if the base state is independent of time - in other words the base flow is {\em statistically} steady.

Even in the cheapest 2D situation, the CE2 stability problem  requires handling $5 N_x$ correlation matrices of size $N_y^2$ all linked through the mean equation. This leads to a matrix of size $ \left( N_y+5 N_x N_y^2\right)^2$  
or $\approx 25N_x^2 N_y^4$ elements which is out of reach even for modest resolutions. Given this, we explore a route to potentially still capture the essence of the CE2 statistical stability approximation without the considerable cost. To do this, we discuss a connection back to the equations of motion  which plausibly retains the  steadiness of the stability problem.

%
%

\subsection{\label{II.D}Steady statistics and simplications}

\rk{A statistically steady base flow has steady mean $U$ and cumulants $\bC$ with their Fourier components  $\bC^{mn}(1,2)$ also steady under ensemble averaging. This ensures that the associated stability problem  (\ref{dC_1})-(\ref{C-U1}) has temporally-constant coefficients and is therefore a (conceptually at least) simple eigenvalue problem.} 
%
%
%
\rk{In what follows below, we will not attempt to solve this directly as it is too unwieldy. Instead, a smaller, more practical QL stability problem is sought as a good proxy for it. The key in doing this is identifying a suitable base velocity field around which to develop a QL-type stability problem. A straightforward approach is to find the `best' rank-1 approximation of each $\bC^{mn}$ and use the  associated velocity field, $\bfu^{mn}_0(y)$, as representative of the base flow. This can be accomplished by minimising the Frobenius matrix norm of the difference},
    \begin{align}
    ||\bC^{mn}-\bfu^{mn}_0 \otimes \bfu^{*mn}_0||^2_F &:=
    \sum_{i,j=1}^3 \sum_{p,q=1}^N
    (\,C^{mn}_{i,j}(p,q)-[\bfu^{mn}_0(y_p)]_i[\bfu^{*mn}_0(y_q)]_j\,) \nonumber \\
    &\hspace{1.75cm}\times (\,C^{*mn}_{ij}(p,q)-[\bfu^{*mn}_0(y_p)]_i[\bfu^{mn}_0(y_q)]_j\,)
     \nonumber\\
     & \hspace{-0.5cm}=\left\|\left(\begin{array}{ccc}
     C^{mn}_{11} & C^{mn}_{12} & C^{mn}_{13} \\
     C^{mn}_{21} & C^{mn}_{22} & C^{mn}_{23} \\
     C^{mn}_{31} & C^{mn}_{32} & C^{mn}_{33} 
\end{array} \right) - 
\left[\begin{array}{c} \tu^{mn}_{(1:N)}\\ \tv^{mn}_{(1:N)} \\ \tw^{mn}_{(1:N)}
\end{array} \right]
\left[\begin{array}{ccc} \tu^{*mn}_{(1:N)} &\tv^{*mn}_{(1:N)} & \tw^{*mn}_{(1:N)}
\end{array} \right]
\right\|^2_F
    \end{align}
where e.g. $\tu^{mn}_{(1:N)}:=[ \tu^{mn}(y_1)\, \,\, \tu^{mn}(y_2)\,\, \cdots\,\, \tu^{mn}(y_N)]^T$. Since $\bC^{mn}=[\bC^{mn}]^H$ (the Hermitian conjugate) and positive definite, the required $\bfu^{mn}_0$ is the leading right eigenvector of $\bC^{mn}$ associated with the largest eigenvalue $\sigma_1$ scaled so that $|\bfu^{mn}_0|=\sqrt{\sigma_1}$.

\rk{We now discuss how the QL stability problem based upon the $\bfu^{mn}_0$ relates to the CE2 stability problem based on $\bC^{mn}=\bfu^{mn}_0 \otimes \bfu^{*mn}_0$. Possible disturbances partition into two types: Type A where the disturbance has energy in Fourier pairings not excited in the base flow and Type B where the disturbance has energy in Fourier pairings which are a subset of those present in the base flow  i.e. $\bd \bfu^{mn}$ is only non-zero if $\bfu_0^{mn}$ is  (see {Appendix \ref{appendixA}} for more detail). The former (Type A) case is straightforward (since $\bd \bC={\bf 0}$) so we focus here on the Type B situation. The QL stability problem - hereafter the $\delta$QL problem - is}
\begin{align}
\partial_t \bd \bfu^{mn} &= \optL^{mn}(U) \bd \bfu^{mn} +\J^{mn}(\dd U) \bfu_0^{mn} \qquad \qquad \forall (m,n) \neq (0,0) \label{dQL}\\
\partial_t \dd U &= \frac{1}{Re}\dd U_{yy} +\dd G - \sum_m \sum_n \partial_y 
\biggl\{
\tu_0^{mn} \dd \tv^{-m-n}
+\dd \tu^{mn} \tv_0^{-m-n}
\biggr\}
\end{align}
where
\begin{align}
\optL^{mn}(U)\, \bfu^{mn} &:=  \frac{1}{Re} [\,\partial_y^2-m^2\alpha^2 -n^2 \beta^2\,]\bfu^{mn} 
-\left[\begin{array}{c} im\alpha \\ \partial_y \\ in\beta \end{array} \right] \tilde{p}^{mn} -im \alpha{U} \bfu^{mn} - \tilde{v}^{mn} U_y \bhx, \label{L}\\
\J^{mn}(\dd U) &:=\lim_{\epsilon \rightarrow 0} \, \frac{1}{\epsilon}[\,\optL^{mn}(U+ \epsilon\dd U)-\optL^{mn}(U)\,]=\optL^{mn}(U+\dd U)-\optL^{mn}(U)  \label{J}
\end{align}
(as $\optL^{mn}$ is affine in $U$). This has temporally-constant coefficients and therefore admits an eigenfunction of the form
\begin{equation}
(\delta U(y,t), \, \bd \bfu(x,y,z,t))
=\biggl( \delta \hat{U}(y) e^{\lambda t}, \,\sum_m \sum_n \biggl[ 
\bd \bfu^{mn}(y,t):=\bd \hat{\bu}^{mn}(y) e^{\lambda t}
\biggr]e^{i(m \alpha x+ n \beta z)} 
\biggr)
\end{equation}
\rk{The key point is that this has a direct equivalent in the CE2 problem (\ref{dC_1})-(\ref{C-U1}) where the same eigenvalue $\lambda$ has the  corresponding eigenfunction $\bC^{mn}:= \bfu_0^{mn} (y_1,t)\otimes \bd \bfu^{-m-n}(y_2,t)+ \bd \bfu^{mn} (y_1,t)\otimes \bfu_0^{-m-n}(y_2,t) =\hat{\bC}^{mn}(y_1,y_2)e^{\lambda t}$. Given this, instability within the  much-more-tractable $\delta$QL problem is therefore sufficient to conclude statistical instability within $\delta$CE2}. Going further to claim that the stability within the $\delta$QL problem also implies stability within the $\delta$CE2 problem seems very likely but is not assured (see Appendix \ref{appendixA}).  With this one caveat, we nevertheless assume that the stability of the (much) smaller QL system is a proxy for the statistical stability of the CE2 system: in particular, stability in the $\delta$QL system is taken to imply statistical stability within $\delta$CE2. This allows us to generate an extended version of Orr-Sommerfeld analysis but, before pursuing this in the next section, we make a remark and an observation. 

The remark is that certain notational liberties have been taken here to keep the discussion as clear as possible. For example, the operator $\optL^{mn}$ strictly maps an incompressible flow field to another which involves a supplementary scalar field (the pressure) entering into the definition. As is well known, this is determined by imposing incompressibility. An implicitly-incompressible representation for the velocity field could be used (e.g. reducing the problem down to just using wall-normal velocity and vorticity) to avoid this wrinkle but staying with primitive variables makes the various manipulations as clear as possible.

The observation is that the special form of the nonlinearity in the QL and CE2 formulations means that different wavenumber pairings only interact through the mean flow equation. As a result eigenvalues can be sought within any subset of the wavenumbers possible and these are still valid within the full system of wavenumbers i.e. this is not a truncation merely a subclass of disturbances. In particular, for CE2 the simplest perturbation of the base state $(U,\bC)$ just consists of perturbations in  one wavenumber pairing and the mean flow, 
\begin{equation}
(\,\dd \hat{U}, \,\,\bd \hat{\bC}^{mn}\,) \,e^{\lambda t}.
\end{equation}
Even this leads to an eigenvalue matrix calculation of size $(N_y+9N_y^2) \times (N_y+9N_y^2)$ in 3D.  The equivalent QL perturbation
has to consider
\begin{equation}
(\,\dd \hat{U}, \,\,\bd \bhu^{mn},\,\,\bd \bhu^{-m-n}\,) \,e^{\lambda t}
\end{equation}
and requires an eigenvalue matrix calculation of size $(N_y+8N_y) \times (N_y+8N_y)$ which is {\it much} smaller (or $(N_y+6N_y)\times (N_y+6N_y)$ in 2D - see Appendix \ref{appendixB}).

%
%
\section{\label{III}Extended Orr-Sommerfeld analysis}

We refer to the $\dd$QL problem which includes some information on the second-order-flow statistics of the base flow as `Extended Orr-Sommerfeld' stability analysis. The approach is as follows.

\begin{enumerate}
    \item {\em Estimate the base flow statistics} $(U, \bC)$   
    
    
\item {\em Approximate each base cumulant component tensor $\bC^{mn}$ as rank 1 to obtain a representative physical base field  $\bfu^{mn}_0(y)$.}
    
\item {\em Solve the Extended Orr-Sommerfeld eigenvalue problem for small perturbations $(\delta U, \,\bd \bfu, \,\dd \tp)$ which is}
\begin{equation}
\begin{aligned}
    \partial_t\delta U&=\frac{1}{Re}\partial^2_y \delta U+\delta G - \sum_{m,n}
    \partial_y \left(\dd \tu^{mn} \tv^{-m-n}_0 +\tu_0^{mn}\dd \tv^{-m-n} \right)
    \end{aligned}
    \label{eq:EOS_mean}
\end{equation}
coupled with the fluctuation equation for every Fourier mode $(m,n)$
\begin{gather}
    \partial_t{\bd \bfu}^{mn}  = \frac{1}{Re} \nabla^2 \bd  \bfu^{mn}-\bm{\nabla} \dd \tp^{mn} - im U \bd \bfu^{mn} - \dd \tv^{mn}U_y\bm{\hat{x}}
        \textcolor{blue}{-im{\delta U} \bfu_0^{mn} }
    \textcolor{red}{-\tv_0^{mn} \delta U_y \bm{\hat{x}}}, \label{eq:EOS_fluct}   \\
    im \alpha\, \delta\tu^{mn}+\partial_y\, \delta \tv^{mn}+in\beta\, \delta \tw^{mn} = 0    
    \label{eq:EOS_incomp}
    \end{gather}
(coloured terms depend on $\dd U$ to be discussed later).
\end{enumerate}

This  eigenvalue problem has size $ (N_y+4 N_x N_y N_z)^2 \approx  16N_x^2N_y^2N_z^2$ which is probably impractical for all but the smallest systems since all the wavenumber pairings are coupled through the mean equation. A natural way to simplify the calculation is to only include a targeted subset of the wavenumber pairings and, going further, to only consider one wavenumber pairing. This latter approximation is obviously the most extreme but is also closest in spirit to the original Orr-Sommerfeld analysis - we call this {\em `minimally Extended Orr-Sommerfeld'} (mEOS) analysis.

%
%
\subsection{\label{III.A}minimally Extended Orr-Sommerfeld equations (mEOS)}
Here, the sum over wavenumber pairings in (\ref{eq:EOS_mean}) is removed leaving the mean flow equation (\ref{eq:EOS_mean})   
%
coupled with just one Fourier mode equation (\ref{eq:EOS_fluct}).
%
This eigenvalue problem focusses on the coupling between an individual Fourier mode and the mean flow and its size is now $ (5 \times N_y)^2 = 25 N_y^2$ so comparable to OS analysis. Just as for OS analysis, it needs to be repeated for each Fourier wavenumber of interest. 

%
%
\subsection{\label{III.B}infinitely-Extended Orr-Sommerfeld equations (iEOS)}

Before going on to test EOS and mEOS on some statistically steady flows, it's worth briefly discussing how far this approach of using the cumulant framework to further motivate more sophisticated versions of Orr-Sommerfeld analysis can go. In fact, it turns out that only one further enhancement is possible and then only for CE$\infty$. This is because any intermediate closure will lead to an equation which when `unwrapped' at the highest level contradicts those obtained at lower cumulant orders. 
To see this, consider CE3, the next closure  after CE2 in which $\bC^{(4)}$ is ignored. Unwrapping the statistical equation for $\bC$ will recover the Navier-Stokes equations whereas unravelling the $\bC^{(3)}$ equations will not as the nonlinear term leading to $\bC^{(4)}$ has been dropped (any other closure will also suffer this inconsistency). The only way to avoid this is: 1) to only unwrap one cumulant equation - so CE2 which leads to EOS, or 2) ensure that all unwrapped equations are the same  which means no closure and CE$\infty$. In this latter case, the  Navier-Stokes equations re-emerge which, under perturbation around a steady base state, lead to the linearized Navier-Stokes equations as the ultimate extension of OS analysis, i.e.
\begin{equation}
\begin{aligned}
    \partial_t\delta U&=\frac{1}{Re}\partial^2_y \delta U+\delta G - \sum_{m,n}
    \partial_y \left(\dd \tu^{mn} \tv^{-m-n}_0 +\tu_0^{mn}\dd \tv^{-m-n} \right)
    \end{aligned}
    \label{eq:iEOS_mean}
\end{equation}
and for each wavenumber pair $(m,n)$
\begin{gather}
    \partial_t{\bd \bfu}^{mn}  = \frac{1}{Re} \nabla^2 \bd  \bfu^{mn}-\bm{\nabla} \dd \tp^{mn} - im \alpha U \bd \bfu^{mn} - \dd \tv^{mn}U_y\bm{\hat{x}}
    -im \alpha {\delta U} \bfu_0^{mn} 
    -\tv_0^{mn} \delta U_y \bm{\hat{x}}  \nonumber  \\
    \hspace{8cm}-[\bd \bfu \cdot \bnab \bfu_0+\bfu_0 \cdot \bnab \bd \bfu]^{mn} \nonumber\\
    im \alpha \, \delta \tu^{mn}+\partial_y \delta \tv^{mn}+in \beta \, \delta \tw^{mn} = 0    
    \label{eq:iEOS_fluct}
    \end{gather}
This could be called an {\em `infinitely-Extended Orr-Sommerfeld'} problem (or iEOS for short) as there is no further extension possible with the cumulant expansion framework considered here.  It's worth remarking that the extra term added in iEOS  would get dropped in any minimal version where the individual wavenumber pairings are considered separately: i.e. minimalizing iEOS is just mEOS. There is, however, considerable potential in avoiding this drastic reduction in favour of retaining small subsets of interacting wavenumber pairings such as triads which interact through the newly present term. This or iEOS will not be pursued further here to avoid overcomplicating the discussion.

%
%
 \begin{figure}
 \centerline{
 \includegraphics[width=0.5\textwidth]{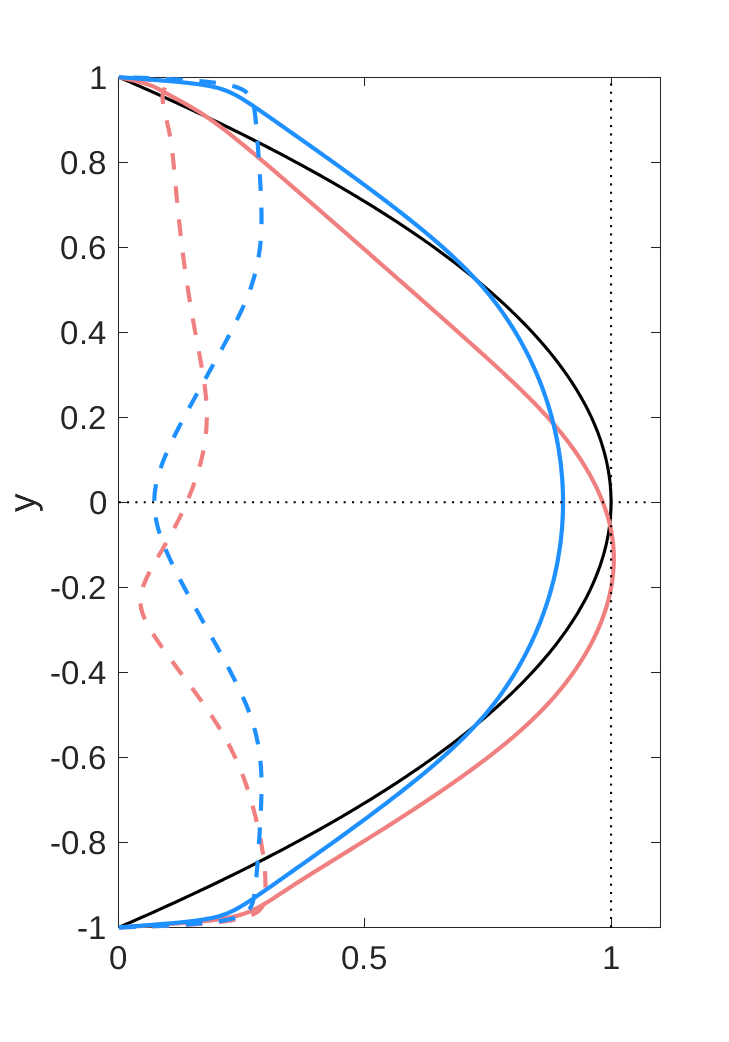}%
  \includegraphics[width=0.5\textwidth]{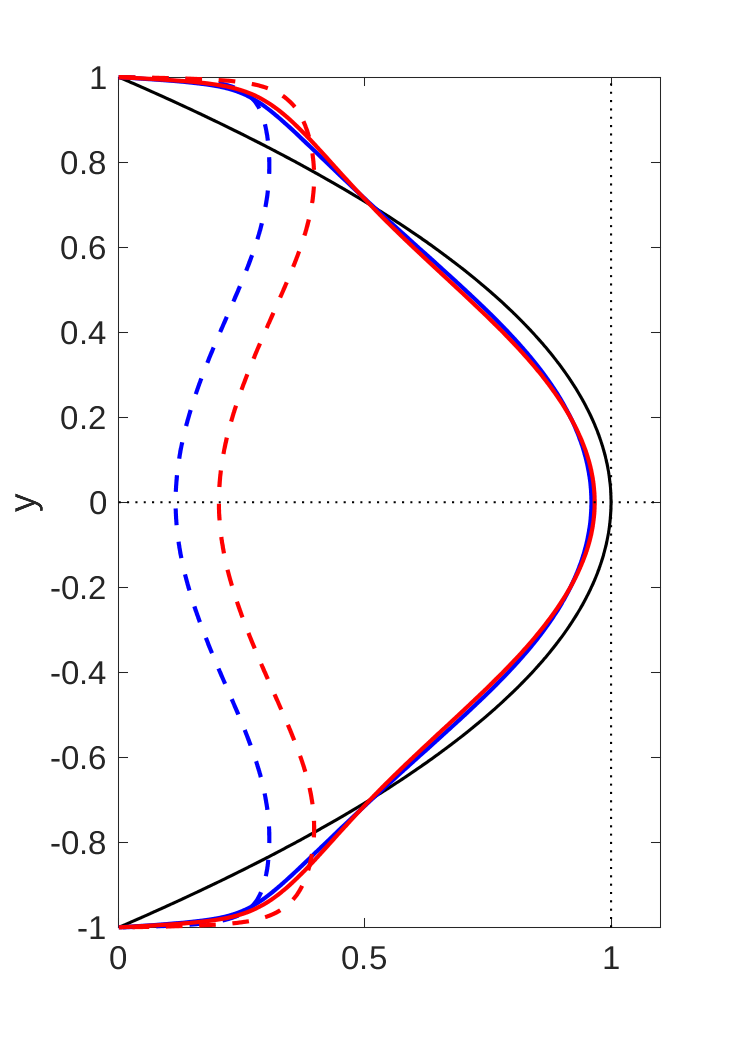}}%
 \caption{\label{fig:meanprofs} Mean velocity profiles (solid lines) and streamwise root-mean-squared velocity profiles (dashed lines) for the four 2D channel test states at $Re=36,300$. Left: states with an applied pressure gradient: symmetric S (light blue) and asymmetric A (pink). Right: body forced states F1 (blue) and F2 (red). Black solid line shows the laminar parabolic profile in both plots for reference.}
 \end{figure}


%
%
\section{\label{IV}Application}

We now test the extended stability approaches, EOS and mEOS, upon turbulent states obtained by direct numerical simulations of 2D channel flow. 
At high enough $Re$, the expectation is that the flow will be statistically stationary \citep{Malkus56}.
The simulations are performed using the open-source partial differential equation solver Dedalus \citep{dedalus}.  The 2D flow is simulated using a vorticity-streamfunction formulation so that $(u,v):=(\psi_y,-\psi_x)$ and
\begin{equation}
 \omega_t + \psi_y \omega_x - \psi_x \omega_y = \frac{1}{Re} \nabla^2 \omega - \partial_y f(y,t), \quad
\omega:=- \nabla^2 \psi 
\label{eq:vorticity2D}
\end{equation}
where the flow is driven by a streamwise body force $f(y,t)\bhx$ so that the volume flux is fixed. $Re$ is defined as in \S \ref{II.A} so the following conditions are imposed 
\begin{equation}
\psi\left(-1\right)=0, \quad \psi\left(1\right)=\tfrac{4}{3}, \quad
\psi_y\left(-1\right)=0, \quad \psi_y\left(1\right)=0.
\label{flux}
\end{equation}
with the latter two reflecting the presence of non-slip walls.
Following earlier work \citep{Falkovich_2018}, the length of the channel $2L_x$ is set to 4 times its height ($L_x=4$), $Re=36,300$ and computationally,  $1024$ Fourier modes are used to discretize in the $x$ direction and $256$ Chebyshev modes in $y$ (see \cite{MK21} for details). 

The streamwise body force is defined in terms of a profile function $g(y)$ as follows
\begin{equation}
f(y,t):=G(t)\left(g(y)-1\right)
\end{equation}
where setting $g=0$ recovers the usual $y$-independent applied pressure gradient $-G(t)$. For this situation it is already known that there is bistability with two statistically steady states possible in 2D channel flow at $Re=36,300$ \citep{MK21}: a state which is statistically symmetric about the channel midplane  - the symmetric state S -  and another which is statistically asymmetric - the asymmetric state A: see Fig. \ref{fig:meanprofs}(left). Choosing non-zero $g(y)$ is a device to diversify the test states available with two extra examples generated by using profiles $g_1:=(1-y^2)^6$ (state F1) and $g_2:=\cos \tfrac{3}{2} \pi y^2$ (state F2).
%

%
%
%
 \begin{figure}
 \centerline{
 \includegraphics[width=0.5\textwidth]{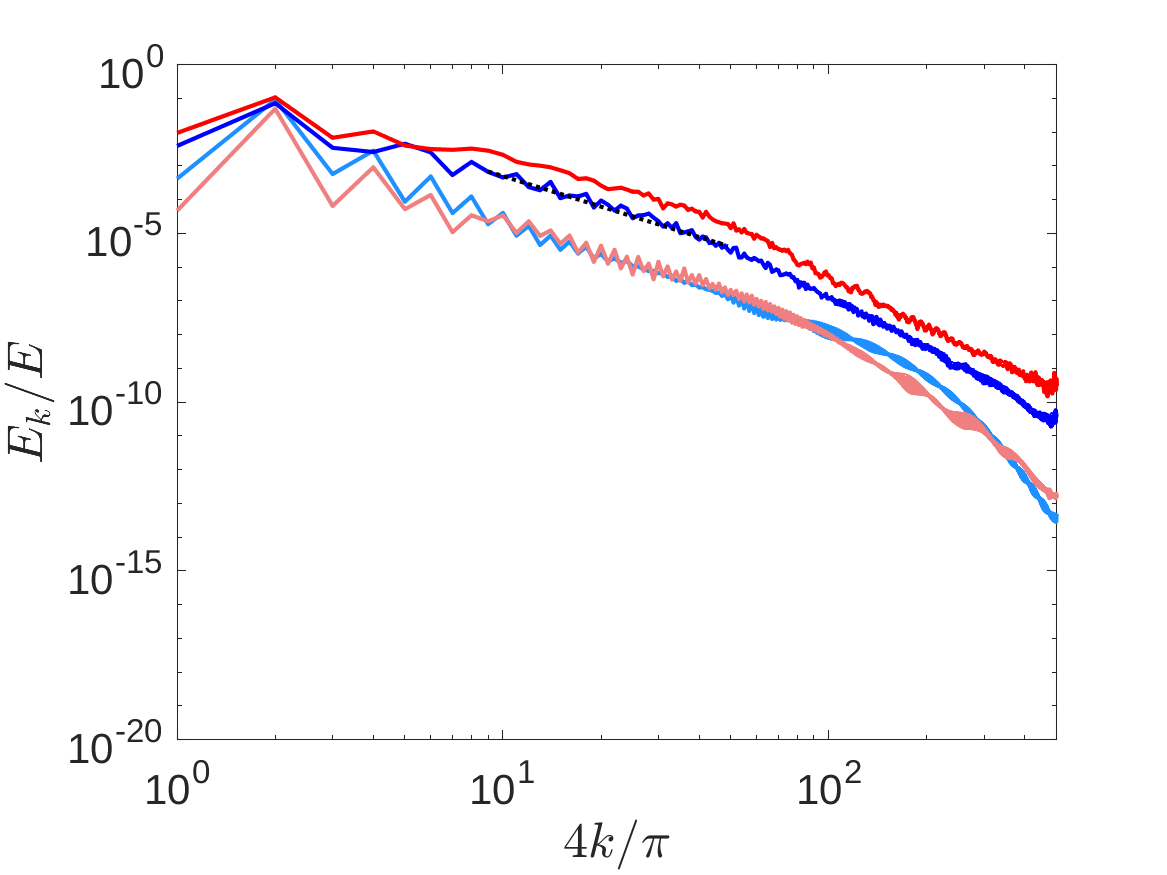}%
  \includegraphics[width=0.5\textwidth]{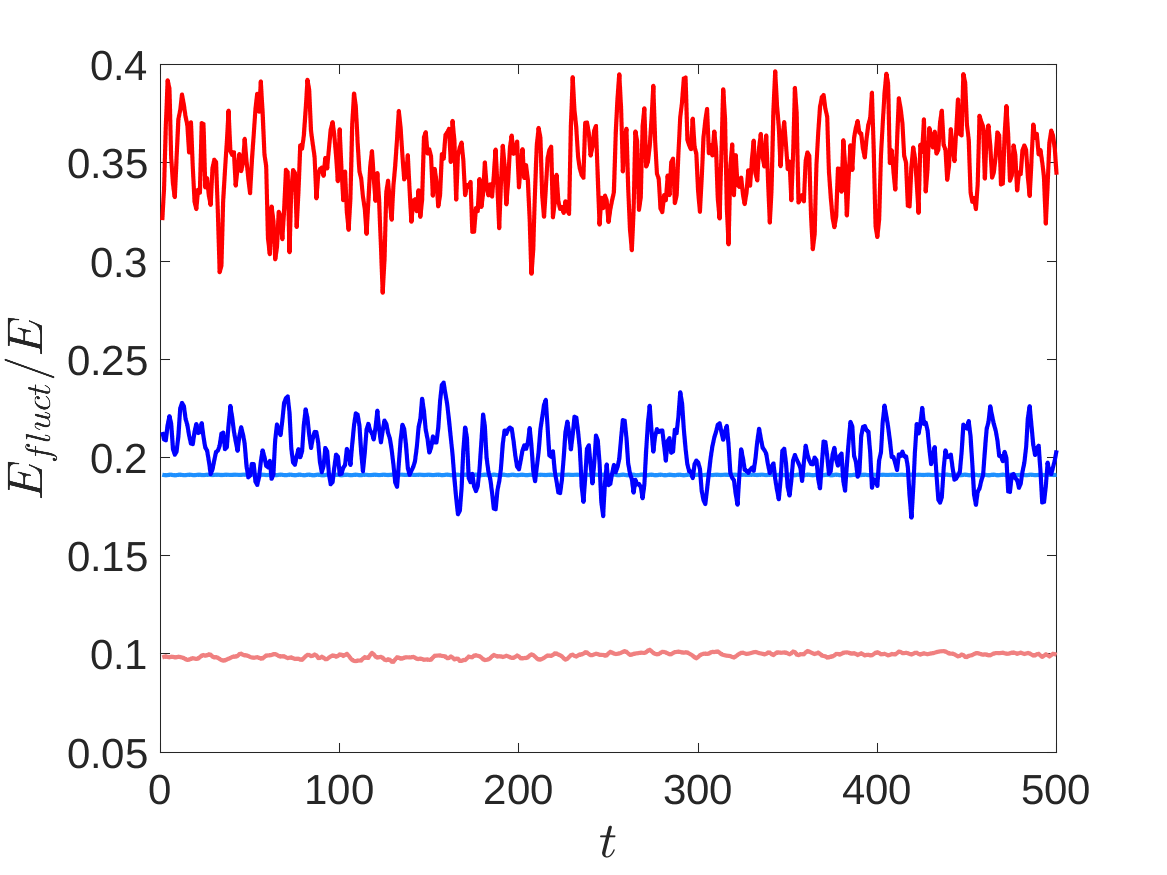}}%
 \caption{\label{fig:energydistr} Fluctuation energy as a ratio to the  total flow energy for the four 2D channel test states at $Re=36,300$: symmetric S (light blue), asymmetric A (pink), F1 (blue) and F2 (red). Left: decomposition by streamwise wavenumber $k$, right: temporal fluctuations over a typical time-averaging window.  }
 \end{figure}

%
%
\subsection{Statistically stable test states}

\rk{A spatial average over the streamwise direction and temporal averaging was used as a proxy for the ensemble-averaged statistics discussed in deriving EOS (these are equivalent for a statistically steady flow over a large enough domain and long enough time).}
The mean profiles for states S, A, F1 and F2 are shown in Fig. \ref{fig:meanprofs} along with their streamwise root-mean-squared velocity profiles, all obtained by time-averaging over $10^4$ time units.
Their respective power spectra are shown in Fig. \ref{fig:energydistr} (left). While the dominant wavenumber is the same for all four states ($k_d=2\alpha=\pi/2$), significant differences are seen at neighbouring streamwise wavenumbers $2k_d$ and $3k_d$ for example. Typical temporal variations of the fluctuation field energy $E_{fluct}$ compared to the total energy of the flow $E$ - Fig. \ref{fig:energydistr} (right) - show the desired  wide variety of mean fluctuation energy and fluctuation amplitudes. In particular, the amplitude of the fluctuations in state F2 are  $\approx 60\%$ of the mean and only $\approx 30\%$ for state A.

For each state the second-order cumulant matrices were computed every unit of time and then averaged over a period of $10^3$. During the process of calculating EOS and mEOS eigenvalues, correlation matrices time-averaged over different time windows t = 500, 1000, 1500 were used as well as correlation matrices time-averaged with different time-steps (0.5 and 1) for one time window t=500. The qualitative results were all the same, with only minor quantitative differences which did not affect the relative positions of eigenvalues obtained by different methods. The time-averaged cumulant matrix was then diagonalised and the flow field corresponding to the leading (real) eigenvalue used to typify the base fluctuation field. The implications of the time-averaging to the rank of the matrix are discussed in \S\ref{V.A}.

%
%
\begin{figure}
\centerline{\includegraphics[width=0.5\textwidth]{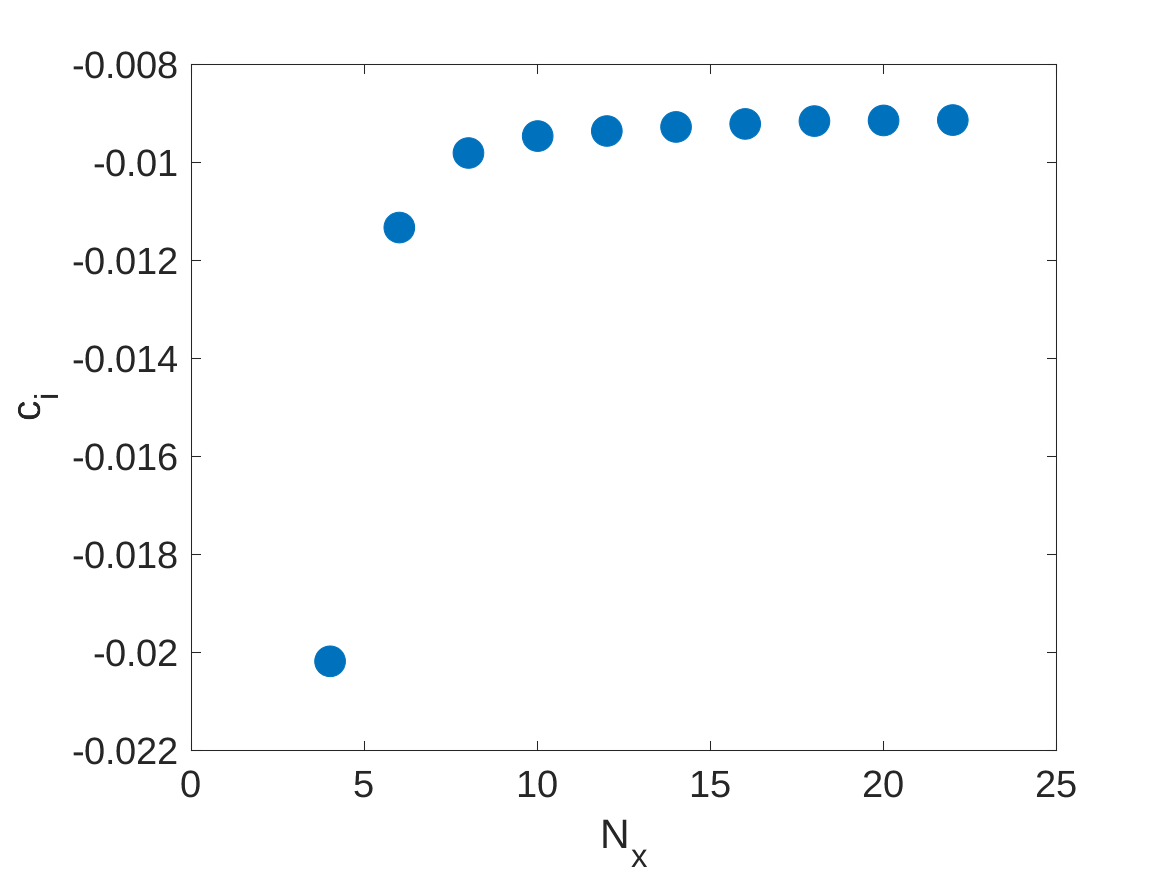}
\includegraphics[width=0.5\textwidth]{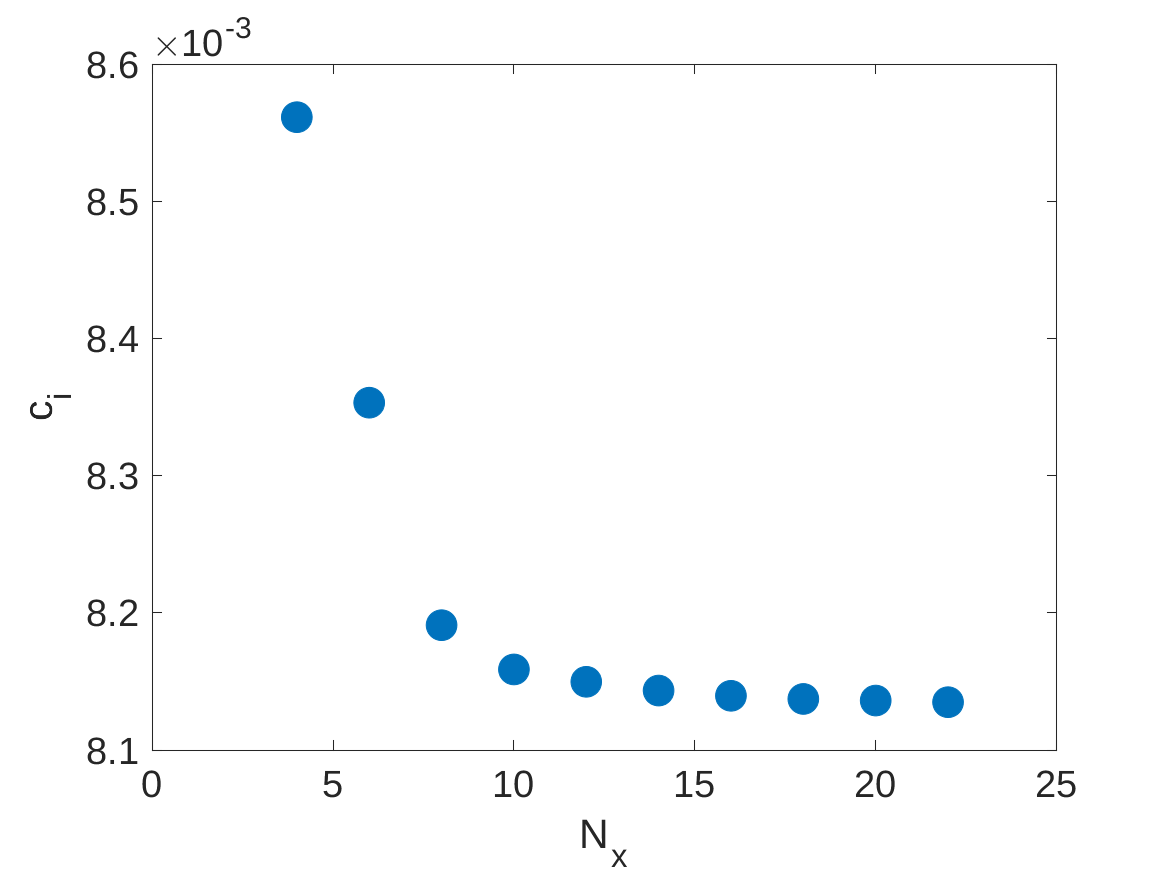}}%
\caption{\label{fig:eval_convergence} Typical eigenvalue convergence with respect to the total number of the streamwise wavenumbers $N_x$ included in the EOS model, shown for $k=2$ (left) and $k=3$ (right) for the F2 state.
}
\end{figure}

The EOS problem - equations (\ref{eq:EOS_mean})-(\ref{eq:EOS_fluct}) - and the mEOS problem were posed as generalised matrix eigenvalue problems of matrix size  $\approx \left( 4 N_x N_y\right)^2$ and $(5N_y)^2$ respectively.
For EOS, $N_x=20$ streamwise wavenumbers were used as a balance between accurate eigenvalue convergence and computational accessibility: e.g. see Fig. \ref{fig:eval_convergence}. Wall-normal resolution of the eigenvalue problem was also varied to ensure eigenvalue convergence in the mEOS case. Doubling the wall-normal resolution from  the DNS resolution of $256$ Chebyshev modes to $512$ produced less that a $10^{-8}$ relative error in the eigenvalues so all stability calculations were done using the DNS wall-normal resolution.

%
%
\subsection{\label{IV.C}Comparisons}

\begin{figure}
 \centerline{
 \includegraphics[width=0.5\textwidth]{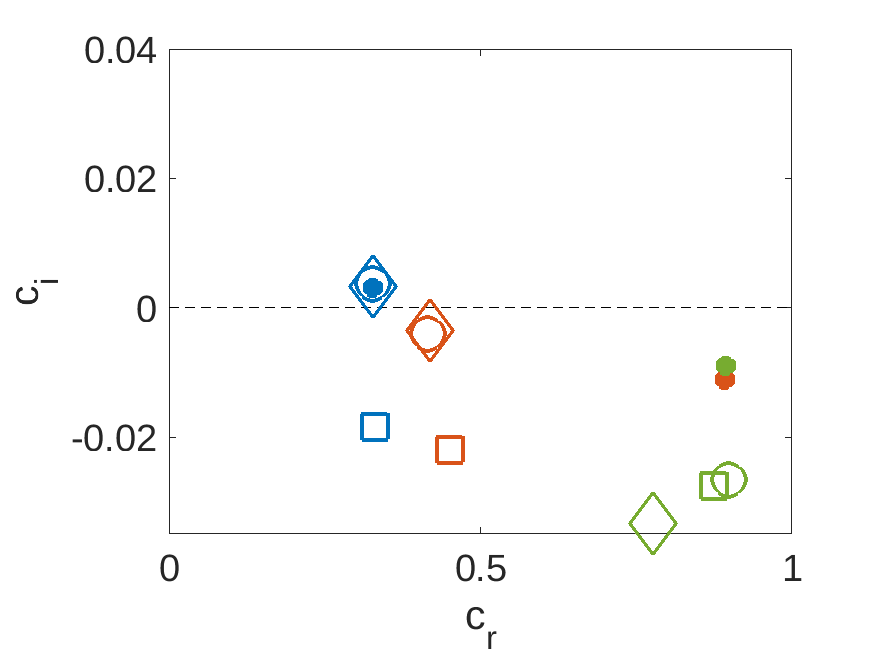}%
 \includegraphics[width=0.5\textwidth]{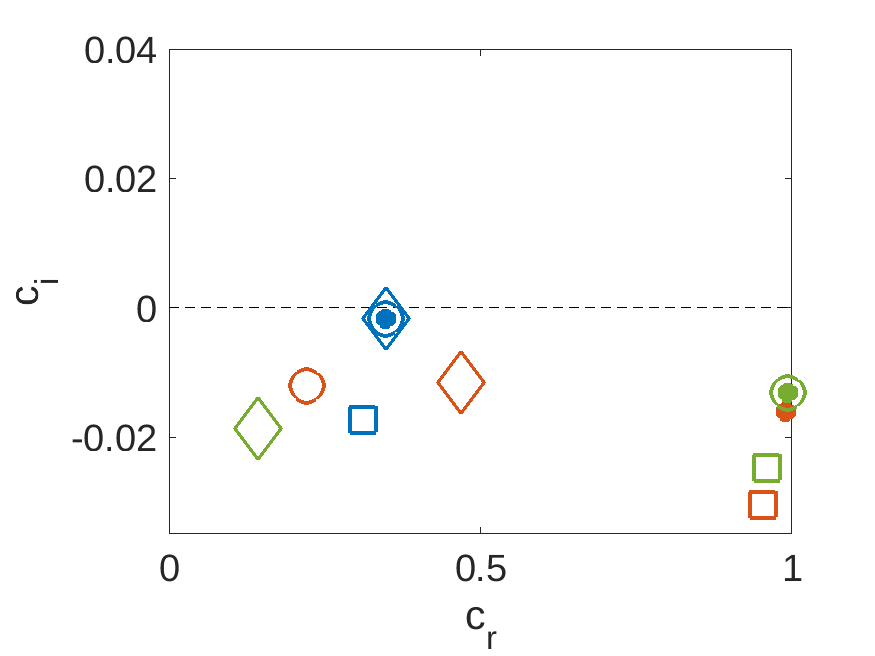}}
 \centerline{
  \includegraphics[width=0.5\textwidth]{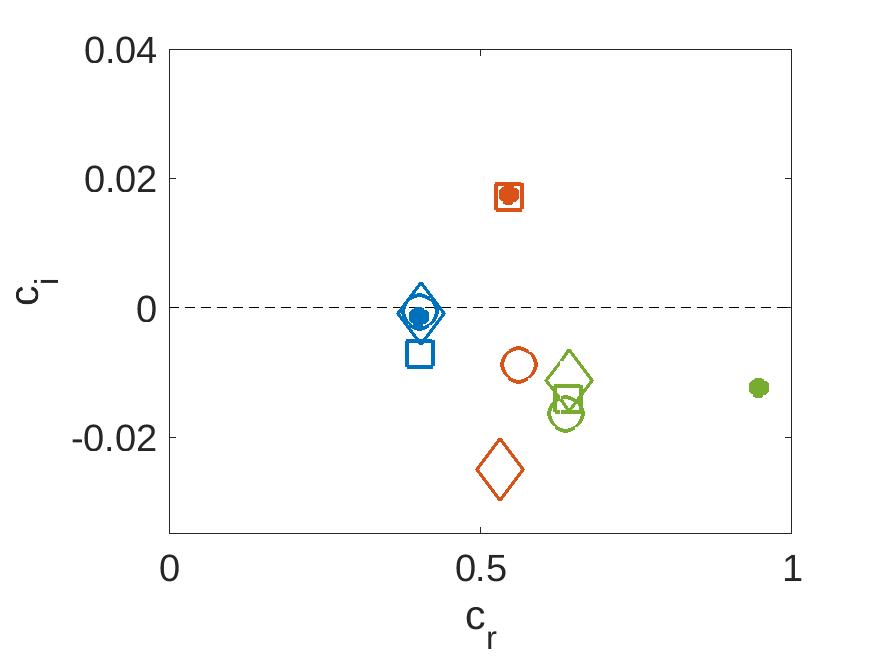}%
   \includegraphics[width=0.5\textwidth]{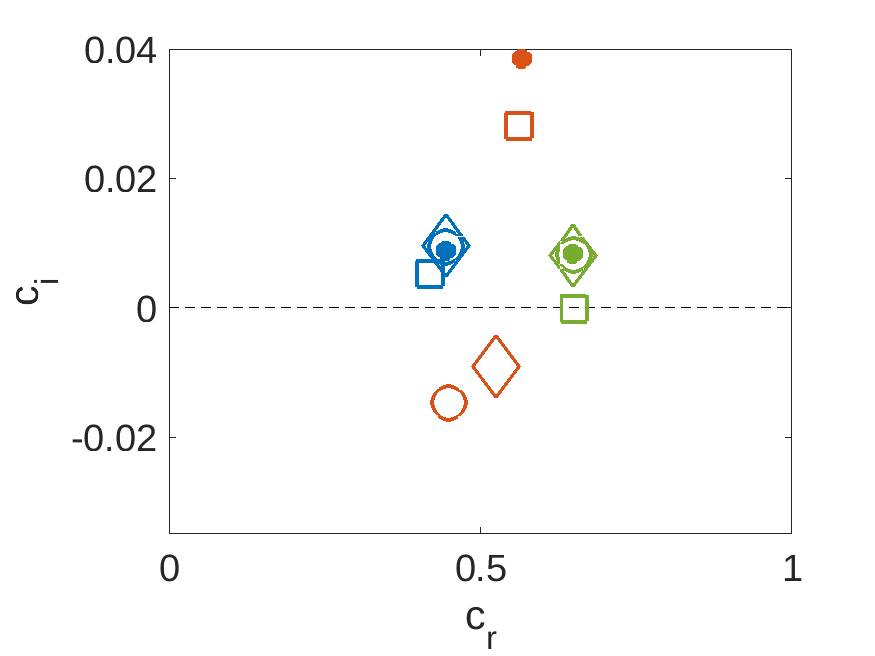}}%
 \caption{\label{fig:EOSvsNOS} 
 Comparison of the eigenvalues $\sigma=-im\alpha(c_r+i c_i)$ - so growth rate is $m \alpha c_i$ - for OS ($\bullet$), OS with eddy viscosity ($\square$), EOS ($\Diamond$) and mEOS analysis ($\medcirc$). Only the leading eigenvalues for {\color{blue}$m=1$ (blue)}, {\color{BrickRed}$m=2$ (red)} and {\color{OliveGreen}$m=3$ (green)} are shown. 
 Top left: symmetric S, top right: asymmetric A, bottom left: F1, bottom right: F2 states. 
 }
 \end{figure}

Here we compare the eigenvalues obtained by standard (OS), extended (EOS) and minimally extended Orr-Sommerfeld (mEOS) stability analyses on the four different turbulent states. Since all our test cases are {presumed} statistically steady, for a model to be good at predicting statistical stability of the turbulent state, we expect all the eigenvalues to be stable, i.e. to have a negative real part. For the symmetric (top left, figure \ref{fig:EOSvsNOS}) and asymmetric (top right, figure \ref{fig:EOSvsNOS}) states, we observe no significant difference in the leading eigenvalues between the three stability models. While all three models predict the asymmetric state to be statistically stable, they also all predict statistical instability for the symmetric state. For these test states, the leading eigenvalue is associated with the first streamwise wavenumber (marked blue in the plot). We observe some changes for the other eigenvalues which we do not examine further as they are all stable and do not affect characterization of statistical stability of the turbulent states.  

\begin{figure}
\centerline{\includegraphics[width=0.5\textwidth]{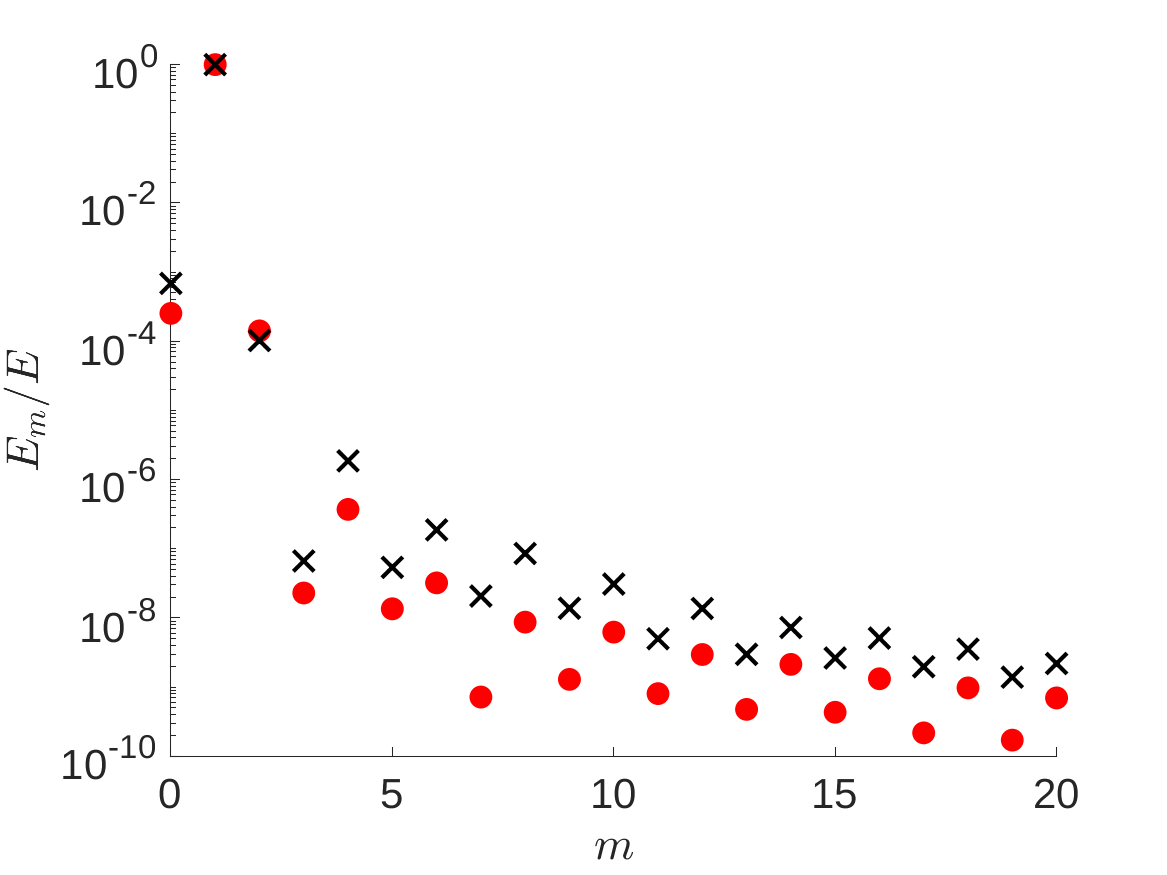}
\includegraphics[width=0.5\textwidth]{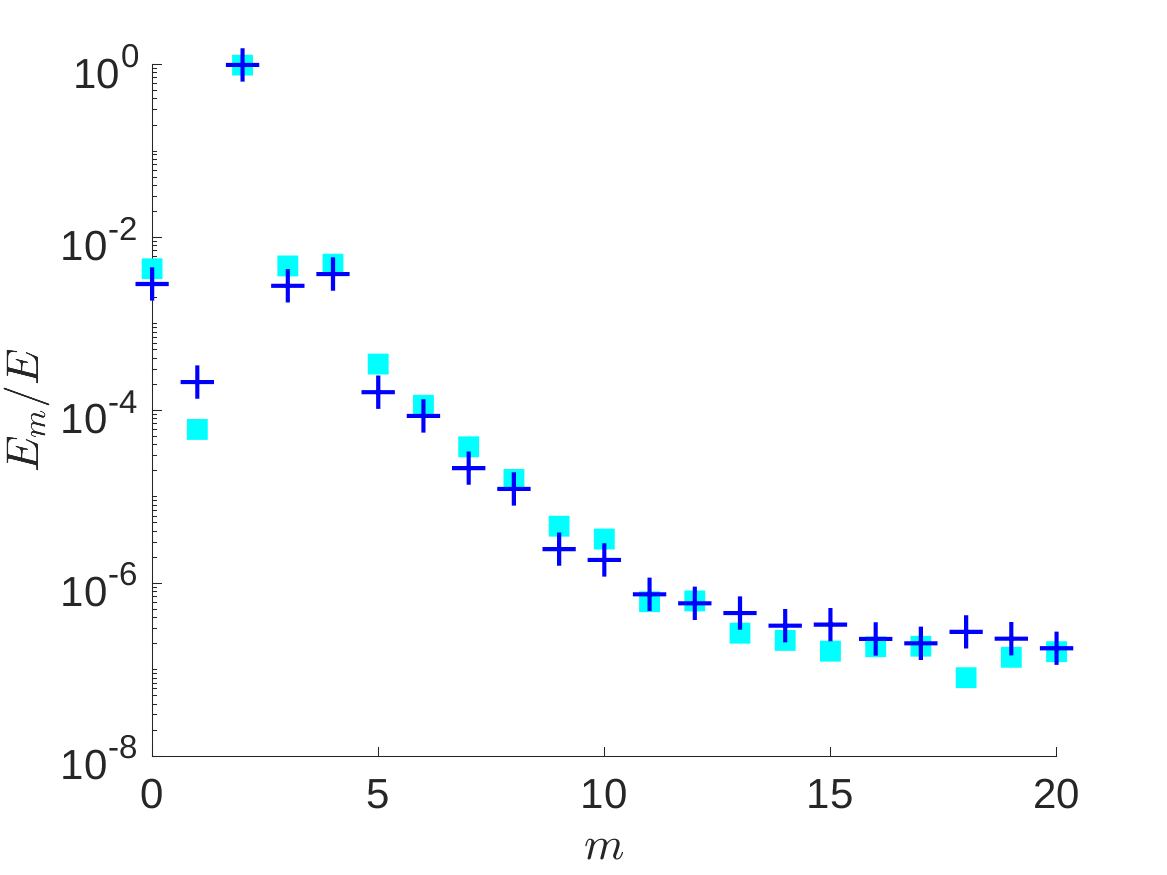}}
\caption{
\label{fig:evect_spectrum} 
Power spectra for the EOS eigenvectors for the leading eigenvalues. Left: symmetric S state (\mbox{\boldmath $\times$}) and asymmetric A state ({\color{red}$\bullet$}) with $m=1$;
Right: states  F1 ({\color{cyan}\mbox{$\blacksquare$}}) and F2 ({\color{blue}\mbox{\boldmath $+$}}) with $m=2$. The mean velocity perturbation ($m=0$) is at least an order of magnitude smaller than the leading wavenumber contribution to the power spectrum in all 4 cases.
}
\end{figure}

Since all streamwise wavenumbers are coupled in the extended model, the leading wavenumber for the eigenvalue is assigned by examining the power spectrum of the corresponding eigenvector. Example power spectra for leading eigenvalues are shown in figure \ref{fig:evect_spectrum} where it is apparent that the mean velocity component of an eigenvector is at least an order of magnitude smaller than the leading wavenumber contribution.

%
%

The leading eigenvalue for the body-forced states states F1 (bottom left, figure \ref{fig:EOSvsNOS}) and F2 (bottom right, figure \ref{fig:EOSvsNOS}) is associated with the second streamwise wavenumber (marked red in the plot). For both F1 and F2 states, this eigenvalue is unstable in the standard stability analysis but becomes stable in EOS. The stabilisation effect is also captured by mEOS but isn't so strong. While the statistical stability of the F1 state is predicted correctly by the extended model, the improvement is only partial for the F2 state. In addition to the unstable leading eigenvalue which is stabilised by the extended model, there are two subsequent unstable eigenvalues corresponding to the first and third streamwise wavenumbers (marked blue and green on the plot). Similar to the symmetric state case, these two unstable eigenvalues see no improvement towards stability when the extended stability models are applied. 

We also repeat the standard OS analysis with eddy viscosity as described in section \ref{II.A} (marked using squares in figure \ref{fig:EOSvsNOS}). While eddy viscosity has a general stabilising effect on the OS eigenvalues, this effect is rarely enough to stabilise the unstable OS eigenvalues. Significant improvement is only seen for the symmetric S state (top left plot) where the unstable eigenvalue corresponding to $m=1$ is stabilised, while the unstable eigenvalue corresponding to $m=3$ for the F2 state (bottom right plot) moves to just $\mathcal{O} (10^{-4})$ under the instability line. However, the most unstable eigenvalues corresponding to $m=2$ for the states F1 and F2 where significant improvement is seen using the extended OS analyses, are not notably affected by the eddy viscosity.
Note, to avoid the singularity in the expression for eddy viscosity (\ref{eq:eddy_viscosity}) for the asymmetric state, the mean velocity profile $U$ was symmetrised by $\frac{1}{2}\left(U(y)+U(-y)\right)$ and the Reynolds stress $\overline{ \tilde{u} \tilde{v}}$ was anti-symmetrised by $\frac{1}{2}\left( \overline{ \tilde{u} \tilde{v}}(y) - \overline{ \tilde{u} \tilde{v}}(-y) \right)$.

%
%

So, in summary, for each unstable eigenvalue observed in our test cases, the standard and extended models either agree on their statistical stability prediction, or the extended model shifts the prediction towards statistical stability. The stabilising effect is qualitatively captured by both EOS and mEOS with the mean velocity components making a non-zero power contribution to the eigenvectors of the extended model.  We now examine why the extended stability models sometimes perform better than OS and sometimes make very little improvement.

%
%
\begin{table}
\begin{center}
\begin{tabular}{@{}crccr@{}}
 \quad & \hspace{1.5cm}  &\hspace{1cm} & \quad & \hspace{1.5cm}  \\ \hline \hline
 & & & & \\
\multicolumn{2}{c}{F2 state: leading eigenvalue for $m=2$}             &     &  
\multicolumn{2}{c}{S state: leading eigenvalue for $m=1$} \\  
 & & &  &\\ \cline{1-2} \cline{4-5}     
 & & &  &\\
$\sigma_{OS}$  &  0.06058 +    0.8888i  & & $\sigma_{OS}$ & 0.0024849 +    0.25601i\\
$\sigma_{OS}+\textcolor{blue}{\D \sigma_A}+\textcolor{red}{\D \sigma_S}$ &    -0.16937+     0.80087i
& &
$\sigma_{OS}+\textcolor{blue}{\D \sigma_A}+\textcolor{red}{\D \sigma_S}$ & 0.0029878 +     0.25637i \\     
$\sigma_{mEOS}$ &  -0.02294 +     0.7043i
& & 
$\sigma_{mEOS}$ & 0.0028772 +    0.25653i \\
               &   &   &    &                   \\
$\textcolor{blue}{\D \sigma_A}$ &    \textcolor{blue}{0.02738+0.05860i}
&  &  
$\textcolor{blue}{\D \sigma_A}$ &    \textcolor{blue}{0.00011499-0.0000561i}\\
$\textcolor{red}{\D \sigma_S}$  &    \textcolor{red}{-0.25733-0.02933i} &  &  
$\textcolor{red}{\D \sigma_S}$  &    \textcolor{red}{0.00038787+0.0004194i}\\
              &     &  &                                       \\ \hline \hline
\end{tabular}
\end{center}
\caption{Perturbation analysis results for the leading eigenvalue in the F2 state (left) and the symmetric S state (right). The first order change in the eigenvalue caused by the advection ({\color{blue}$\D \sigma_A$}) and shear ({\color{red}$\D \sigma_S$}) are shown in the minimal extended Orr-Sommerfeld model (mEOS). Unperturbed (standard Orr-Sommerfeld) eigenvalues are given by $\sigma_{OS}$ and  mEOS eigenvalues are given by $\sigma_{mEOS}$. The Table shows that F2 changes to being stable when using mEOS whereas S remains unstable. }
\label{tb:perturbationanalysis}
\end{table}

%
%
  \begin{figure}
  \centerline{
 \includegraphics[width=0.8\textwidth]{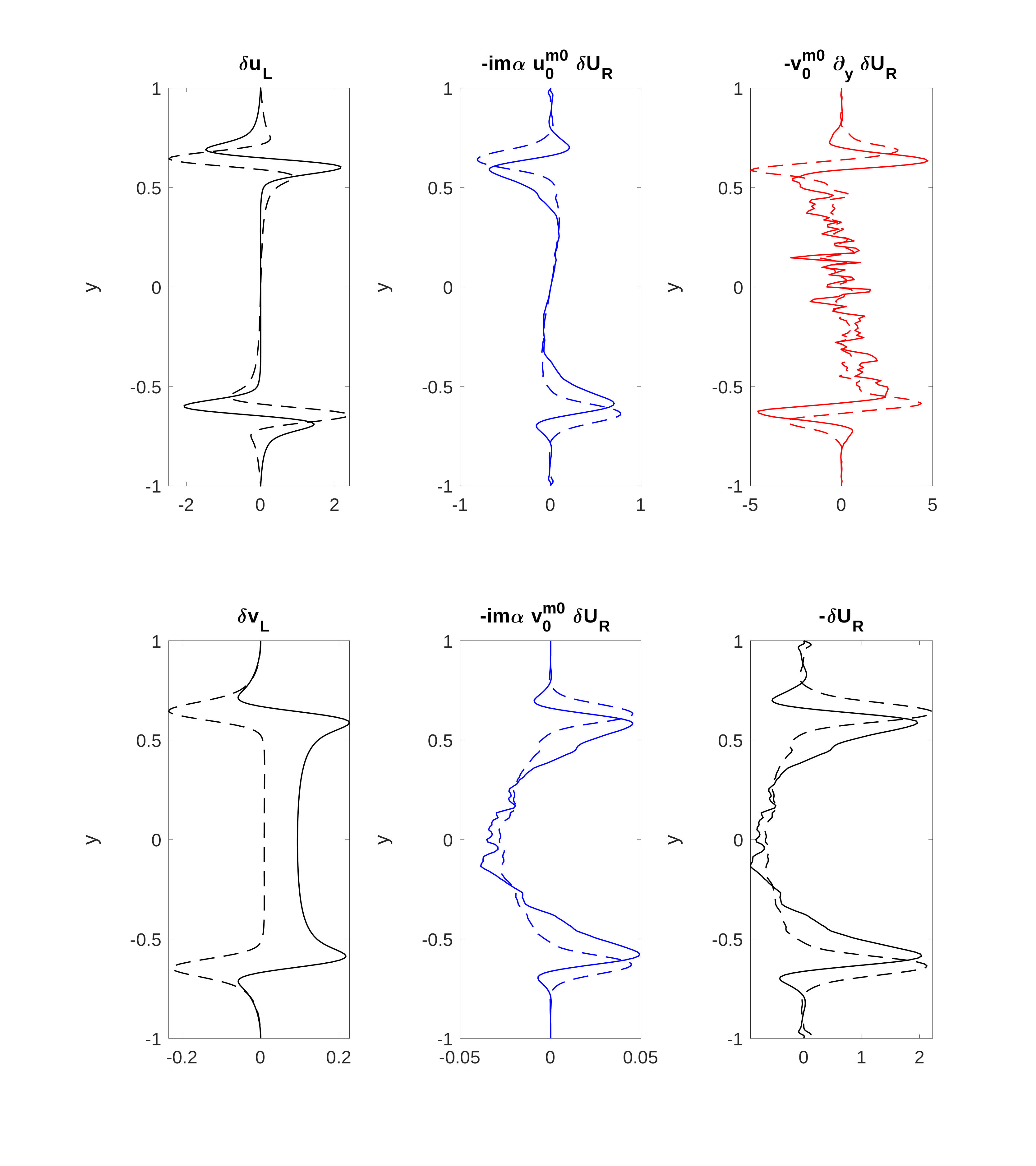}} 
 \caption{\label{fig:perturbation_analysis_F2} \vm{Perturbation analysis results for F2 state. Top row: streamwise component of the unperturbed left eigenvector $\delta u_{sL}$, streamwise component of the perturbed advection term $-im\alpha u_{0s}^{m0} \delta U_{sR}$ and streamwise component of the perturbed shear term $-v_{0s}^{m0} \partial_y \delta U_{sR}$.
Bottom row: wall-normal component of the unperturbed left eigenvector $\delta v_{sL}$, wall-normal component of the perturbed advection term $-im\alpha v_{0s}^{m0} \delta U_R$ and mean profile component of the unperturbed right eigenvector $-\delta U_{sR}$. Solid (dashed) lines indicate real (imaginary) parts of the vectors. Subscript $(\cdot)_s$ refers to the sine components of the eigenfunctions as explained in {Appendix \ref{appendixB}}. }}
 \end{figure}
 
 %
 %
 \begin{figure}
 \centerline{
 \includegraphics[width=0.8\textwidth]{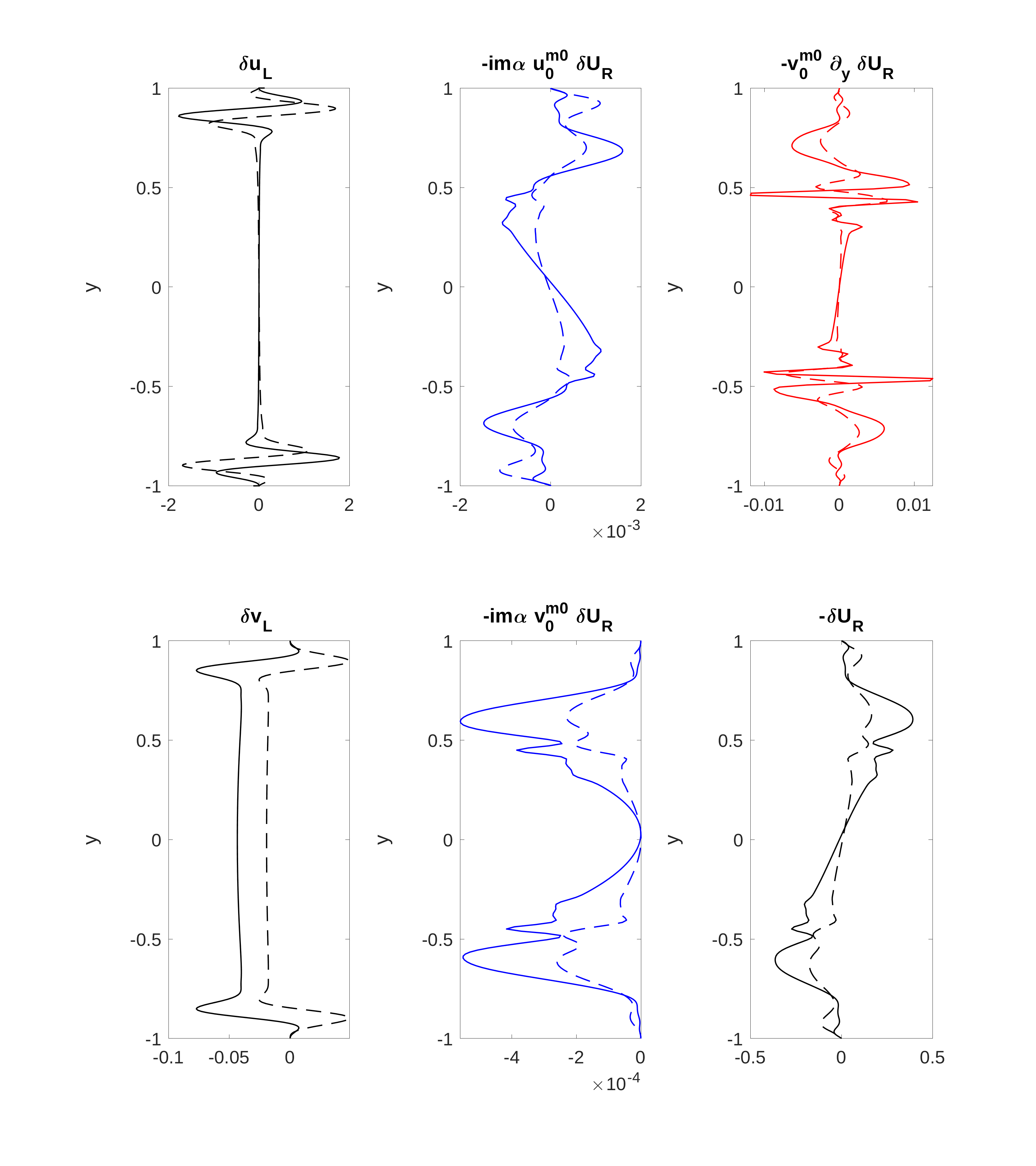} }
 \caption{\label{fig:perturbation_analysis_S} \vm{Perturbation analysis results for S state. Top row: streamwise component of the unperturbed left eigenvector $\delta u_{cL}$, streamwise component of the perturbed advection term $-im\alpha u_{0c}^{m0} \delta U_{cR}$ and streamwise component of the perturbed shear term $-v_{0c}^{m0} \partial_y \delta U_{cR}$.
Bottom row: wall-normal component of the unperturbed left eigenvector $\delta v_{cL}$, wall-normal component of the perturbed advection term $-im\alpha v_{0c}^{m0} \delta U_R$ and mean profile component of the unperturbed right eigenvector $-\delta U_{cR}$. Solid (dashed) lines indicate real (imaginary) parts of the vectors. Subscript $(\cdot)_c$ refers to the cosine components of the eigenfunctions as explained in {Appendix \ref{appendixB}}. }}
 \end{figure}

%
%
\subsection{Eigenvalue perturbation analysis}

{In an attempt to rationalise} the improvement (stabilisation) or not  in mEOS of the leading eigenvalues, {we carry out} an eigenvalue perturbation analysis. While this  neglects the interactions between different streamwise wavenumbers, it includes the minimal improvement that our models offer: adding the mean velocity perturbation into the stability consideration. This leads to an additional equation which governs the evolution of the mean velocity perturbation, and two extra terms - advection and shear - which affect the fluctuation perturbations. While formally limited to only small effects, the perturbation analysis provides a framework to study the structures of the mean velocity perturbations, base fluctuations and fluctuation perturbations and how their interactions affect the eigenvalues. 

The unperturbed mEOS eigenvalue problem is the OS problem (equation (\ref{eq:EOS_fluct}) for one Fourier pairing without the $\dd U$ coloured terms) {\em and} (\ref{eq:EOS_mean}) (we are imagining a small number $\varepsilon$ inserted in front of the coloured terms and developing a perturbation expansion in this  but actually $\varepsilon=1$). The latter is passive: it sets $\dd U$ but there is no feedback to the OS equation. Discretized, it reads
\begin{equation}
    A \bm{y_R} = \sigma B \bm{y_R} 
\end{equation}
with matrices $A,B$, eigenvalue $\sigma$ and right eigenvector $\bm{y_R}:=[\dd U\,\, \bd \bfu^{m0}]^T$. The coloured $\dd U$ terms in (\ref{eq:EOS_fluct}) are then treated as perturbations of $A$  which causes  $(\sigma,\bm{y_R}) \rightarrow (\sigma+\D \sigma, \,\bm{y_R+\D y_R})$ so that
\begin{equation}
    \left(A + {\color{blue}\delta A_A} +{\color{red}\delta A_S}  \right)\left(\bm{y_R} + \bm{\delta y_R } \right) = \left( \sigma +\delta \sigma \right) B \left( \bm{y_R} +\bm{\delta y_R} \right)
\end{equation}
where ${\color{blue}\delta A_A\bm{y_R}}$ corresponds to the advection term ${\textcolor{blue}{-im\alpha \bfu_0^{m0}\dd U_R }}$ and ${\color{red}\delta A_S \bm{y_R}}$ 
to the shear term $\textcolor{red}{-\tv_0^{m0} \partial_y \dd U_R \bm{\hat{x}}}$.
By taking the inner product using the corresponding left eigenvector of the unperturbed system $\bm{y_L}$ (${\bm y_L}^{\dag} A=\sigma {\bm y_L}^{\dag} B$), and considering first order terms only, an expression then follows for the first order perturbation to the eigenvalue $\delta \sigma$: 
\begin{equation}
    {\delta \sigma} = 
    \textcolor{blue}{\delta \sigma_A} +
    \textcolor{red}{\delta \sigma_S}:=
    \frac{\bm{y_L}^{\dagger}\textcolor{blue}{\delta A_A\bm{y_R}}}{\bm{y_L}^{\dagger}B\bm{y_R}}+\frac{\bm{y_L}^{\dagger}\textcolor{red}{\delta A_S\bm{y_R}}}{\bm{y_L}^{\dagger}B\bm{y_R}}.
\end{equation}

We present the two most interesting cases of the eigenvalue perturbation analysis. In the first case, we perform eigenvalue perturbation analysis on the leading eigenvalue of the standard Orr-Sommerfeld analysis for the F2 state, which is seen to stabilise in mEOS and EOS. The second case corresponds to the leading eigenvalue of the symmetric state, which shows no significant difference between the standard and extended approaches. The eigenvalues as well as perturbation analysis predictions are summarised in Table \ref{tb:perturbationanalysis}. The left and right eigenvectors and perturbed terms are shown in figures \ref{fig:perturbation_analysis_F2} and \ref{fig:perturbation_analysis_S}.

For the body-force-driven F2 case, the advection term has a small and positive contribution to the real part of $\delta \sigma$ while the shear term has a large negative contribution to the real part of $\delta \sigma$, as summarised in Table 1.
This latter prediction is only qualitatively correct as the shear term is not a small perturbation giving a much larger predicted decay rate of $-0.169$ than the actual value of $-0.023$ obtained from mEOS. The perturbation analysis, however, indicates that  the shear in the perturbed mean field is causing  the stabilisation.
Looking in more detail at the fields in figure \ref{fig:perturbation_analysis_F2}, it is clear that  the active regions of the left and right eigenvectors coincide. In particular, the shear in $\delta U_R$ coincides with where the  left eigenvector is significant and is of opposite phase giving the large stabilising effect. In contrast, the resulting advection term is not only an order of magnitude smaller than the shear term, but also has the same sign contribution as the left eigenvector, yielding a small destabilizing contribution to the eigenvalue. 

For the symmetric S case, the perturbation analysis shows both advection and shear terms make small destabilizing contributions to the eigenvalue commensurate with mEOS although there is still an error due to the finiteness of the perturbation.  Examining the structure of the left and right eigenvectors shown in figure \ref{fig:perturbation_analysis_S} indicates that they are ill-matched. The left eigenvector is concentrated much closer to the channel walls than the right eigenvector. Moreover, even though the right eigenvector component $\delta U$ has some structure near the wall, the interaction of this component with the perturbed operators moves this structure towards the centre of the wall for the shear term, avoiding any significant interaction with the left eigenvector. The advection term has some overlap with the left eigenvector, but it is also an order of magnitude smaller than the shear term, overall resulting in small destabilizing contributions. It's also worth remarking that even at resolution of $2048$ wall-normal Chebyshev modes, the eigenvectors possess the same small scale structure seen in the figures. 

The `take home' message from this section is that the mean velocity perturbation, even if an order magnitude smaller than the other components of the eigenvector - see Figure \ref{fig:evect_spectrum}, can interact with the base flow fluctuation fields in a way that produces a strong stabilising influence. 


%
%
\section{\label{V}Model limitations}

In this section, we discuss the various  approximations made in deriving EOS and mEOS which include  time-averaging the statistics, approximating correlation matrices as rank 1 and neglecting higher order statistics. {As already mentioned, the channel flow states considered above were assumed statistically steady. Making this assumption is required to apply Malkus's OS analysis in the first place and so subsequently EOS and mEOS, but is only an approximation. Time-averaging the statistics in a small computational box is a commonly-used approach to get a better estimate of the  steady statistics hypothetically generated in an infinite computational box. } 

%
%
\subsection{\label{V.A}Rank 1 approximation of correlation matrices}

\begin{figure}
\centerline{
\includegraphics[width=0.5\textwidth]{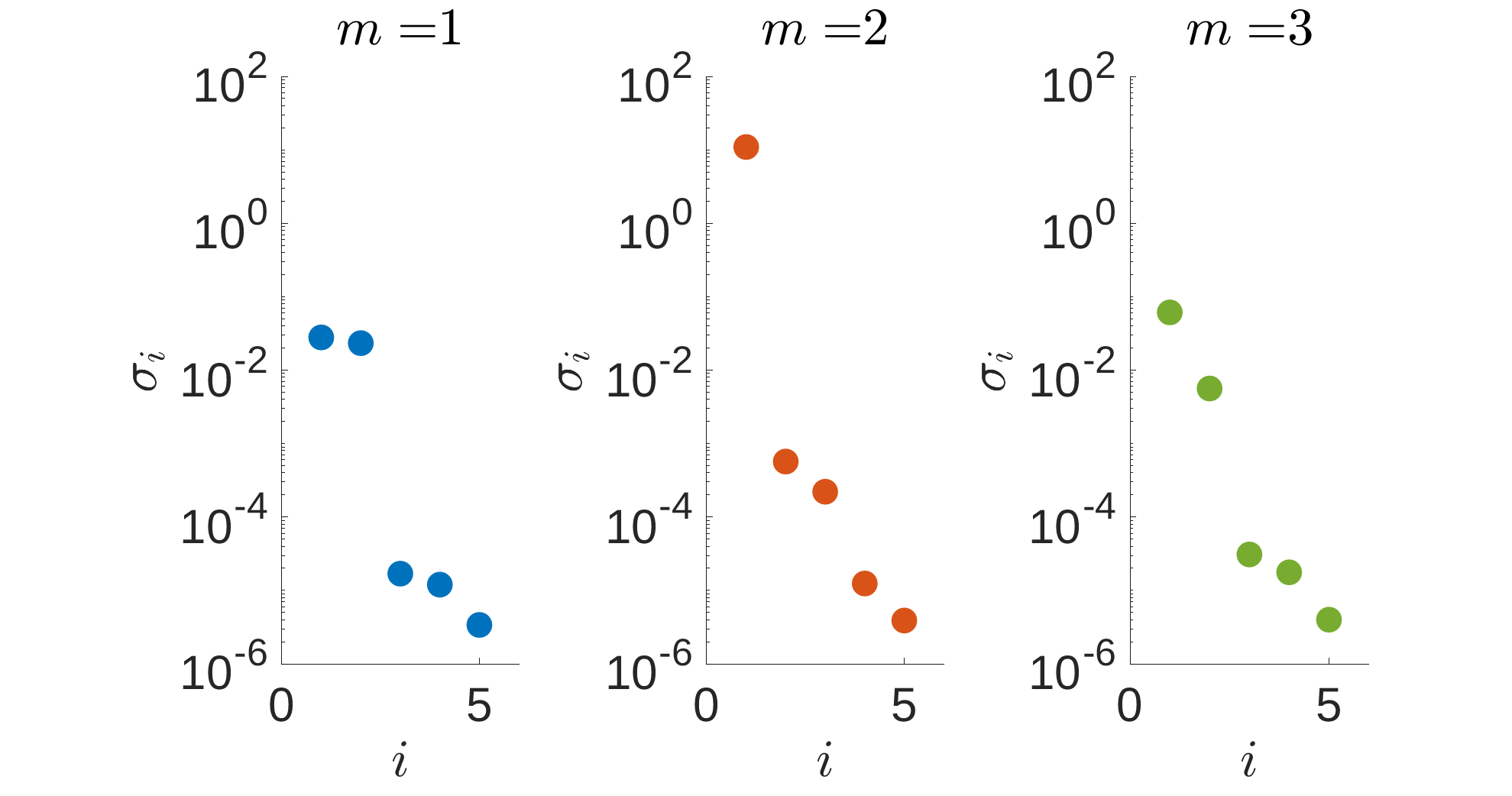}%
\includegraphics[width=0.5\textwidth]{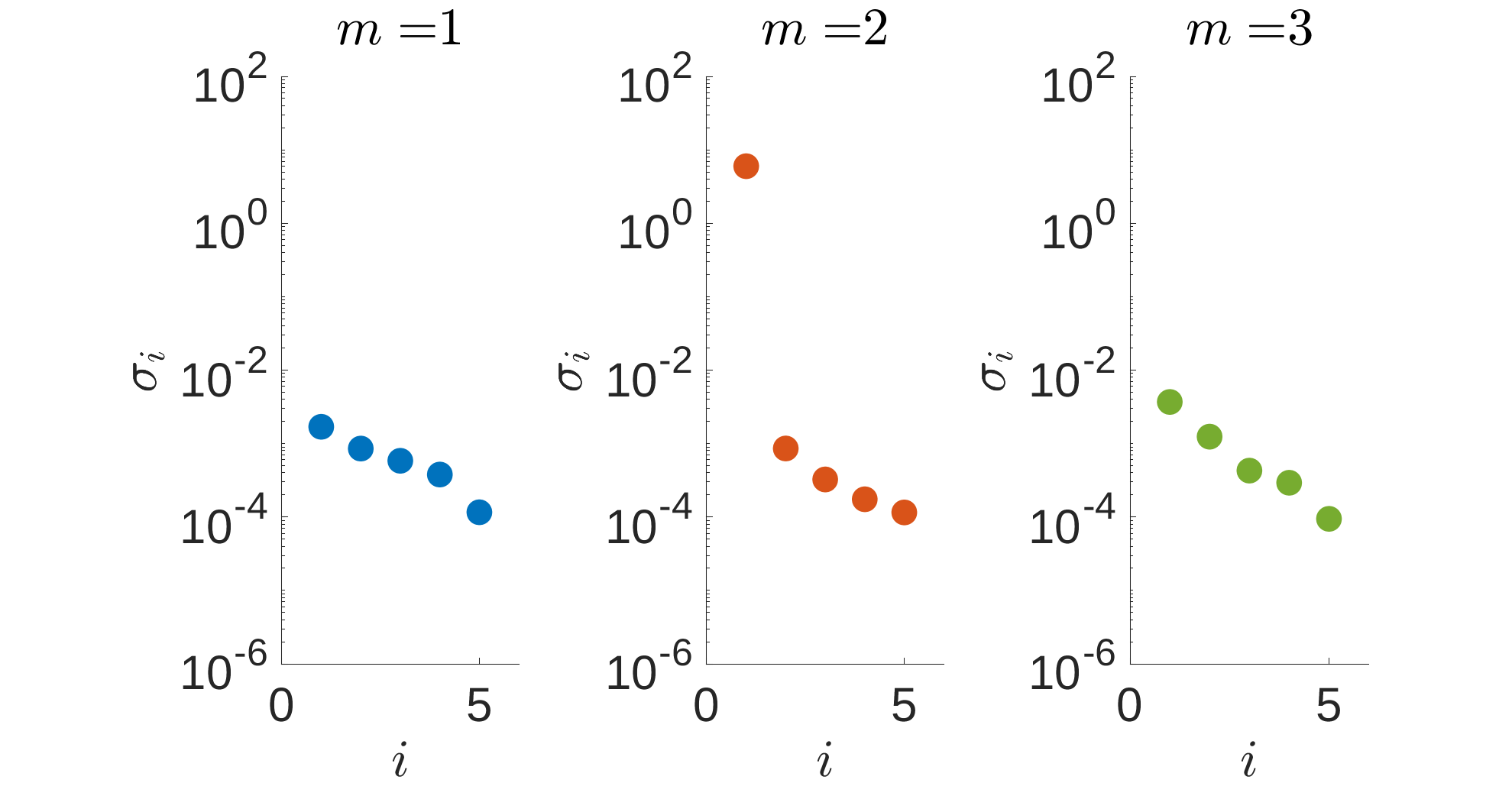}}
\centerline{\includegraphics[width=0.5\textwidth]{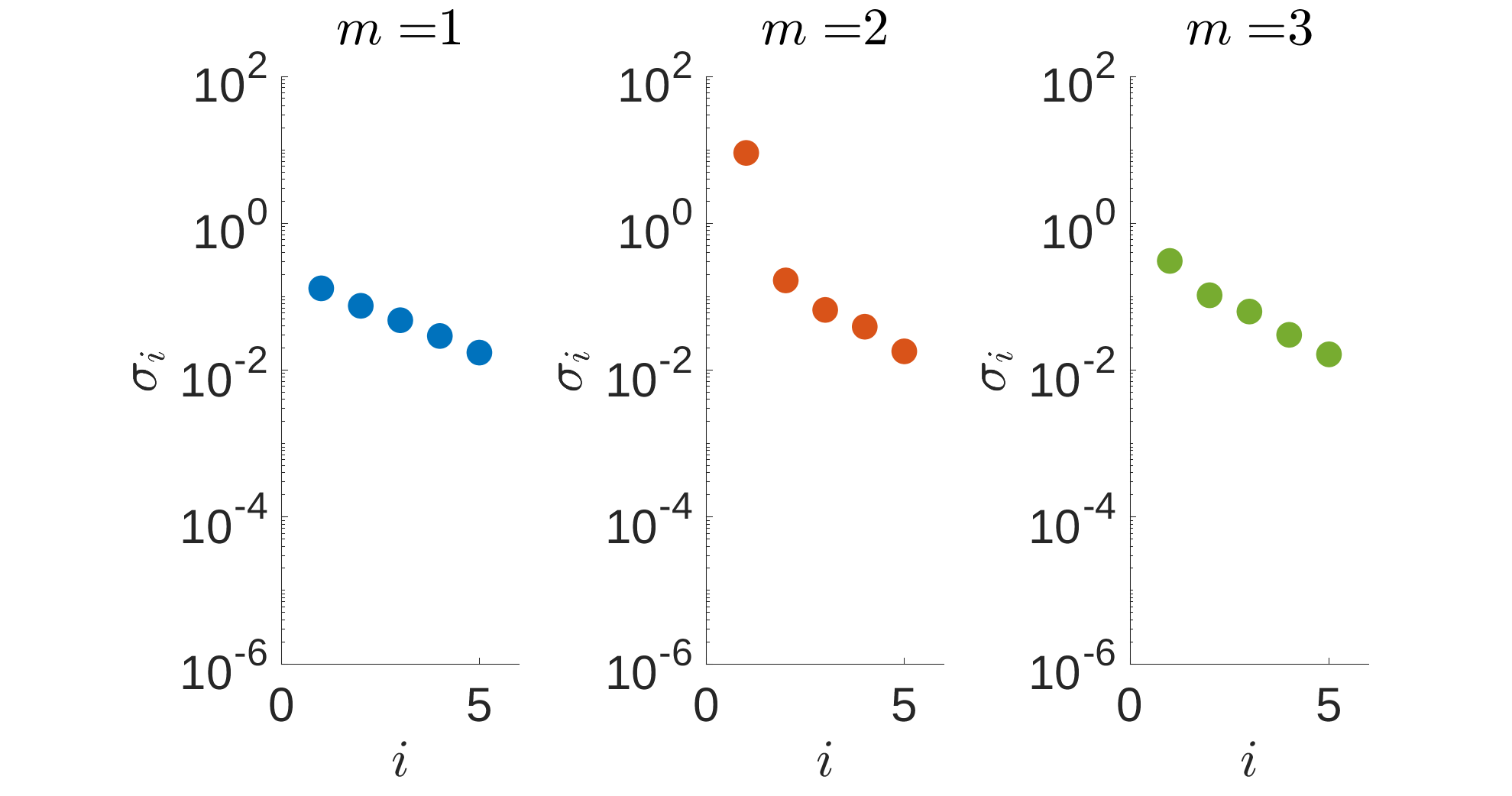}%
\includegraphics[width=0.5\textwidth]{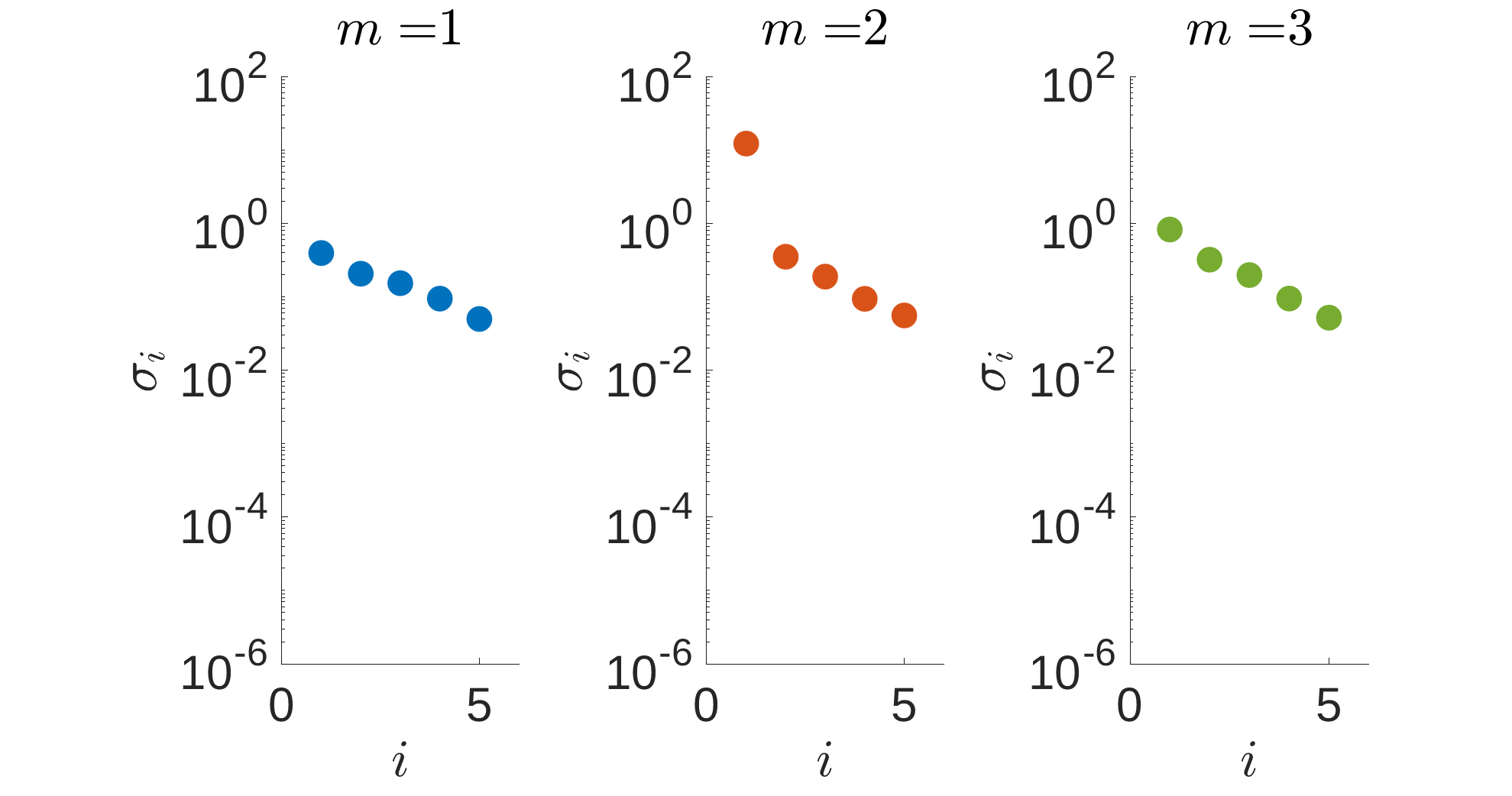}}%
\caption{\label{fig:SVD} Leading singular values $\sigma_i$ of the correlation matrices $\bC^{m0}$ time-averaged over 1000 time units. Shown for {\color{blue}$m=1$ (blue)}, {\color{BrickRed}$m=2$ (red)} and {\color{OliveGreen}$m=3$ (green)}. 
Top left: symmetric S, top right: asymmetric A, bottom left: F1, bottom right: F2 states. Note the gap after the leading singular value for {\color{BrickRed}$m=2$ (red)} across all states confirming rank 1 approximation of the correlation matrix for the cases when mEOS and EOS analyses show stabilisation.}
\end{figure}

Time-averaging the correlation matrix increases its rank from 1 up to potentially its full dimension - see Fig. \ref{fig:SVD} for singular values $\sigma_i$ of correlation matrices time-averaged over 1000 time units. {This is not necessarily a reflection that the statistics are non-stationary as a fluctuation field with two frequencies will give a rank-2 correlation matrix under averaging.} 
For $m=2$ which is the dominant streamwise wave-number in all of the test states, a gap between the leading and the second singular values is larger than an order of magnitude indicating that a rank-1 approximation of the time-averaged correlation matrix may be  reasonable.  On the other hand, time-averaged correlation matrices for $m \in\{1,3\}$ do not show a significant gap between the leading and second singular values. In this case, it is much more likely that some important fluctuation field features might not be accurately represented by the rank 1 representation of the correlation matrix and, as a result, the approach is less justified.

In an attempt to explore a more accurate representation of
$\bC^{mn}=\sum_{i=1} \sigma_i \,{\bf \hat{e}_i} \otimes \, {\bf \hat{e}_i}$ (where
${\bf \hat{e}_i}$ is the $i^{th}$ normalised eigenvector and $\sigma_i$ the largest eigenvalue or singular value of $\bC^{mn}$ as it is Hermitian) beyond just 
the rank-1 approximation $\bC^{mn} \approx \sigma_1 \,{\bf \hat{e}_1} \otimes \, {\bf \hat{e}_1}$ ($\bfu^{mn}_0=\sqrt{\sigma_1} {\bf \hat{e}_1}$), a weighted average of the leading $N$ fields,
\begin{equation}
    \bfu^{mn}_0=\sum_{i=1}^{N} \frac{\sigma_i}{\sum_{i=1}^{N} \sigma_i} \,\sqrt{\sigma_i} {\bf \hat{e}_i},
\end{equation}
was also considered for the F2 state where both $m=1$ and $m=2$ wavenumbers have a non-rank 1 time-averaged correlation matrix $C^{m0}$. However, including $N=1,2,5$ or even $10$ eigenvectors into the expansion showed no significant difference in the leading eigenvalues (less than $1\%$ difference in the absolute value of the eigenvalue). 
Understanding the effect of a rank-1 approximation really requires time-stepping the full statistical equations which is beyond the scope of this work.

%
%
%
%
\subsection{Time-averaged statistics}

\begin{figure}
 \centerline{
 \includegraphics[width=0.8\textwidth]{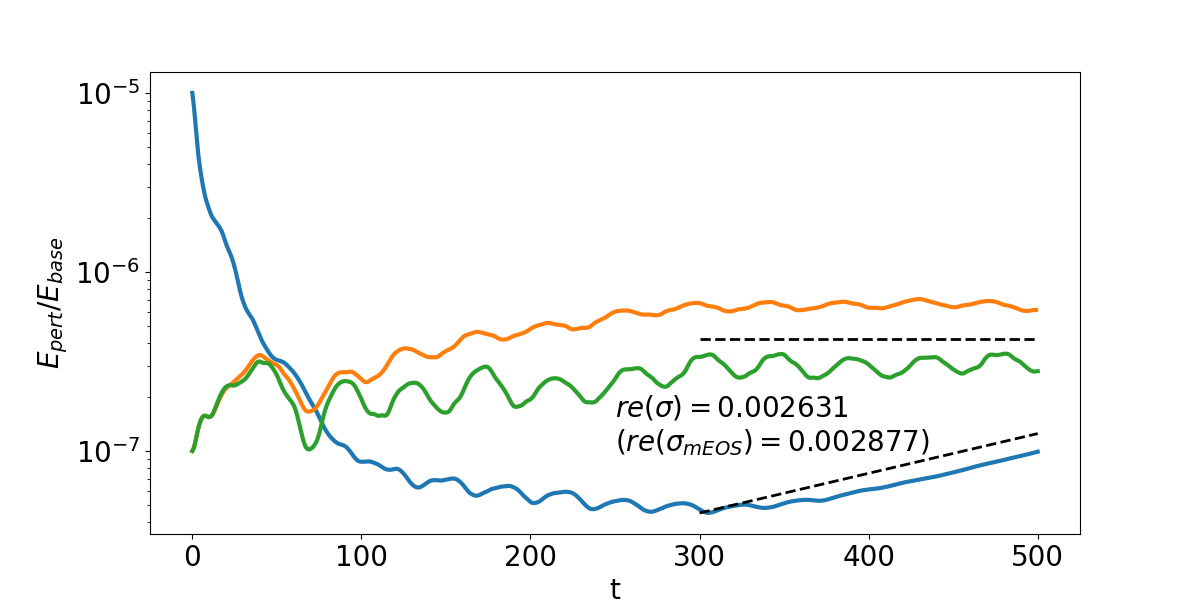} }
 \centerline{
 \includegraphics[width=0.8\textwidth]{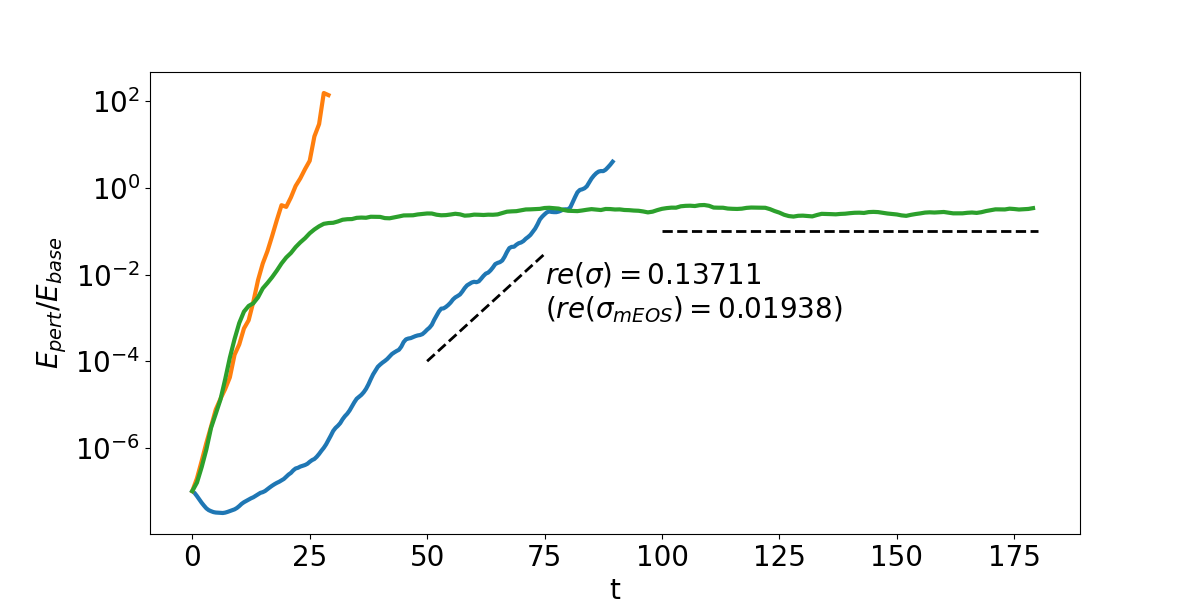}}
 \caption{\label{fig:QLvsDNS} Results of various numerical experiments: time evolution of the perturbation energy $E_{pert}$ as a ratio to the base flow energy $E_{base}$ averaged over multiple DNS runs. Perturbations were time-stepped using linearised quasilinear (blue), linearised Navier-Stokes (orange) or full non-linear Navier-Stokes (green) equations. For each case, base flow was time-dependant and generated using full Navier-Stokes direct numerical simulations. Results shown for the symmetric S (top) and body-force driven BF2 (bottom) states. Black dashed lines are provided as guides to emphasise the zero gradient of the saturated states and to indicate the growth rate $re(\sigma)$ of the unstable states. For comparison, the real part of the leading mEOS eigenvalue $re(\sigma_{mEOS})$ is given for each of the states.}
 \end{figure}

\begin{figure}
 \centerline{
    \includegraphics[width=0.33\textwidth]{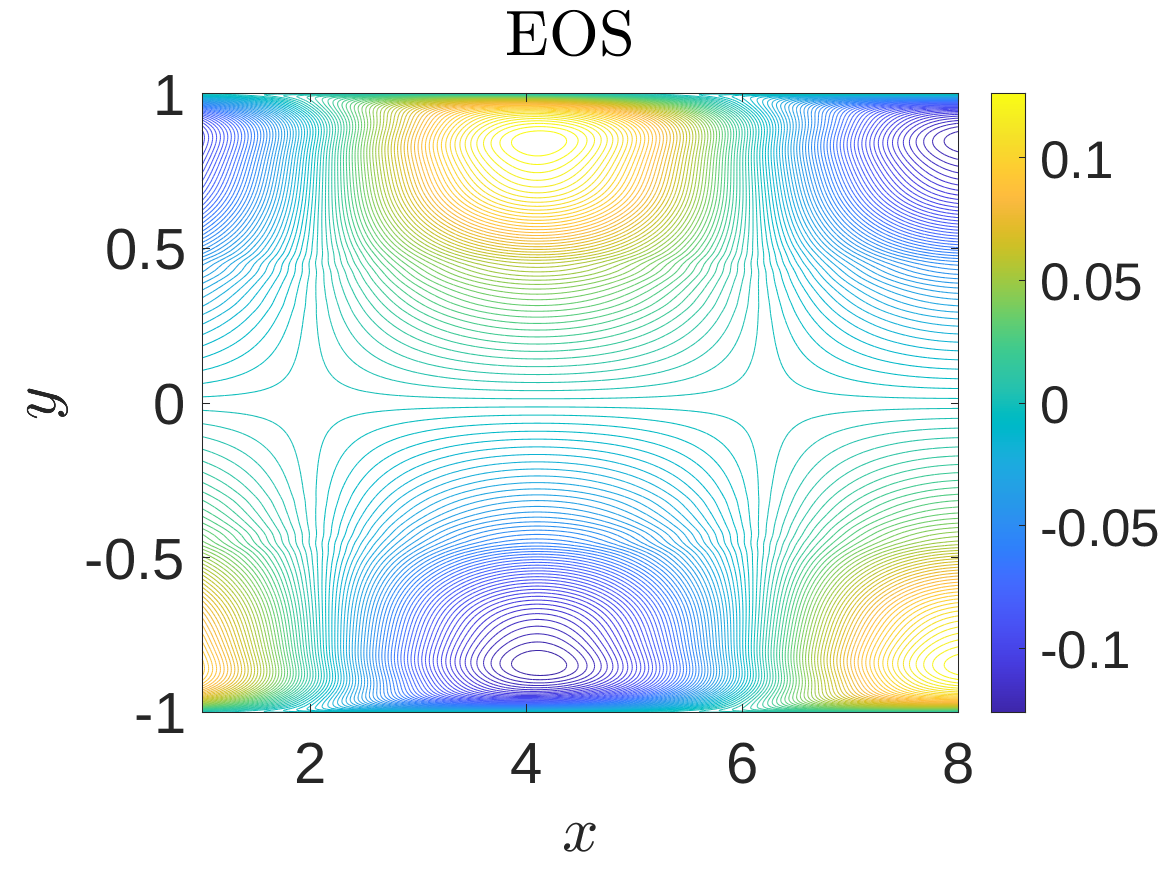} 
    \includegraphics[width=0.33\textwidth]{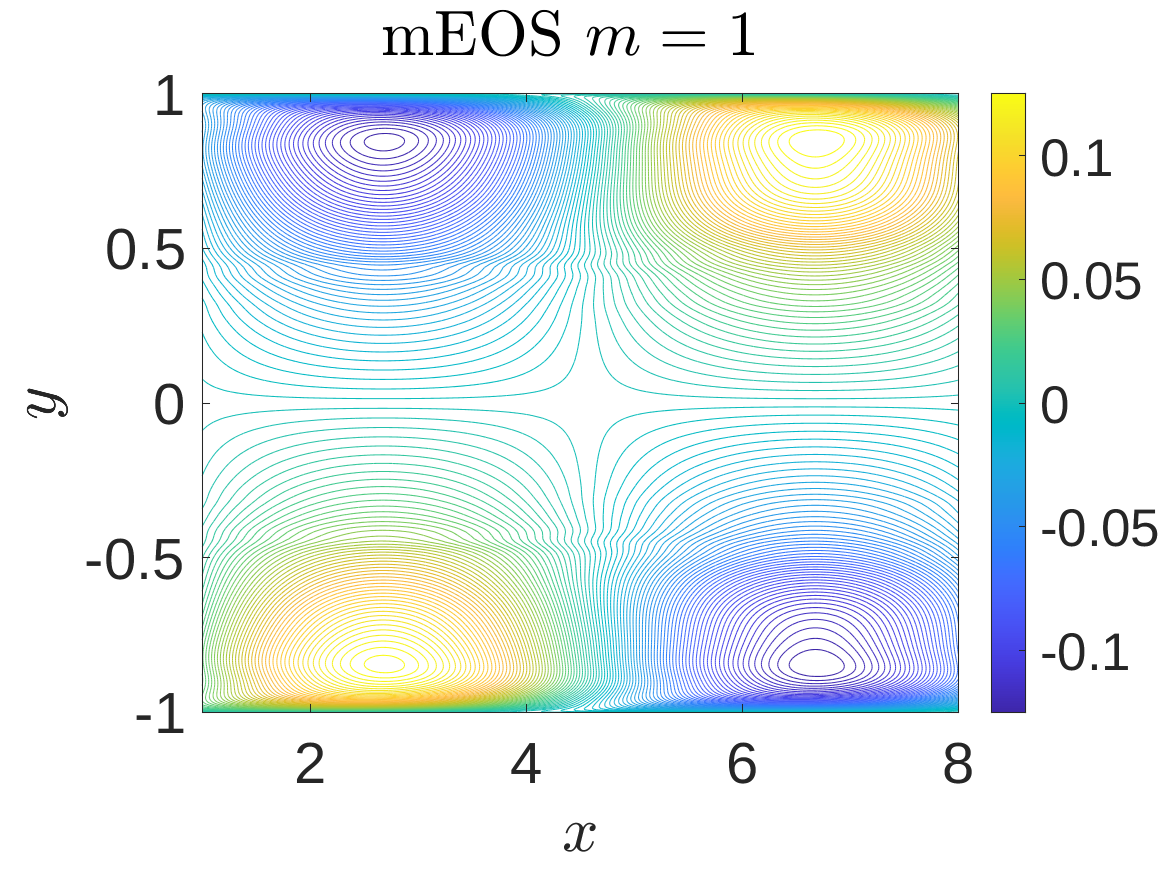}
    \includegraphics[width=0.33\textwidth]{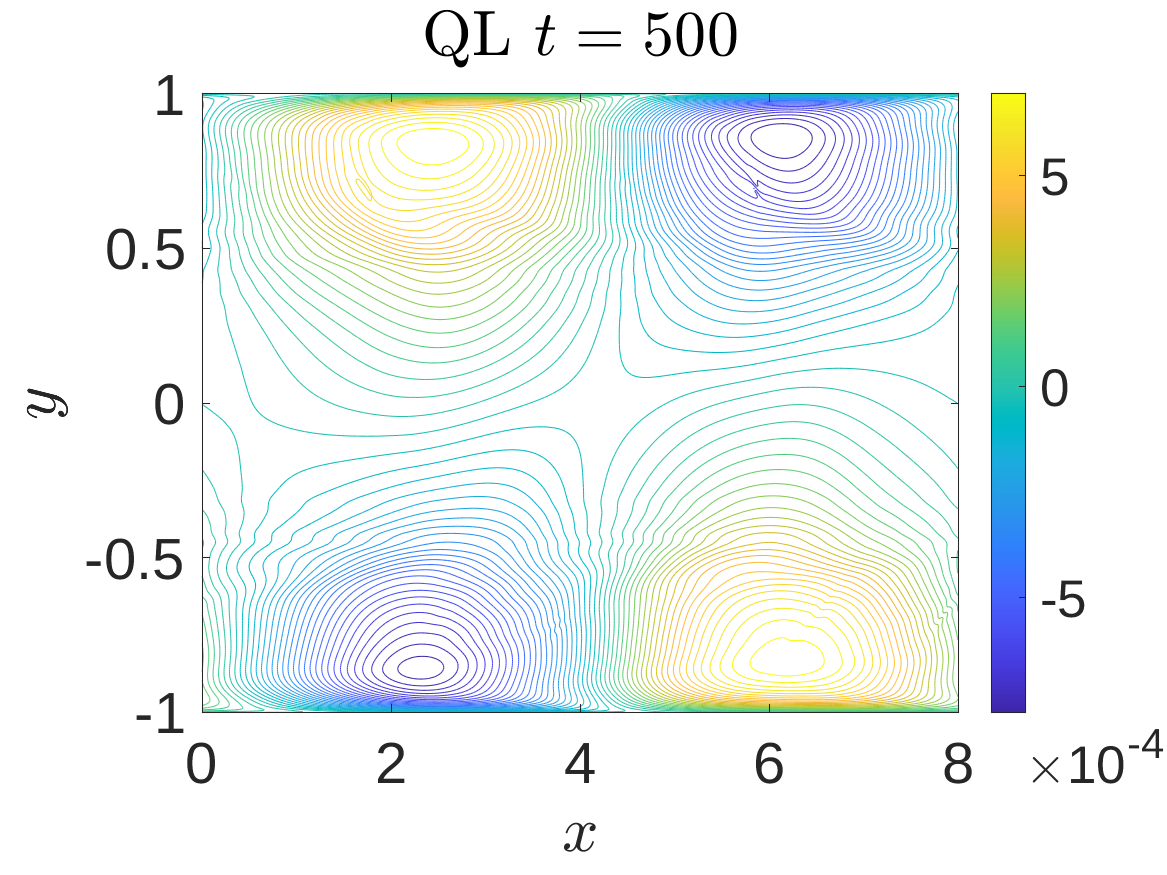} }
    \caption{\label{fig:11} Comparison of the leading eigenvectors from the EOS (left), mEOS $m=1$ (middle) analyses and an instantaneous snapshot of the velocity perturbation field in the QL experiment at $t=100$ (right) for the symmetric (S) state. Only the streamwise velocity component is shown. There is a good correspondence in the leading eigenvector across EOS and mEOS analyses and the QL experiment using a time-dependent base flow. }
 \end{figure}

To understand the effect of time-averaging the statistics in the Extended Orr-Sommerfeld stability models, numerical experiments were performed in which the perturbations to the turbulent base flow were exposed to a time-varying base flow. 
This experiment was first performed for the symmetric (S) state where the unstable Orr-Sommerfeld eigenvalue remains unstable in EOS and mEOS.
From the DNS data, 10 turbulent flow histories were taken of length $\Delta t=500$ separated by at least 1000 time units to avoid statistical dependence. Each was taken as a turbulent, time-dependent base flow. 
Two random perturbations with energy $E_{pert} = 10^{-5} E_{base}$ of the base flow were then added to each of the 10 base flow histories (making a total of 20 cases) 
and then their evolution computed over a $\Delta t=500$ time window using the QL equations. For $(\delta U, \bd\bfu^{m0})$, these read
\begin{align}
  \partial_t\delta U&=\frac{1}{Re}\partial^2_y \delta U+\delta G -[\,\bfu_0(t).\bnab \bd \bfu+ \bd\bfu.\bnab \bfu_0(t)\,]^{00} \\
  \partial_t{\bd \bfu}^{m0}  &= \frac{1}{Re} \nabla^2 \bd  \bfu^{m0}-\bm{\nabla} \dd \tp^{m0} - im U \bd \bfu^{m0} - \dd \tv^{m0}U_y\bm{\hat{x}}
-im{\delta U} \bfu_0^{m0} -\tv_0^{m0} \delta U_y \bm{\hat{x}}, \quad  m>0
\label{QL}
\end{align}
so that the perturbation-base flow fluctuation interactions are not included in the perturbation fluctuation equation (compare with (\ref{linearNS}) below where they are retained). 
Averaging the perturbation energy growth over the 20 different simulations revealed an unstable mode with exponential growth after initial transients decayed. The growth rate of this unstable mode was found to be $\sigma=0.00263$ which is within $5\%$ of the growth rate predicted by the Extended Orr-Sommerfeld analysis - see Fig. \ref{fig:QLvsDNS}. The most unstable eigenvectors of the EOS and mEOS analyses agree and are observed as the growing structure in this QL experiment - see Fig. \ref{fig:11}. We conclude that time dependence of the base flow is not enough to recover the statistical stability of the symmetric state S.

%
%
%
%

%
%
\subsection{Importance of higher order statistics}

To evaluate the need of flow statistics beyond the second order to recover statistical stability of the symmetric (S) state, we repeated the numerical experiment described above but this time the perturbation fields were time-stepped using the linearised Navier-Stokes equations where perturbation-base flow fluctuation interactions are now included and the base flow is time varying: 
\begin{align}
    \partial_t{\bd \bfu}^{m0} & = \frac{1}{Re} \nabla^2 \bd  \bfu^{m0}-\bm{\nabla} \dd \tp^{m0} - im U(t) \bd \bfu^{m0} - \dd \tv^{m0}U_y(t)\bm{\hat{x}}
-im{\delta U} \bfu_0^{m0}(t)\nonumber \\
& \hspace{2cm}-\tv_0^{m0}(t) \delta U_y \bm{\hat{x}}  
-[\,\bfu_0(t).\bnab \bd \bfu+ \bd\bfu.\bnab \bfu_0(t)\,]^{m0}.
    \label{linearNS}
    \end{align}
 We repeat the experiment with initial $E_{pert} = 10^{-5} E_{base}$ and $E_{pert} = 10^{-7} E_{base}$ obtaining the same qualitative results. When the energy growth of the perturbation field is time-averaged over 20 different runs, the perturbation growth is seen to saturate after an initial transient growth - see Fig. \ref{fig:QLvsDNS} (top). This suggests that the higher order statistics ignored in QL are important here  for recovering statistical stability of the symmetric (S) turbulent state. 

%
%
 \begin{figure}
 \centerline{
    \includegraphics[width=0.5\textwidth]{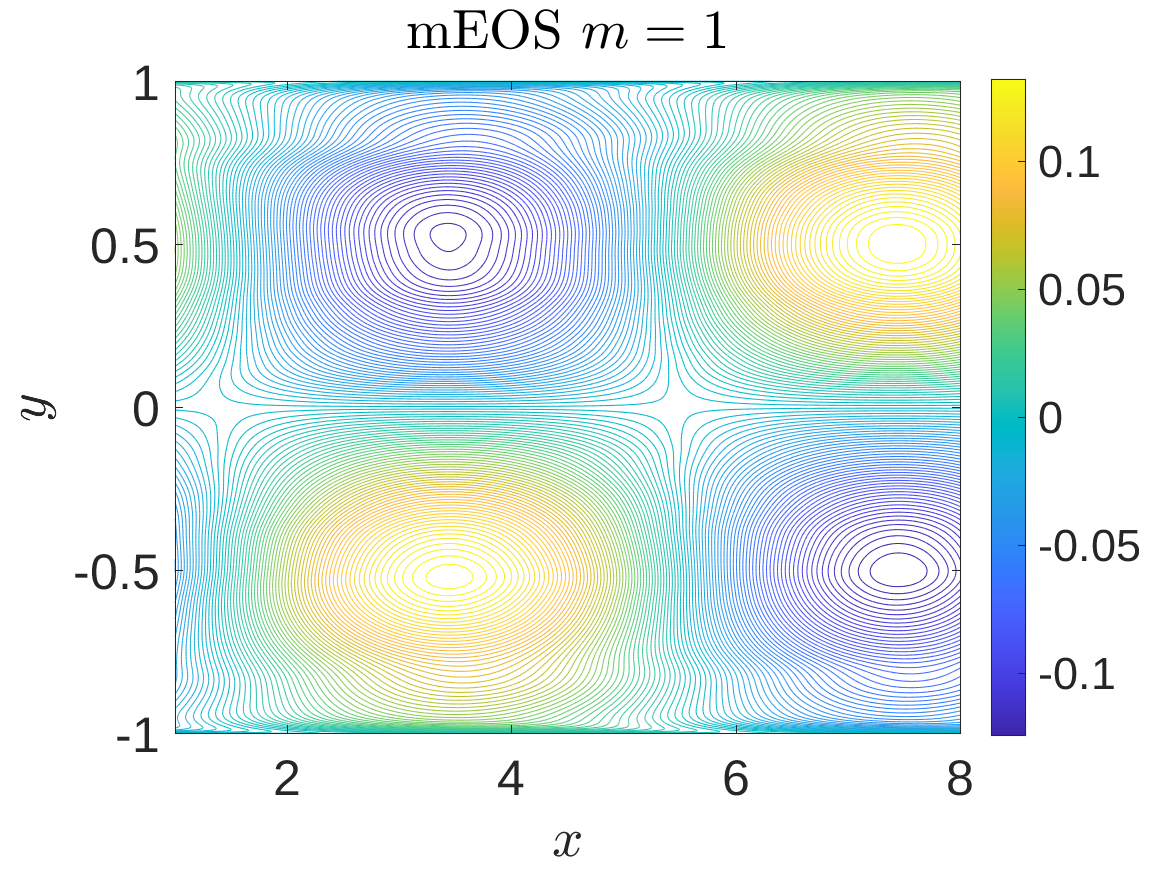} 
    \includegraphics[width=0.5\textwidth]{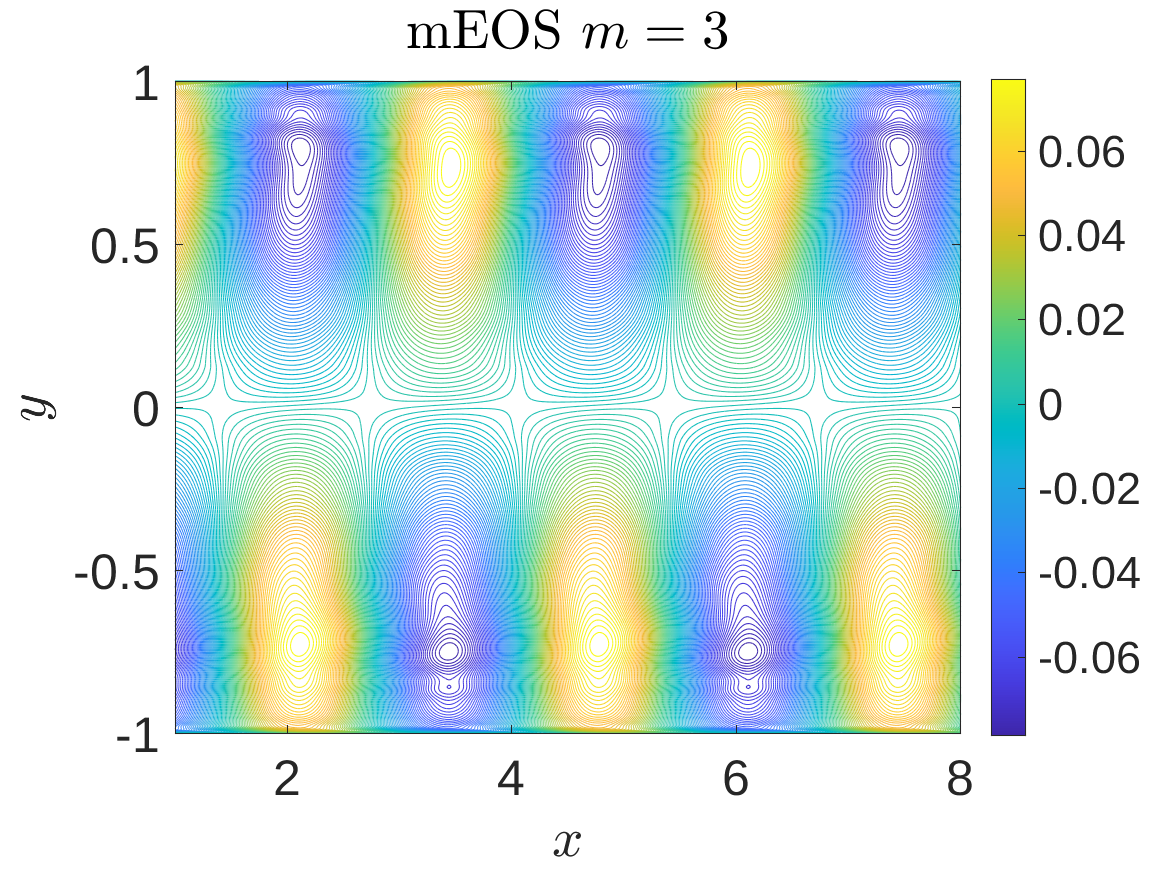}}
   \centerline{ \includegraphics[width=0.5\textwidth]{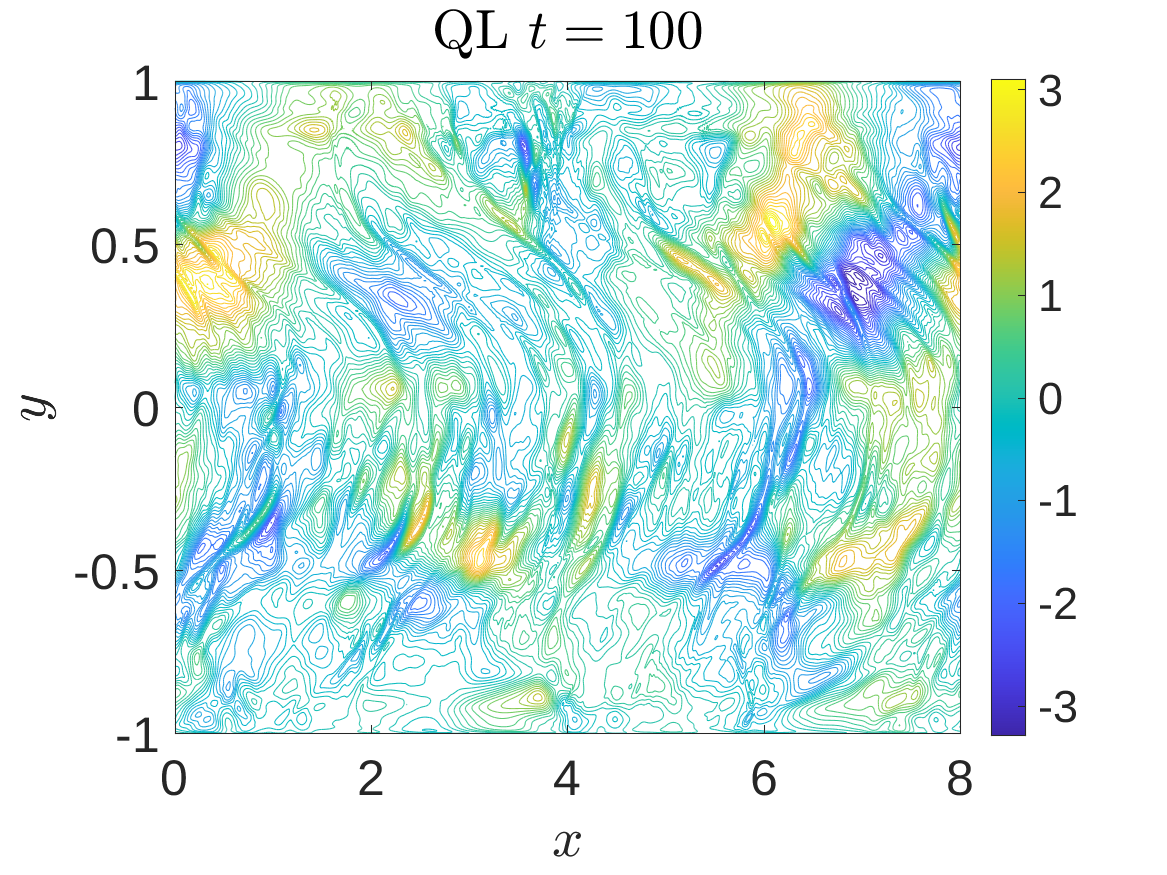} \includegraphics[width=0.5\textwidth]{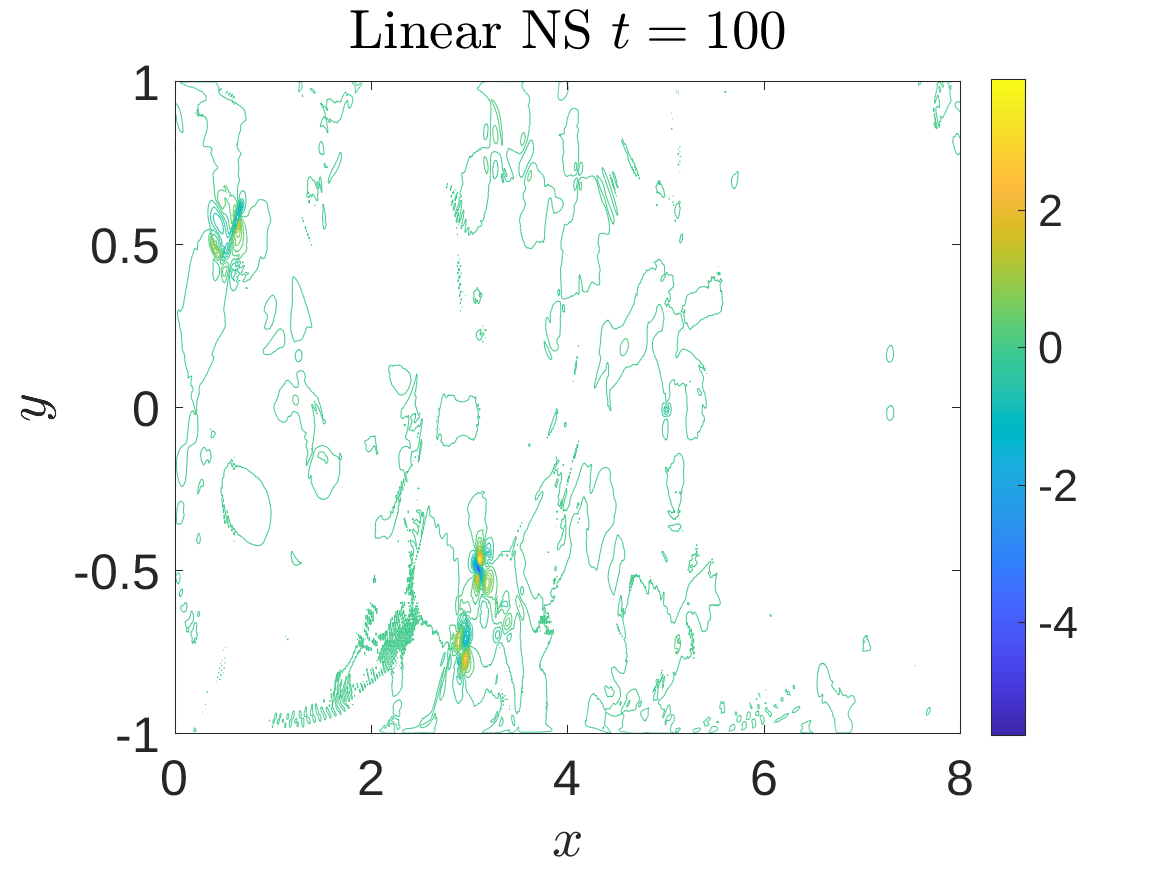} }
    \caption{\label{fig:12} Comparison of the leading eigenvectors from the mEOS $m=1$ (top left) and mEOS $m=3$ (top right) analysis with the instantaneous snapshots of the perturbation fields obtained from the QL experiment (bottom left) and linear NS experiment (bottom right) for the F2 state. The structures of the leading mEOS eigenvectors seem unrelated to the  growing perturbations in the QL or LNS regimes. }
 \end{figure}

%
%
\subsection{Non-linear effects}

Extended Orr-Sommerfeld stability analysis applied to the body force F2 state shows a significant stabilisation of the eigenvalue corresponding to the second streamwise wavenumber, but shows no significant difference for the other two unstable eigenvalues of the standard Orr-Sommerfeld approach.  Repeating the numerical experiments described above applied to the F2 state shows that both the quasi-linear (QL) equations {\em and} the  linearised Navier-Stokes (LNS) equations yield rapid exponential growth of the perturbations applied to the time-dependent base flow: see Figure \ref{fig:QLvsDNS} (bottom). Periodically renormalising the growing perturbation in the latter LNS case and continuing the simulation shows sustained growth over a period of O(1000) time units (not shown). This behaviour is in contrast to what was found for state S where the perturbation saturated (see Figure \ref{fig:QLvsDNS} (top)\,). On the face of it, this experiment seems to show that state F2 is linearly unstable but this does not necessarily follow. Time-stepping a perturbation using the linearised Navier-Stokes equations (or in the tangent space of a turbulent attractor) can continually sample the stretching or contracting dynamics {\em along} the attractor which can dominate the anticipated decaying behaviour perpendicular to the attractor. It seems stretching along the attractor dominates for state F2 but not for state S.
A comparison of the leading growing structures in Figure \ref{fig:12} is consistent with this: the QL and LNS flows bear no similarity with the predictions of mEOS. These LNS calculations for the state F2 highlight perfectly the difficulty in assessing the (linear) statistical stability using the Navier-Stokes equations touched on in the introduction.

%
%
 \begin{figure}
 \centerline{
    \includegraphics[width=0.8\textwidth]{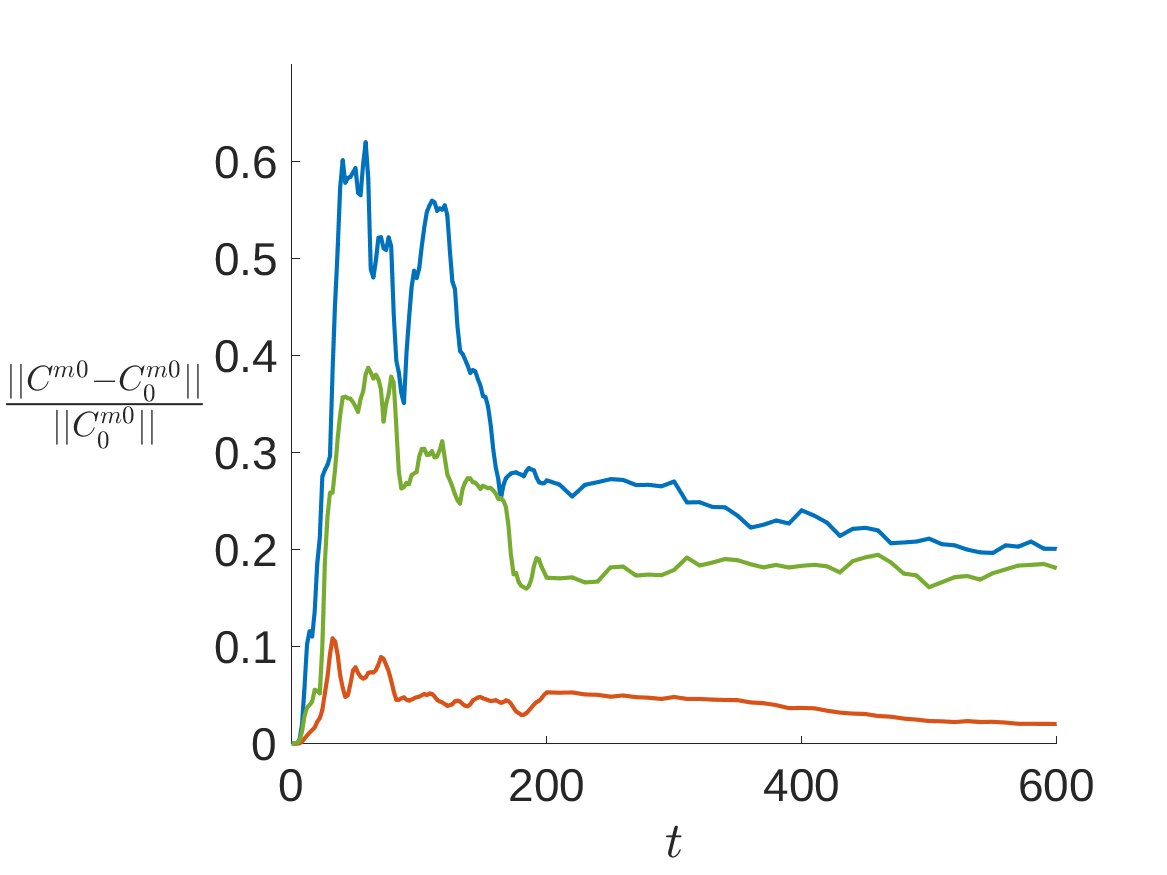}}
    \caption{\label{fig:13} Time evolution of the relative difference between correlation matrices of the full flow $C^{m0}$ and base flow $C^{m0}_0$ shown for {\color{blue}$m=1$ (blue)}, {\color{BrickRed}$m=2$ (red)} and {\color{OliveGreen}$m=3$ (green)}. The correlation matrices were time-averaged over a rolling time window to display the statistical behaviour. Shown for 
    the full non-linear NS experiment  
    where the difference slowly approaches $0$ signifying the recovery of the same statistical state. }
 \end{figure}

\begin{figure}
 \centerline{
    \includegraphics[width=0.5\textwidth]{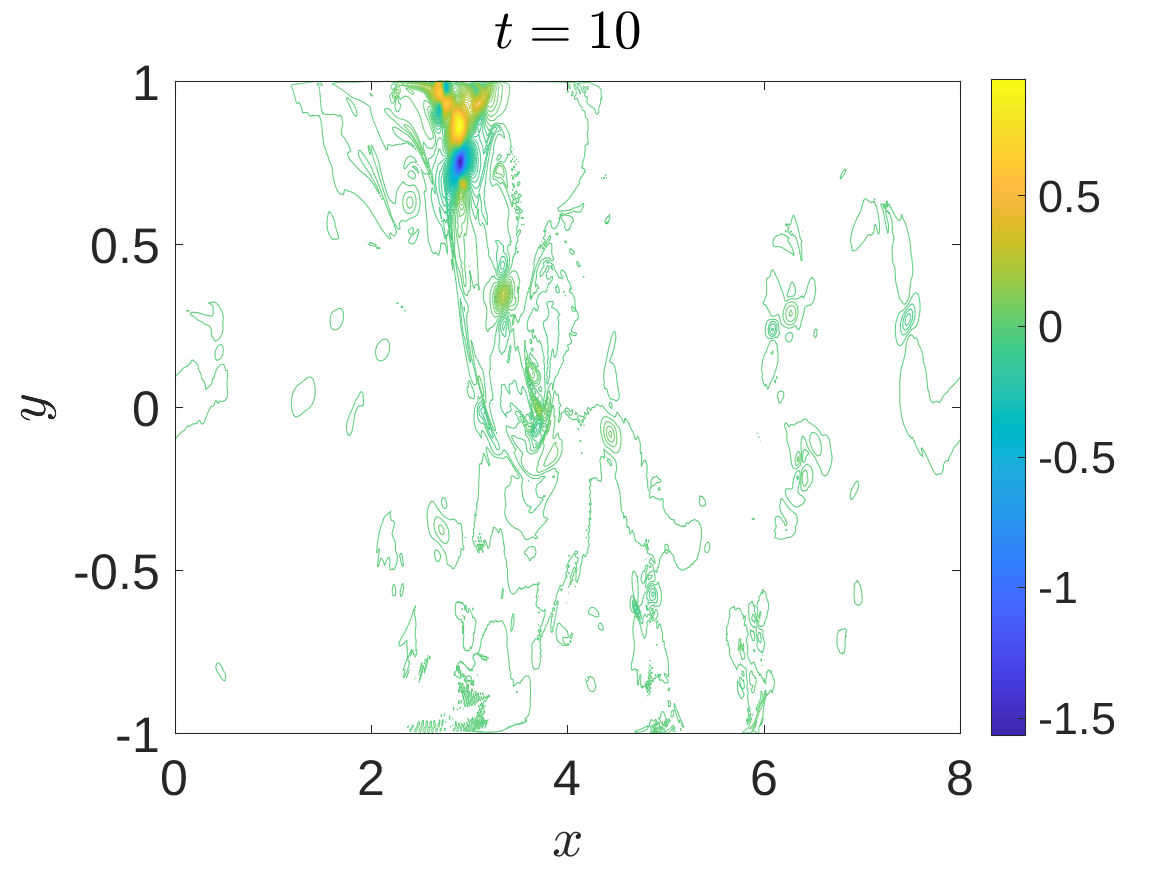} 
    \includegraphics[width=0.5\textwidth]{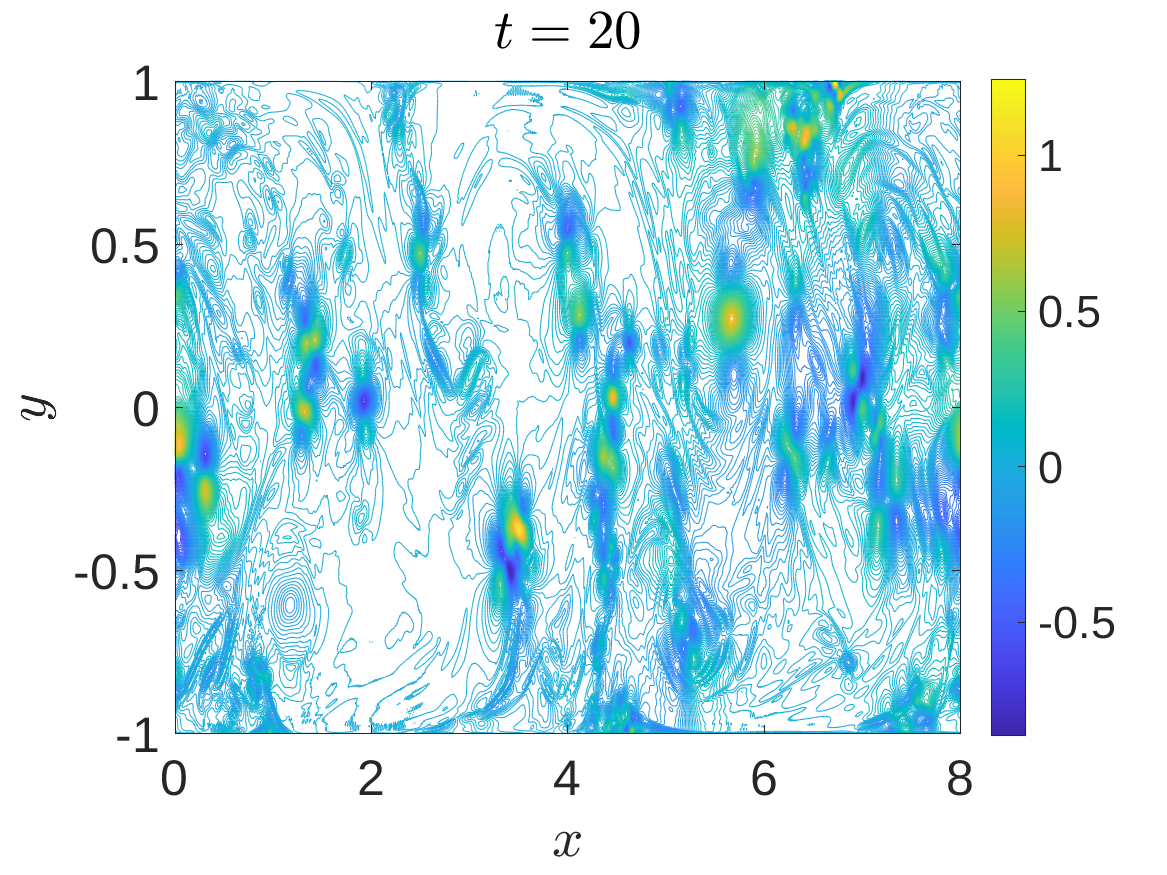}}
    \centerline{
    \includegraphics[width=0.5\textwidth]{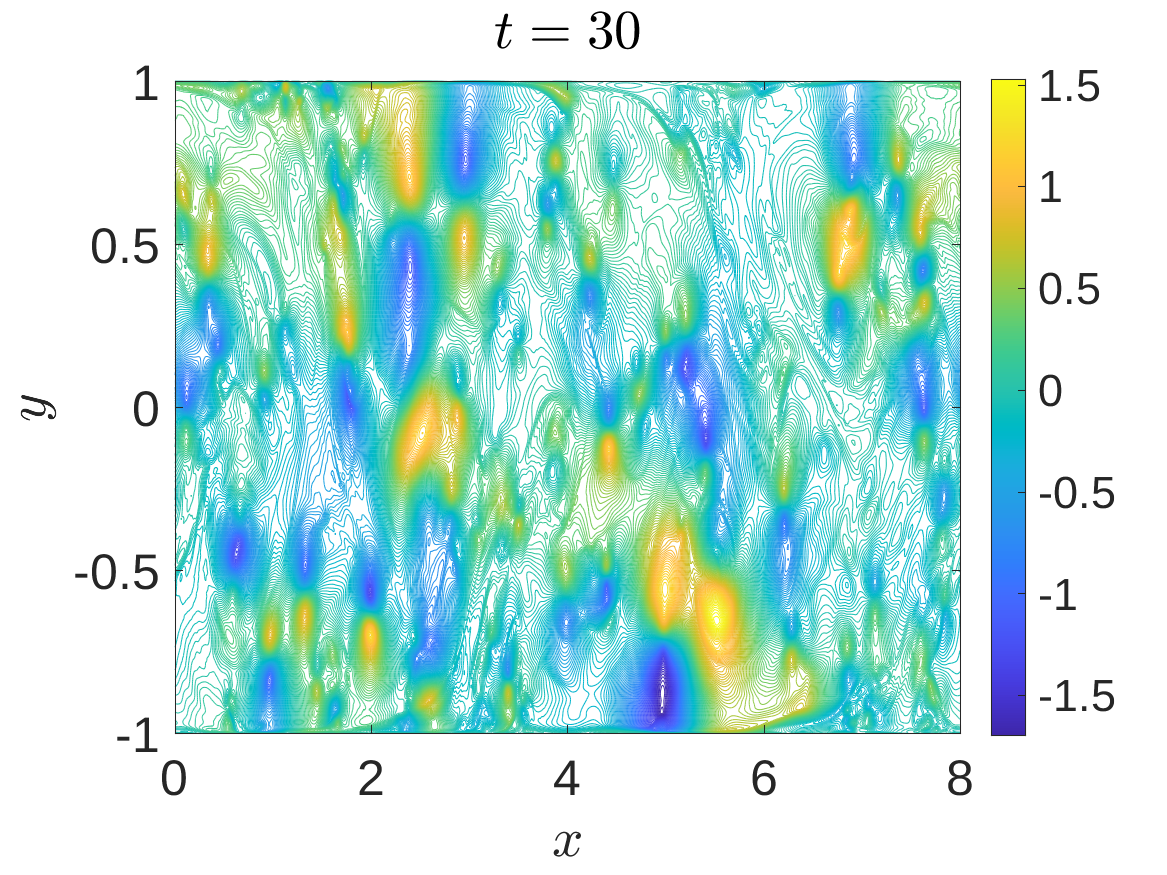} 
    \includegraphics[width=0.5\textwidth]{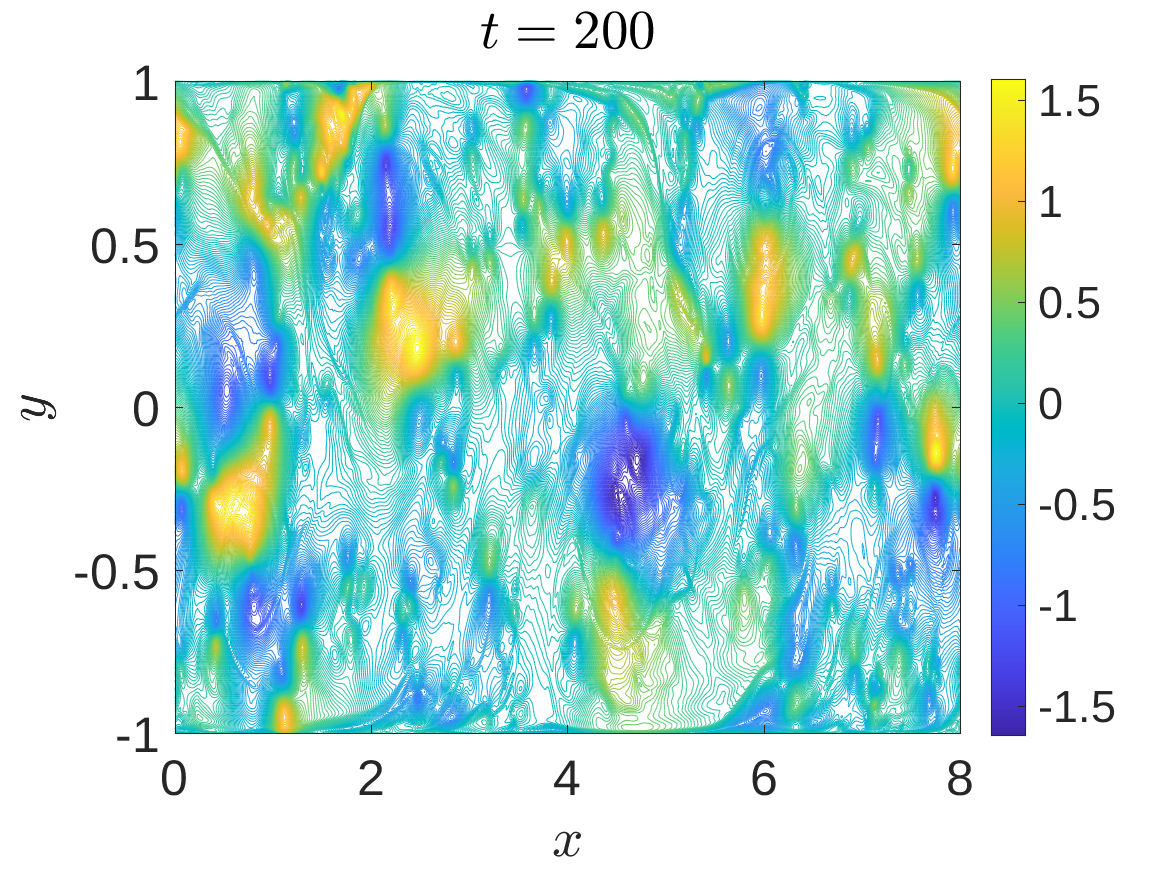}}
    \caption{\label{fig:14} Snapshots of the streamwise velocity component of the perturbation field taken from the full non-linear Navier-Stokes experiment with the F2 state. At time $t = 10$ the unstable structure is still growing until time $t = 20$ when it reaches its maximum size and starts to spread out through the channel, at time $t = 30$ the perturbation saturates and shows no qualitative change at a much later time $t = 200$.}
 \end{figure}

Repeating the experiment (with $E_{pert} = 10^{-5} E_{base}$) but now time-stepping the perturbation field with the full Navier-Stokes equations, we see the perturbation energy saturate after an initial transient: see  Fig. \ref{fig:QLvsDNS} (bottom). Importantly, and as anticipated, the statistics of the turbulent state F2 recover albeit slowly. Figure \ref{fig:13} shows the relative difference between the correlation matrices averaged over a rolling time window between the perturbed and unperturbed F2 state. The statistical differences ebb away after an initial transient of growth. Since F2 is  statistically linearly stable, this initial growth indicates that the statistical evolutionary operator linearised around the state F2 fixed point has to be non-normal. Finally, Fig. \ref{fig:14} shows the evolution of the streamwise perturbation velocity field. We see that the highly-localised structure visible at $t=10$ spreads in the channel at $t=20$ and reaches its saturated state at $t=30$ which does not change qualitatively even at much longer times $t=200$. It takes much longer for the statistics to recover suggesting the adjustment along the attractor happens much quicker than normal to it: see Figure \ref{fig:13}.

%
%


\section{\label{VI}Discussion}

%
%

\subsection{Summary}

We first summarise what has been done in the paper. The motivating objective has been to develop a theoretical approach which can assess the statistical stability of a turbulent state and goes beyond just examining the mean flow as Malkus did in the 1950s \citep{Malkus56}. That such an approach should exist in some form is predicated on the fact that the statistics of a  disturbed turbulent flow always return to their undisturbed values if the disturbance is not too large regardless of whether it decays or not. That is, a turbulent flow is assumed {\em linearly} statistically stable (note this doesn't preclude the possibility of the statistical disturbance initially increasing before ultimately decaying away as the linear statistical operator may be non-normal). The way forward is to include more statistical information than just the mean of the flow in the analysis and the simplest such is the 2nd rank cumulant or 2-point equal-time correlations of the flow fluctuations. To do this, we have proposed to examine the statistical stability of the turbulent flow from within a cumulant equation framework truncated to just consider these statistical quantities (called CE2). Even in CE2, however, the ensuing statistical spectral problem is so unwieldy that a substitute has had to be sought. Here the well-known connection between CE2 and the Quasilinear (QL) equations can be exploited and we have argued that a suitably-designed QL spectral problem defined in \S\ref{III} can act as a good proxy to detect statistical instability. The key in designing this QL problem is to translate a statistically-steady base flow in CE2 into a representative {\em steady} base flow in the QL problem with no pretence that this  base flow satisfies the QL equations itself {(the QL equations offer an  approximation of uncertain accuracy to the Navier-Stokes equations after all).}

The issue of how the statistical spectral problem in CE2 is related to the QL spectral problem is an intriguing one which seems not to have been discussed before. It is important because it underpins the tacit assumption that a CE2 simulation starting from given initial conditions should exactly shadow (statistically) the equivalent QL simulation given one is derivable from the other \citep[e.g.][]{Marston08,Tobias11, Tobias13}. However, this ignores the difference in dimensionality between the two formulations and the possibility of triggering unstable `non-physical' perturbations in the larger CE2 problem (meaning perturbations which cannot exist in the QL setting). In Appendix \ref{appendixA} we have tried to clarify the different types of instability which exist in these formulations - Type A and B - and how they are related. In particular, for Type A instabilities, the spectral QL problem captures the most unstable disturbance possible in the CE2 system and so is an accurate proxy for CE2: CE2 can only be Type A unstable if QL is. For Type B, the situation is less certain. While it remains the case  that any Type B instability in QL is mirrored in CE2 and CE2 will likely have further Type B instabilities, it is not clear  whether CE2 can have a Type B instability and QL not. Given the as-yet-unreported consequence of this - CE2 could suffer a bifurcation away from QL - it is tempting to assume that this is not possible and then that QL also is a good proxy for Type B perturbations. With this one caveat, probing the QL spectral problem - named here Extended Orr-Sommerfeld analysis (EOS) - then appears the best way to examine the statistical stability of a flow where the  2-pt correlations as well as the mean profile of the flow are incorporated.

Despite improving upon the CE2 spectral problem, EOS is still unfortunately quite costly to implement so a reduced version was proposed in \S\ref{III.A} - minimally Extended Orr-Sommerfeld analysis (mEOS) - which has a similar cost to the usual Orr-Sommerfeld analysis of the flow mean and is the most straightforward extension of doing linear stability analysis only around the mean flow. Just as in OS-analysis one chooses a particular Fourier mode $(m,n)$ to analyse but now extra information about the corresponding base Fourier mode is included. mEOS was found to capture the qualitative effect of EOS with either both producing an improved estimate of statistical stability (e.g. the $m=2$ mode for state F1: see figure \ref{fig:EOSvsNOS} bottom left) or offering no correction of the OS eigenvalue (e.g. the $m=1$ mode for state F1). 
Reassuringly, both approaches never made a poorer statistical stability prediction than OS analysis at least over the turbulent channel flow states tested. 
\rk{mEOS works comparatively well because the structure of the key eigenmodes in the EOS problem tend to consist of one dominant (primary) fluctuation field interacting with a mean flow perturbation (see figure \ref{fig:evect_spectrum}). This is presumably because the comparatively small size of the cumulants or base fluctuation fields relative to the mean profile (e.g. indicated by the singular values in figure \ref{fig:SVD}) means that only the few fluctuation fields close to marginality on the mean flow play a role. This may well change with increasing $Re$ but the relative size of the cumulants should also increase possibly acting to deprioritise this near-marginal stability property. Clearly higher $Re$ flows warrant further study. }

\noindent
%
%
%
%
%
%
%
%

\subsection{Implications}

%
%
By incorporating 2-pt fluctuation correlations, EOS and its slimmed down version, mEOS, represent an attempt to generate a better way to assess the statistical stability of a turbulent flow than just considering the mean profile. Malkus (1956) wanted to use this to constrain the set of realisable turbulent mean profiles and it still remains an interesting requirement given the paucity of predictive alternatives. The concept of marginal stability still periodically rears its head in the literature to post-rationalise turbulent mean profiles (e.g. see \cite{Brauckmann17} and references therein) and EOS represents a finer tool for this. As noted in the introduction, the mean flow is increasingly taken as the starting point for resolvent flow analysis. To move this towards a full wall-bounded turbulence theory, a way has to be found to close the cycle between the fluctuation field and the mean. Just maybe, requiring statistical stability with respect to the dominant fluctuation fields could help constrain this.

%
%
A byproduct of this work is that it has also suggested an improved approach for predicting coherent structures in turbulent flows. {Coherence here means weakly damped or growing fluctuation {\em and} mean disturbances which emerge from the linear operator built around the base flow rather than, say, POD modes derived from the 2-pt correlations which have no direct relation to the underlying equations}. Initial work \citep{CrightonGaster76,Gaster85,Roshko93} linearising around the turbulent mean proved successful in free shear flows but less so when viscosity plays a leading role. Admittedly, EOS or mEOS needs extra (second cumulant) information about the base flow which adds cost to the analysis  but the potential for better predictions is clear. A popular (cheap) alternative approach has been to apply OS analysis around the mean profile which incorporates an eddy viscosity to acknowledge the presence of the base fluctuations \citep{ReynoldsHussain72, Sen00, Sen07,DelAlamo06, Cossu09,HwangCossu10a, HwangCossu10b}.  This eddy viscosity is taken to be that which is needed to sustain the turbulent mean profile and so makes no attempt to incorporate fluctuation information for each  excited wavenumber of the base flow as in EOS.  \rk{In the well-known triple expansion of Hussain \& Reynolds \citep{Hussain70}, the coherent wave part ($\tilde{f}$ in their equation (1.1)) is the equivalent  of $\bd \bfu^{mn}$ which there is excited by a wavemaker and then studied as it travels downstream in a turbulent flow. Orr-Sommerfeld analysis was found unable to reconcile what was seen experimentally but EOS should do better. }

%
%
The cumulant framework examined here to motivate EOS was used only to the first non-trivial level where only the first and second cumulants are retained. Whether this represents a good approximation or not to modelling the turbulent flow itself is a question not considered here. Rather the focus is on the associated spectral problem characterising the statistical stability {\em within} this framework {\em taking} the turbulent base flow as a given {from DNS or experiment}. Clearly considering higher cumulants beyond the the second would be better but the computational cost of handling the ensuing spectral problem becomes prohibitive. Interestingly, `projecting down' to an underlying Navier-Stokes-type problem fails immediately unless no closure at all is performed which means CE$\infty$ is considered. This gives what we have called the infinitely Extended Orr-Sommerfeld analysis (iEOS) which is as costly as EOS and unfortunately does not have a `minimal' version. Again, only targeting certain subsets of fluctuation fields - e.g. key triads suggested by resolvent flow analysis - may yet make this tractable.

\subsection{Future perspectives}

In terms of future work, an obvious focus is to improve the main approximation in EOS which is  deriving a representative base flow. Here, we have used the leading singular vector of the second cumulant for each excited (vector) wavenumber but other strategies involving information from higher order cumulants are certainly possible (leaving aside the cost of computing these...). This would seem sensible in the case of iEOS which is a proxy for CE$\infty$ where {\em all} the cumulants are retained and the only other cumulant system beyond CE2 which allows a Navier-Stokes-type proxy. Pragmatically, a steady base flow has been sought which leads to a (conceptually) simple eigenvalue problem. Trying to go beyond this profoundly complicates matters.

One interesting future direction hinted at in the introduction is to relax the averaging procedure  to just being over the streamwise direction or, more generally, redefine what constitutes the `mean'. A key feature we have been able to exploit is the connection between CE2 and the QL equations which still holds if the mean is more flexibly defined 
to include some wavenumber subset of the fluctuation field. The then `generalised' QL equations \citep{Marston16} have the advantage of including more of the nonlinearity in the Navier-Stokes equations and can interpolate between the QL equations and the full Navier Stokes equations depending on how the mean is defined. A minimal mean extension would be to include one spanwise wavenumber field whereas including {\em all} fluctuation fields with no streamwise variation is equivalent to a mean defined by only streamwise averaging. There are many theoretical reasons for considering the latter, where the mean is $U(y,z)\bhx$ rather just $U(y)\bhx$, in wall-bounded flows (e.g. the SSP/VWI mechanism \cite{Waleffe97, Hall10}) but capturing it experimentally is very time-consuming task and using it reduces the predictive power of resolvent flow analysis. Even in computations, the spanwise homogeneity needs to be broken otherwise there is no reason to suppose the mean flow should not be spanwise-invariant (and then the spanwise structure of the mean profile needs to be divorced from the source of spanwise inhomogeneity). Nevertheless, this choice is starting to  receive attention at least at the level of doing OS analysis on $U(y,z)\bhx$ (see \cite{Lozano-Duran21} and references herein) and this analysis can be extended as envisaged here.

Many challenges remain. Statistical stability is an important fundamental feature of turbulent flows yet is difficult to assess and hence to exploit. Hopefully, the work reported here represents a useful but admittedly small step forward.


\appendix
\section{\label{appendixA}Relationship between dynamic stability in QL and statistical stability in CE2}

In this appendix we argue that the dynamical stability problem in QL based on $\bfu_0^{mn}(y)$ can be used to predict statistical instability within  CE2 based upon $\bC^{mn}(1,2)= \bfu_0^{mn} (y_1) \otimes \bfu_0^{-m-n} (y_2)$. This is because any instability present in QL has an equivalent in CE2. We have  been unable to establish the converse however plausible: a statistical instability in CE2 does not necessarily imply a dynamical instability in QL. Despite this lack of complete equivalence, the main text takes the stability problem in QL to be a good proxy of that in CE2.


There are two types of instability in QL and CE2. To identify them, we define two sets: $\Lambda$ which contains all the Fourier modes that contain energy in the base state $\bfu_0$, and $\delta \Lambda$ which does the same for the perturbation field $\bd \bfu$, i.e.
\begin{equation}
\Lambda:=\biggl\{(m,n)\,\, \biggl| \int^1_{-1} |\,\bfu_0^{mn} \,|^2 \, dy > 0 
\biggr. \biggr\}
\quad \& \quad
\dd \Lambda:=\biggl\{(m,n)\,\, \biggl| \int^1_{-1} |\,\bd \bfu^{mn}\,|^2 \, dy > 0 
\biggr. \biggr\}.
\end{equation}
There are then two types of eigenfunction for the linear stability problem in QL (the $\dd$QL problem):
\begin{itemize}
\item Type A: $\dd \Lambda \subseteq \Lambda'$, the complement of $\Lambda$ which is the set of all possible {\em un}excited Fourier modes in $\bfu_0$.
\item Type B: $\dd \Lambda \subseteq \Lambda$ when there is an exact equivalent eigenfunction for $\dd$CE2 (the statistical stability problem in CE2).  
\end{itemize}
 Type A is the most straightforward to understand as there is no feedback of the fluctuation field onto the mean in $\dd$QL (the feedback is quadratic in the fluctuation disturbance). In this case, the stability problem simplifies to an Orr-Sommerfeld-type problem on the base mean profile. As a result, a Type A eigenfunction $\dd \bfu_A$ in $\dd$QL consists of just one wavenumber pair $(\hat{m},\hat{n})$. The situation is similar in CE2 with the corresponding eigenmatrix $\dd C_A=|\dd \bfu_A|^2$ but with two notable  differences. The first is the corresponding eigenvalue in CE2 is $2\Re e(\lambda_A)$ where $\lambda_A$ is the QL-eigenvalue (\,$\partial_t |\dd \bfu_A|^2=(\lambda_A \bfu_A)\bfu_A^*+\bfu_A(\lambda^*_A \bfu_A^*)=(\lambda_A+\lambda_A^*)|\dd \bfu_A|^2$\,).
 The second is that a mean flow perturbation is forced, 
 \begin{equation}
 {\color{blue}\dd U_A} = \biggl[2\Re e({\lambda_A})-\frac{1}{Re} \partial_y^2\biggr]^{-1}
 \partial_y \dd C^{\hat{m}\hat{n}}_{A\,12}(y,y)
 \label{A:forced mean}
 \end{equation}
 (in the simpler constant-pressure gradient situation), which in turn forces perturbation components across {\it all} wavenumber pairings in $\Lambda$ since
 \begin{align}
 \dd C_{A\,ij}^{mn}(1,2)=& {\cal L}(\lambda_A)^{-1} \biggl[ 
 im [\,{\color{blue}\dd U_A(2)}- {\color{blue}\dd U_A(1)}\,] C_{ij}^{mn}(1,2)
 - {\color{blue}\partial_y\dd U_{A}(2)} C_{i2}^{mn}(1,2) \delta_{1j}
                         \nonumber \\
                         & \hspace{4cm}-{\color{blue}\partial_y \dd U_{A}(1)} C_{2j}^{mn}(1,2) \delta_{i1}
 \biggr]
 \label{A:forced waves}
 \end{align}
 where ${\cal L}(\lambda_A)$ is the spatial operator defined by the system (\ref{dC_1}) and (\ref{dC_incompress}) with $\partial_t$ replaced by $\lambda_A$. Importantly if a Type A QL eigenvalue is unstable - i.e. $\Re e({\lambda_A})>0$ then so is the corresponding CE2 eigenfunction. 

The eigenvalue problem for Type B is more complicated since it involves  multiple wavenumber pairs coupled to a mean field perturbation. However, it is straightforward to see that a Type B eigenfunction in $\dd$QL,
\begin{equation}
\biggl[\,\dd U_B(y,t), \,\bd \bfu_B(x,y,z,t)\,\biggr]=\biggl[\,\dd U_B(y),
\sum_{(m,n) \,\in \,\dd \Lambda \,\subseteq \,\Lambda} 
\bd {\bf v}_B^{mn}(y) e^{i(m \alpha x+n \beta z)}\, \biggr] \,\,e^{\lambda_B t}
\end{equation}
with eigenvalue $\lambda_B$ corresponds  to the eigenfunction 
\begin{align}
\biggl[\,\dd U_B, \,\dd \bC_B(y_1,y_2)=&\sum_{(m,n) \,\in \,\dd \Lambda \,\subseteq \,\Lambda} \dd \bC^{mn}_B(y_1,y_2):= \bfu_0^{mn}(y_1) \otimes \dd \bfu_B^{-m-n}(y_2) \nonumber \\
& \hspace{4cm}+ \dd \bfu_B^{mn}(y_1)\otimes \bfu_0^{-m-n}(y_2) \,\biggr] \,e^{\lambda_B t}
\end{align}
with same eigenvalue in the CE2 system. 
There are no hybrid eigenfunctions in $\dd$QL with $\dd \Lambda \not \subseteq \Lambda$ as those wavenumber pairs not in $\Lambda$ simply decouple even if they have the same eigenvalue (which is then degenerate).

%
%
It is worth briefly illustrating this partitioning into Type A and Type B perturbations in a simple example. Let $q(x,t):=\mq(t)+\fq(x,t)$ over the periodic domain $x\in [-\pi,\pi]$ where $\overline{(\,\cdot\,)}={1 \over 2\pi}\int^{\pi}_{-\pi} (\,\cdot\,) \,dx $ so $\overline{\fq}=0$ (and we assume $\fq(-x)=\fq(x)$ for simplicity) then consider the QL-like system
\begin{align}
\dot{\mq}+\frac{1}{Re}\mq -\frac{3}{Re^2}&=-2\overline{q^{\prime 2}}=-\sum_{n=1}^\infty  q_n^{\prime 2} \qquad {\rm where} \quad \fq:=\sum_{n=1}^\infty \fq_n(t) \cos nx \label{QLmodel_1}\\
\dot{q_n^\prime}                                &=\biggl[\mq-\frac{n^2}{2Re}\biggr]\fq_n \qquad n=1,2,\ldots \label{QLmodel_2}
\end{align}
where the mean flow is forced by an `imposed pressure gradient' $3/Re^2$, there are the usual dissipation terms and  quadratic feedback of the fluctuations onto the mean.  There are multiple steady states: we choose to linearise around
\begin{equation}
\mq=\frac{2}{Re}, \quad \fq_2=\frac{1}{Re}, \quad \fq_n=0\quad n \neq 2.
\end{equation}
which leads to the linear perturbation equations
\begin{align}
\dot{\delta \mq}                                            &=-\frac{1}{Re}\delta \mq -2\fq_2 \delta \fq_2, \\
\dot{\delta q_2^\prime}                                &=\fq_2 \delta \mq,\\
\dot{\delta q_n^\prime}                                &=\biggl[\mq-\frac{n^2}{2Re}\biggr]\delta q_n^\prime  \qquad n \neq 2.
\end{align}
These highlight the difference between  `new' wavenumbers ($n \neq 2$) and `already-present' wavenumbers ($n=2$). Using the base flow and restricting to the perturbation to just $n \in \{1,2\}$, then
\begin{equation}
\frac{d}{dt}\left[\begin{array}{c}   
\delta \mq\\ \delta q_1^\prime \\ \delta q_2^\prime 
\end{array}
\right]
=
\left[\begin{array}{ccc}   
-\frac{1}{Re} & 0 & -2q_2^\prime \\
0 & \mq-\frac{1}{2Re} & 0 \\
q_2^\prime & 0 & 0 
\end{array}
\right]
\left[\begin{array}{c}   
\delta \mq\\ \delta q_1^\prime \\ \delta q_2^\prime 
\end{array}
\right]
=
\frac{1}{Re}\left[\begin{array}{rrr}   
-1 & 0 & -2 \\
0 & \frac{3}{2} & 0 \\
1 & 0 & 0 
\end{array}
\right]
\left[\begin{array}{c}   
\delta \mq\\ \delta q_1^\prime \\ \delta q_2^\prime 
\end{array}
\right] \label{QLmodel}
\end{equation}
The associated {\it statistical equations} for $\mq$ and $C:= \overline{q^{\prime 2}}$ are 
\begin{align}
\dot{\mq}+\frac{1}{Re}\mq  &=\frac{3}{Re^2}-2C=\frac{3}{Re^2}-\sum_{n=1}^\infty C_n \qquad {\rm where} \quad C_n:=q_n^{\prime 2}\\
\dot{C_n}                                &=2\biggl[\mq-\frac{n^2}{2Re}\biggr]C_n \qquad n=1,2,\ldots
\end{align}
The basic state in this formulation is $\mq=2/Re$ and $C_2=1/Re^2$ (all other $C_n=0$) and the linear `statistical' stability equations (again just for $n \in\{1,2 \}$) are
\begin{align}
\dot{\delta \mq}                                            &= -\frac{1}{Re}\delta \mq -\delta C_1-\delta C_2, \\
\dot{\delta C_2}                                            &=2 C_2 \delta \mq,\\
\dot{\delta C_1}                                            &=2\biggl[\mq-\frac{1}{2Re}\biggr]\delta C_1. 
\end{align}
This leads to the problem
%
%
%
\begin{equation}
\frac{d}{dt}\left[\begin{array}{c}   
\delta \mq\\ Re\delta C_1 \\  Re\delta C_2 
\end{array}
\right]
=
\frac{1}{Re}
\left[\begin{array}{rrr}   
-1 & -1 & -1 \\
0 & 3 & 0 \\
2 & 0 & 0 
\end{array}
\right]
\left[\begin{array}{c}   
\delta \mq\\ Re\delta C_1 \\ Re\delta C_2 
\end{array}
\right]
\label{CE2}
\end{equation}
which is similar but different to that in (\ref{QLmodel}). Ignoring the factor of $1/Re$, the `QL' eigenvalues are $\lambda_A=3/2$ (Type A), $\lambda_B=\half(-1+i\sqrt{7})$ and $\lambda_B^*$ (Type B)  with corresponding eigenvectors (respectively)
\begin{equation}
\left[
\begin{array}{c}
0 \\ 1 \\ 0
\end{array}
\right], \quad 
\left[
\begin{array}{c}
\lambda_B \\ 0 \\ 1
\end{array}
\right]
\quad \& \quad
\left[
\begin{array}{c}
\lambda_B^* \\ 0 \\ 1
\end{array}
\right]
\end{equation}
whereas the  `CE2' eigenvalues are $2\Re e(\lambda_A)=3$ (Type A), $\lambda_B$ and $\lambda_B^*$ (Type B) with eigenvectors (respectively)
\begin{equation}
\frac{1}{\lambda_A^2+\lambda_A+2}\left[
\begin{array}{c}
-2 \\ \lambda_A^2+\lambda_A+2 \\ -\lambda_A
\end{array}
\right], \quad 
\left[
\begin{array}{c}
\half \lambda_B \\ 0 \\ 1
\end{array}
\right]
\quad \& \quad
\left[
\begin{array}{c}
\half \lambda_B^* \\ 0 \\ 1
\end{array}
\right]
\end{equation}
This example shows: a) the partitioning of the perturbations into Type A and Type B, and that b) CE2 has double the corresponding Type A QL eigenvalue (as it is real) and the corresponding CE2 eigenfunction contains other forced components beyond the new wavenumber $n=1$. However, these extra components, which get excited in QL at (next) quadratic order, are passive and don't affect the eigenvalue.
%
%

Now we discuss how the  $\dd$QL problem is a strict subset of the $\dd$CE2 problem. Things are clearest for the $\dd$CE2 Type A case where the eigenvalue problem is solely determined by the statistical stability equation for the 2nd rank cumulant,  
\begin{align}
\partial_t \,\dd C_{ij}^{mn}(1,2)  &=  
[\bfu^{mn}(y_1,t)]_i \biggl[ \optL^{mn}_2(U) \bd \bfu^{-m-n}(y_2,t) \biggr]_j
                    +\biggl[ \optL^{mn}_1(U) \bd \bfu^{mn}(y_1,t) \biggr]_i                                        [\bfu^{-m-n}(y_2,t)]_j
                       \label{dC_3}\\
 0 &= \left[\begin{array}{c} 
im \alpha\\ \partial_1 \\ in \beta \end{array}\right]_i \dd C_{ij}^{mn}(1,2)
\label{C_incompr}
\end{align}
Now if ${\bf e}^{(p)}$ is the $p^{th}$ eigenfunction to the Orr-Sommerfeld problem with corresponding eigenvalue $\lambda_p$, it is straightforward to see that $\delta C_{ij}(1,2) = {\bf e}^{(p)}_i(1) {\bf e}^{*(p)}_j(2)$ is a eigenfunction of (\ref{dC_3})-(\ref{C_incompr}) with the corresponding eigenvalue $\lambda_p+\lambda_p^*=2\Re e(\lambda_p)$ as already mentioned (just above (\ref{A:forced mean})\,). Less clear but still true is the fact that $\delta C_{ij}(1,2) = {\bf e}^{(p)}_i(1) {\bf e}^{*(q)}_j(2)$ is also an eigenfunction with eigenvalue $\lambda_p+\lambda_q^*$ which has no counterpart in the Orr-Sommerfeld problem as it does not correspond to one physical flow field (note: 1. it has also forced components via equations (\ref{A:forced mean}) and (\ref{A:forced waves}); and 2. that for complex $\lambda_p+\lambda_q^*$, $\delta C_{ij}$ is not Hermitian but the average with its Hermitian conjugate - an eigenfunction with  eigenvalue $\lambda_p^*+\lambda_q$ - is). 
This has the consequence that CE2 has at least the number of unstable directions as the OS problem and probably more since an unstable eigenvalue linearly combined with a weakly stable eigenvalue could produce a new unstable direction. That is, if the OS problem is unstable so is CE2 and vice versa. 
%
%
A simple example makes this clear: consider the QL (fluctuation) system $\dot{x}=2x$ and $\dot{y}=-y$ which has 1 stable and 1 unstable direction. The equivalent CE2 system where $C_{xx}:=x^2;\, C_{xy}=C_{yx}:=xy$  and $C_{yy:}=y^2$ (and the symmetry $C_{xy}=C_{yx}$ built in)
\begin{equation}
\partial_t \left[ \begin{array}{c}  
C_{xx} \\ C_{xy} \\ C_{yy}
\end{array} \right]= 
\left[ \begin{array}{rrr}  
4 & 0 &  0  \\
0 & 1 &  0  \\
0 & 0 & -2  
\end{array} \right] 
\left[ \begin{array}{c}  
C_{xx} \\ C_{xy} \\ C_{yy}
\end{array} \right]
\end{equation}
has 2 unstable and one stable. The point is in CE2, it is possible to have a perturbation of form $[0\, \,  1\,\, 0]^T$ in the extra unstable direction but this perturbation has no equivalent in the OS problem as it is inconsistent with any one choice of  $x$ and $y$.
This feature holds for any Type A perturbation in CE2 as long as the corresponding velocity is a vector  i.e. there is non-trivial structure in the cross-stream direction (this was absent by design in the previous example (\ref{QLmodel_1})-(\ref{QLmodel_2})\,). This is also clear from the dimensions of the respective problems - the QL eigenvalue problem is $O(N_y \times N_y)$ whereas the corresponding CE2 problem is $O(N_y^2 \times N_y^2)$ so it is clear there have to be  $O(N_y^2-N_y)$ additional eigenvalues in $\dd$CE2. 

Importantly, in Type A perturbations, it is clear that that any additional instabilities in $\dd$CE2 are dependent on the existence of an instability in $\dd$QL {\it and} are also always weaker (ie. their growths rates are smaller) than this QL instability. By exactly similar reasoning, there is also the possibility of additional unstable Type B perturbations in $\dd$CE2. However, here it is less straightforward to deduce their reliance on a QL instability being present or that this dominates their growth rates due to the mean flow perturbation feeding back onto the fluctuations equations. Nevertheless, this seems a plausible assumption: if it is not true, CE2 could exhibit a Type B bifurcation when QL does not. Computations to illustrate some of these features have now been done \citep{Nivarti22}.

In summary, the instability of the QL system is sufficient to conclude the same for the larger CE2 problem whereas concluding stability of CE2 given the stability of QL is true for Type A perturbations but remains a conjecture for Type B perturbations.


\section{\label{appendixB}Solving the eigenvalue problem}

Here we explain how we implemented mEOS and EOS calculations as an eigenvalue problem in 2D channel flow. We expand fluctuation field in terms of sine and cosine modes:
\begin{equation}
\bfu(x,y,t):= \sum_{m=1}^{N_x} \left( \bfu^{m0}_c(y,t) \cos{(m \alpha x)} + \bfu^{m0}_s(y,t) \sin{(m \alpha x)} \right)
\end{equation}
The non-linear term in the mean equation then becomes:
\begin{equation}
     ( \,\overline{\tilde{u} \tilde{v}} \,)_y= \frac{1}{2} \partial_y \sum_{m=1}^{N_x} \left( \tu^{m0}_c \tv^{m0}_c  + \tu^{m0}_s \tv^{m0}_s  \right).
\end{equation}
The Extended Orr-Sommerfeld equations then take the following form:
\begin{equation}
\begin{aligned}
    \partial_t\delta U&=\frac{1}{Re}\partial^2_y \delta U+\delta G - \frac{1}{2} \partial_y \sum_{m=1}^{N_x} \left( \dd \tu^{m0}_c \tv^{m0}_c + \tu^{m0}_c \dd \tv^{m0}_c + \dd \tu^{m0}_s \tv^{m0}_s + \tu^{m0}_s \dd \tv^{m0}_s  \right),
    \end{aligned}
    \label{eq:EOS_mean_B}
\end{equation}
where the scalar variable $\delta G$ is used to impose zero-flux condition on the mean velocity perturbation $\delta U$: $  \int^1_{-1} \,\delta U \, dy =0$,  
and for each wavenumber $(m,0)$:
\begin{gather}
    \partial_t{\dd \tu}^{m0}_c  = \frac{1}{Re} \left( \partial^2_y - m^2 \right) {\dd \tu}^{m0}_c - m {\dd \tp}^{m0}_s -  m U {\dd \tu}^{m0}_s - U_y {\dd \tv}^{m0}_c - m \dd U {\tu}^{m0}_{0s} - \dd U_y {\dd \tv}^{m0}_{0c}\\
    \partial_t{\dd \tu}^{m0}_s  = \frac{1}{Re} \left( \partial^2_y - m^2 \right) {\dd \tu}^{m0}_s + m {\dd \tp}^{m0}_c +  m U {\dd \tu}^{m0}_c - U_y {\dd \tv}^{m0}_s + m \dd U {\tu}^{m0}_{0c} - \dd U_y {\dd \tv}^{m0}_{0s}\\   
    \partial_t{\dd \tv}^{m0}_c  = \frac{1}{Re} \left( \partial^2_y - m^2 \right) {\dd \tv}^{m0}_c - \partial_y {\dd \tp}^{m0}_c -  m U {\dd \tv}^{m0}_s  - m \dd U {\tv}^{m0}_{0s} \\
    \partial_t{\dd \tv}^{m0}_s  = \frac{1}{Re} \left( \partial^2_y - m^2 \right) {\dd \tv}^{m0}_s -\partial_y  {\dd \tp}^{m0}_s +  m U {\dd \tv}^{m0}_c + m \dd U {\tv}^{m0}_{0c} \\ 
    m\tu^{m0}_s+\partial_y \tv^{m0}_c = 0  \\
    -m\tu^{m0}_c+\partial_y \tv^{m0}_s = 0   
    \label{eq:EOS_fluct_B}
    \end{gather}
The eigenvalue problem is discretised using Chebyshev extreme points with the same resolution as in base flow simulations. The no-slip boundary conditions on the velocity variables are imposed by replacing the appropriate rows of the matrix.

This eigenvalue problem is real and so eigenvalues are either real or come in complex conjugate pairs where   a real physical velocity field is formed by addition: 
\begin{equation}
\left[ \begin{array}{c}
     \dd U  \\
     \vdots\\
      \bd \bfu^{m0}_c \cos(m\alpha x) \\
      \bd \bfu^{m0}_s \sin(m\alpha x)\\
      \vdots
\end{array} \right] e^{\lambda t} + 
\left[ \begin{array}{c}
     \dd U^*  \\
     \vdots\\
      \bd \bfu^{*m0}_c \cos(m\alpha x) \\
      \bd \bfu^{*m0}_s \sin(m\alpha x)\\
      \vdots
\end{array} \right] e^{\lambda^* t}.            
\end{equation}


\backsection[Acknowledgement]{\rk{We thank one referee for a very helpful report which improved the presentation of the paper.}}

\backsection[Funding]{VKM acknowledges financial support from EPSRC through a studentship.}
\backsection[Declaration of interests]{ The authors report no conflict of interest.}

\bibliographystyle{jfm}
\bibliography{references}
\end{document}